# Nonlinearity-preserving quantum simulation of flows in the early fault-tolerant regime

Hansol Noh[1,*], Unam Oh[1], Sangkyu Baek[1], Doyeol (David) Ahn[1,2,*]

[1] Department of Electrical and Computer Engineering and Center for Quantum Information Processing, University of Seoul, Seoul 02504, Republic of Korea.

[2] Singularity Quantum Inc., Newport Beach, CA, USA.

*Correspondence to Hansol Noh (hs.noh9003@gmail.com) and Doyeol (David) Ahn (dahn@uos.ac.kr).

## Abstract

Simulating nonlinear flow dynamics from unstable growth to stable saturation within the limits of early fault-tolerant quantum computing (EFTQC) remains a challenge. Linear transport of the ensemble density offers a route to quantum simulation without linearizing the dynamics, but its circuits can become costly as the grid is refined. We show that, for low-degree polynomial reduced-order models fitted by sparse identification of nonlinear dynamics (SINDy), Walsh-synthesized Koopman–von Neumann (KvN) transport needs per-step rotations at most polylogarithmic in cells per coordinate, with degree as exponent. A radially reduced quadratic cylinder-wake model approaching saturation fits our EFTQC window: surface-code estimates of 1.62 h on 560,000 qubits, including preparation, propagation and sampling to 1% root-mean-square field error against the continuum radial model. Three-coordinate runs also fit, taking hours to weeks. Four-qubit IBM experiments demonstrate nonlinear radial KvN transport. Balancing accuracy against circuit cost makes polynomial degree a design variable for EFTQC.

Nonlinear dynamics can make individual trajectories sensitive to small uncertainties in initial conditions, so reproducing attractor dynamics can be more informative than tracking a particular trajectory[1]. Ensemble prediction describes how this uncertainty evolves by propagating a probability distribution over possible states[2]. In the wake behind a circular cylinder at Reynolds number 100, disturbances grow from an unstable steady flow until nonlinearity saturates them into stable periodic vortex shedding[3,4]. Capturing both growth and saturation requires retaining the nonlinear dynamics: linearization about the steady flow captures the initial growth but not the amplitude and frequency of the saturated shedding[3]. The ensemble density obeys a linear transport equation whose drift retains these nonlinear dynamics[5,6], so a quantum computer can propagate it without linearizing the flow. The challenge is to do so on early fault-tolerant hardware.

Near-term studies have demonstrated selected flow processes on quantum processors[7,8], and a learned Koopman model implements linear observable evolution in circuits with nonlinear classical encoding and decoding[9]. Quantum lattice Boltzmann and Trotterized transport schemes treat prescribed velocity fields in which each component is independent of its own coordinate[10,11]. Such divergence-free drift excludes contraction of an ensemble onto a stable state. Other schemes measure, reconstruct and re-prepare the distribution between steps[10,12], and this repeated exchange with a classical processor can erode a potential quantum advantage[13,14].

Carleman linearization and KvN embedding offer fault-tolerant routes to nonlinear drift[5,15], with potential exponential advantage over deterministic Eulerian Liouville solvers for sparse KvN Hamiltonians[5]. Grid refinement can increase the step count or operator norm of Carleman solvers[16,17] and the matrix-element scale in sparse KvN Hamiltonian simulation[5]. An upwind KvN circuit using a linear combination of Hamiltonian simulations[18] (LCHS–KvN) suppresses grid-scale oscillations using about $10^8$ single-target multi-controlled gates for ten steps on a 128-cell grid; its authors note that implementation requires fault-tolerant quantum hardware[19].

Bridging the gap between algorithmic advances and end-to-end feasibility in EFTQC requires controlling circuit costs at the target resolution and their growth under grid refinement. Unstable and stable states make both harder: the drift varies across the grid, so each cell needs its own transport rate, and contraction can demand finer grids. Quantum error correction (QEC) supports coherent computation before large-scale fault-tolerant hardware becomes available[20]. Below-threshold error correction has been demonstrated[21], and IBM's roadmap[22] targets a fault-tolerant processor for 2029. Logical-qubit budgets constrain the number of encoded coordinates, and circuit depth limits coherent propagation steps. The resource budget also covers initial-state preparation and the shots needed for accurate observable estimates.

Without model reduction, an end-to-end analysis of Carleman-linearized lattice-Boltzmann dynamics on a spatial grid suggests a modest quantum advantage for selected observables at high error tolerance[17]. Reduced-order models (ROMs) replace the spatial field with a few coordinates and support repeated predictions for flow-control design and parametric studies[3,23,24]. Eliminating the unresolved variables, however, generally introduces memory and fluctuating-force terms[25]. With such model error and uncertain initial conditions, a single trajectory is not a reliable prediction, which motivates ensembles[26]. A KvN embedding propagates such an ensemble directly, on a register whose width is set by the retained coordinates[5,27,28]. Here we reduce the flow before KvN embedding, so the ROM determines the circuit as well as the physical approximation. We show that its polynomial degree sets how rotations per step grow with the grid, so ensemble prediction[2,29–31] on EFTQC hardware calls for a low-degree drift.

Flow equations are naturally low-degree: incompressible Navier–Stokes advection is quadratic, and standard Galerkin projection preserves this polynomial degree[4]. Quadratic and cubic models have been developed for wake dynamics[3], reactor oscillations[32], neuron activation and compressible flow[33], and nonlinear structural vibrations[34], while polynomial maps estimate drag and lift from wake-mode amplitudes[35]. We fit the reduced equations using SINDy with a prescribed low-degree polynomial library[1,36], selecting the degree by balancing model accuracy against KvN circuit cost. We test whether the model retains growth, saturation and an amplitude-dependent oscillation frequency[4,37].

Walsh synthesis encodes polynomial drift in rotation angles, with per-step rotation counts at most polylogarithmic in cells per coordinate at fixed dimension and degree. We encode the amplitude-dependent oscillation frequency in the same way, using the shedding amplitude to control rotations on two phase qubits. Each shot then proceeds coherently from a single ensemble preparation to the final radius and phase measurements used for field reconstruction. We optimize fault-tolerant designs and compare them, for the same task, with a block-encoding circuit that loads coordinates as amplitudes[38]. Surface-code estimates[39–41] cover preparation, propagation and sampling to a target of 1% relative root-mean-square (RMS) field error against the continuum radial model. On IBM hardware, we demonstrate nonlinear radial transport by measuring ensemble observables as the flow approaches an attracting state.

# Results

## Cylinder-wake dynamics and quantum KvN circuit

An ensemble spreads near an unstable fixed point and concentrates near a stable one (Fig. 1a). For the wake, we retain these dynamics with two oscillatory proper orthogonal decomposition (POD) modes that describe shedding and a shift mode that describes the change in mean flow[3]. We propagate an initial ensemble with a circuit derived from the fitted equations, reconstruct the velocity field from its measurements, and use circuit and shot counts to estimate physical qubits and runtime (Fig. 1b–f).

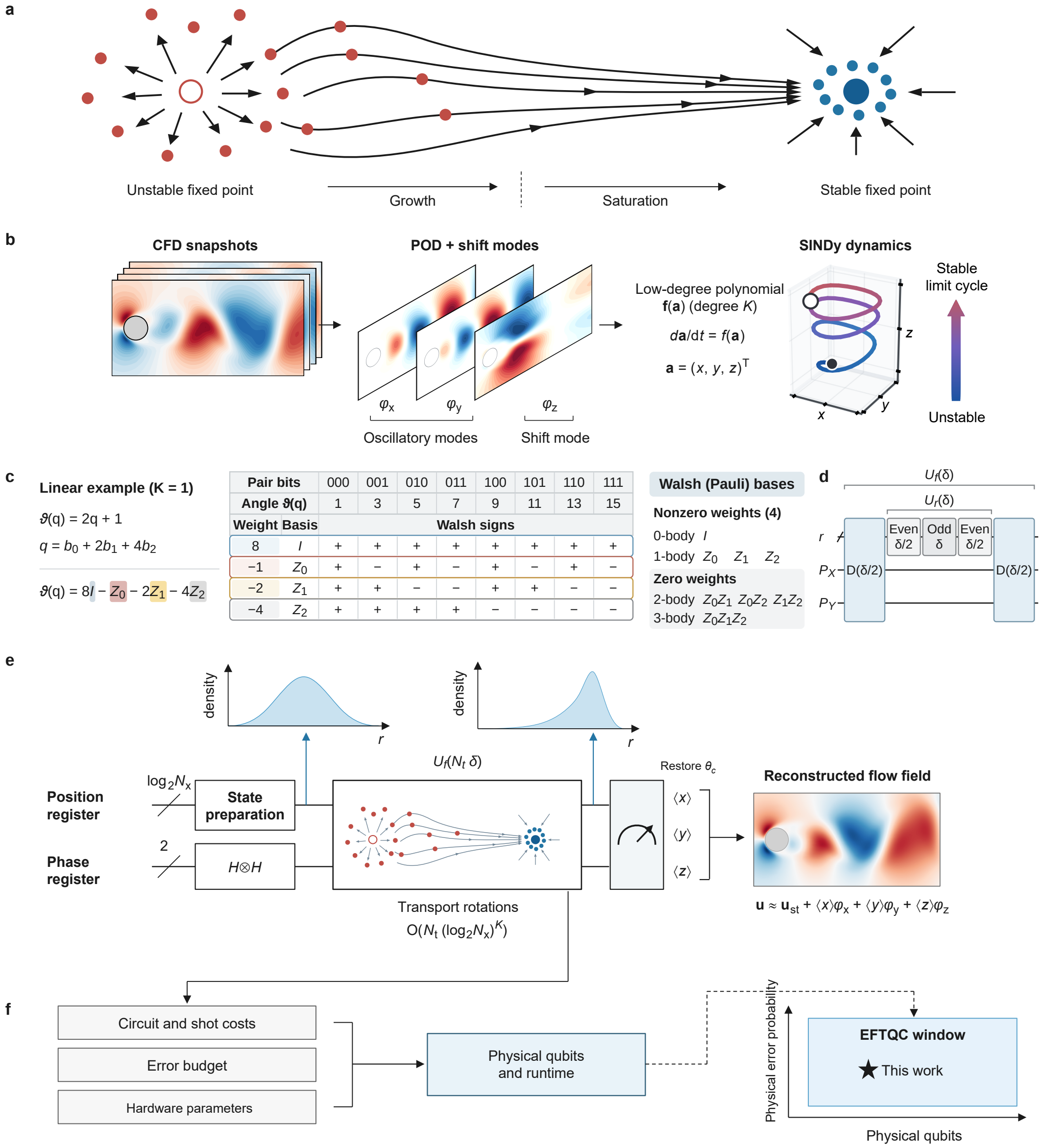

**Fig. 1 | From wake dynamics to fault-tolerant circuit costs.** **a**, Ensemble spreading near an unstable fixed point and concentration near a stable one. **b**, Proper orthogonal decomposition (POD) and sparse identification of nonlinear dynamics (SINDy) yield a three-coordinate polynomial model. **c**, Four Walsh coefficients represent eight rotation angles that vary linearly with the even-pair index on a 16-cell grid. **d**, Phase half steps flank the even–odd–even radial rotations. **e**, Radius and phase readouts yield ensemble-averaged modal coordinates for flow-field reconstruction with the unstable steady-state solution $\mathbf{u}_{st}$ and spatial modes $\varphi_x$, $\varphi_y$ and $\varphi_z$. The known phase $\theta_c$ is restored for the two oscillatory coordinates. **f**, Circuit and shot counts, the error budget and hardware parameters determine the required number of physical qubits and total task time.

We project wake snapshots computed by the finite-element method (FEM) onto the three modes and fit a reflection-symmetric quadratic model with SINDy (Fig. 1b, Methods). The two oscillatory coordinates $x,y$ define the strength of shedding through $r^2 \equiv x^2 + y^2$. In the fitted equations for $x$ and $y$, the nonlinear terms are the products $xz$ and $yz$, with $z$ the shift coordinate. Substituting the fitted closure $z \approx r^2$ makes them cubic[1]. Averaging over phase gives

$$\frac{\mathrm{d}r}{\mathrm{d}t^*} = \mu\left(r - r^3\right), \tag{1a}$$

$$\frac{\mathrm{d}\theta}{\mathrm{d}t^*} = \omega + \gamma r^2. \tag{1b}$$

Here $r$ is the shedding amplitude relative to its saturated value, $\theta$ is its phase and $t^*$ is nondimensional physical time. The fit gives growth rate $\mu = 0.0925$, angular rate $\omega = 0.753$ at zero amplitude and frequency shift $\gamma = 0.302$ at saturation (Supplementary Note 1, Table S1).

We propagate a Gaussian model of initial radius uncertainty[31] under equations (1a) and (1b). The probability density $\rho$ obeys the Liouville equation $\partial_{t^*}\rho = -\nabla \cdot (\mathbf{F}\rho)$. Its $L^2$ norm can change while total probability is conserved, so density evolution can be nonunitary. Since the radial drift vanishes at $r = 0,1$ and the angle is periodic, the continuous KvN evolution is unitary[5,6], with the density given by the squared magnitude of the amplitude:

$$\partial_{t^*}\psi = -\mathbf{F} \cdot \nabla\psi - 1/2\,(\nabla \cdot \mathbf{F})\psi, \tag{2a}$$

$$\rho = |\psi|^2, \tag{2b}$$

where $\psi$ is the amplitude and $\boldsymbol{F}$ the vector field in (1a)–(1b). Contraction towards saturation is therefore compatible with unitary propagation.

We divide the normalized radius interval $[0,1]$ into $N_x = 2^n$ uniform cells and represent them with an $n$-qubit register. Each computational basis state labels one cell, and its squared amplitude gives the cell probability. A real skew-symmetric discretization turns transport into amplitude transfers between neighbouring cells, each a two-state $R_Y$ rotation[12]. Rotations on pairs that share a cell do not commute, so we rotate the even pairs $(0,1), (2,3), \ldots$ and the odd pairs $(1,2), (3,4), \ldots$ in separate layers[12]. Within a layer no cell belongs to two pairs, so its rotations commute. Each time step applies an even half step, an odd full step and an even half step, giving second-order Strang splitting (Methods). A direct operator-norm estimate gives a sufficient step count $N_t = O_\epsilon\left(N_x^{3/2}\right)$; using even–odd cancellations and input regularity, we prove that $N_t = O_\epsilon(N_x)$ suffices at fixed input, duration and state-error tolerance $\epsilon$ (Supplementary Note 15). In each layer, a pair's rotation angle is the drift at its shared face times the layer duration divided by the cell width.

Rotating each pair separately requires $O(N_x)$ rotations per step[12]. To retain potential quantum grid-scaling benefits[5], we use Walsh synthesis[42] to expand the angles in the binary parity basis. One target bit distinguishes the two cells, and the remaining $n-1$ bits $b_i$ encode the pair number $q=\sum_i 2^i b_i$. Subtracting one from each cell number aligns the odd pairs (Methods). The angle of pair $q$ expands as $\vartheta(q)=\sum_S \hat{\vartheta}_S \prod_{i\in S} z_i$, where $S$ ranges over subsets of pair bits and $z_i=1-2b_i$ is the Pauli $Z$ eigenvalue of bit $i$. Within each layer, paired cells meet at $x_q=x_{\text{first}}+2hq$, with cell width $h$ and first pair boundary $x_{\text{first}}$. The angle is proportional to the degree-$K$ drift evaluated at $x_q$, giving a degree-$K$ polynomial in $q$. Since $q$ is a constant plus a weighted sum of $z_i$, each expanded term contains at most $K$ distinct $z_i$. Each nonzero coefficient requires one target rotation, whose sign follows the parity (Fig. 1c). Counting subsets bounds rotations per layer by

$$\sum_{\ell=0}^{\min(K,n-1)} \binom{n-1}{\ell} = O(n^K). \tag{3}$$

The even–odd–even layers of each step therefore require $O((\log_2 N_x)^K)$ radial transport rotations, with $K=3$ for the wake (Supplementary Note 3).

Reconstructing the velocity field requires the mean oscillatory coordinates and the mean shift coordinate. Under the shift closure these follow from radius and phase alone, since $(x,y,z)^{\mathrm{T}}=(r\cos\theta, r\sin\theta, r^2)^{\mathrm{T}}$. The phase governed by (1b) has two parts. One, $\theta_{\mathrm{c}}=\theta_0+\omega(t^*-40)$, is common to the whole ensemble, with $\theta_0=\operatorname{atan2}(y,x)$ taken from the FEM state projected at the initial time $t^*=40$. The other, $\Phi$, accumulates along each radial trajectory. With $\theta=\theta_{\mathrm{c}}+\Phi$, three ensemble averages (moments) $\mathbf{m}=(m_1, m_2, m_3)^{\mathrm{T}}$ give the mean coordinates (Fig. 1e; Methods):

$$\Phi(t^*)=\gamma\int_{40}^{t^*} r(s)^2\, ds, \tag{4a}$$

$$\mathbf{m}=\big(\langle r\cos\Phi\rangle, \langle r\sin\Phi\rangle, \langle r^2\rangle\big)^{\mathrm{T}}, \tag{4b}$$

$$\langle x\rangle = m_1\cos\theta_{\mathrm{c}} - m_2\sin\theta_{\mathrm{c}}, \tag{4c}$$

$$\langle y\rangle = m_1\sin\theta_{\mathrm{c}} + m_2\cos\theta_{\mathrm{c}}, \tag{4d}$$

$$\langle z\rangle = m_3. \tag{4e}$$

Two phase qubits coherently accumulate $\Phi$ through radius-controlled updates around each radial step, without intermediate measurement (Fig. 1d). For this Gaussian input, matrix product state (MPS) preparation has a gate count logarithmic in cell count at fixed bond dimension, MPS layer count and compilation precision[43]. With nine logical qubits, this circuit (Walsh–KvN) takes 33 steps from $t^*=40$ to 49.829, as in the three-coordinate run below. Surface-code estimates fall within our EFTQC window: $10^4$–$10^6$ physical qubits (Fig. 1f).

## End-to-end fault-tolerant resource estimates

We trace the effect of the reduced model's low-degree structure all the way through circuit synthesis and error correction to physical resource requirements[44]. One 33-step attempt on 64 cells uses nine logical qubits and 4,803 arbitrary rotations (Supplementary Note 3). The 1% field-error target takes 815 shots at a reference implementation-error budget of $E_{\max} = 10^{-4}$ per circuit attempt (Methods and Supplementary Notes 9–10). After synthesis, with Clifford gates absorbed into the measurement bases[40], each T or T† gate becomes a Pauli ±π/8 rotation that consumes one T state. These rotations run in a data block of surface-code patches holding the logical qubits; with a buffer for T states from magic-state factories (MSFs), it forms the core (Fig. 2a).

Fewer MSF qubits can increase runtimes[45], but the trade-off is not straightforward. Time per attempt scales with the T-state consumption interval, determined by the longer of two intervals (Fig. 2b): the core interval between consumptions and the supply interval between deliveries. Removing factory units saves qubits and lengthens the supply interval, but costs time only when the supply interval is longer and the core waits. The data block schedules $b_{\text{blk}}$ logical cycles per T state, with $b_{\text{blk}} = 1$ for a fast data block and $b_{\text{blk}} = 9$ for a compact one, which uses fewer qubits[40], and each cycle lengthens with code distance, which also enlarges every patch. At the reference physical error probability, $5 \times 10^{-4}$, the minimum-qubit design with supply enforced (no waiting) uses core distance 25, where one factory unit can keep pace. With waiting allowed, the same unit feeds a distance-19 core at nearly that pace: minimum qubits fall by 27% for only 3.2% more time per attempt (Supplementary Fig. S9c). Code distance thus couples patch size, cycle time and fault rate, making design multi-objective.

We vary data-block layouts, code distances, distillation levels and factory-unit counts from one to the smallest count that eliminates core waiting, retaining architectures Pareto-optimal in qubit count, T-state consumption interval, core error and magic-state error. Since tighter rotation synthesis lowers approximation error but consumes more T states, we select architecture and synthesis precision jointly, minimizing qubits or total task time (Fig. 2c; Methods). At the reference error probability, moving from the minimum-qubit to the fastest radial design increases the physical-qubit count 10.5-fold (53,382 to 559,953) and cuts total task time 12.4-fold, from 20.07 to 1.62 h; both fit our EFTQC window[46,47] (Fig. 3a,b; Table 1 and Supplementary Table S10a). We keep 33 steps to match the three-coordinate run; eight are fastest, 0.464 h on 423,000 qubits (Supplementary Table S9).

This feasibility does not hinge on the error budget (Supplementary Fig. S8a) or the factory model. Across the default and two alternative factories[40,48] with the fast block replacing the compact one, the 33-step radial minimum ranges from 56,100 to 122,000 at the reference error probability, inside our EFTQC window (Supplementary Fig. S8b). The optimum does depend on the circuit. Whether allowing waiting lengthens or shortens an attempt depends on its T count before synthesis and the physical error probability (Supplementary Fig. S9a,b,e). Each circuit therefore requires joint optimization of architecture, supply policy and synthesis precision for its resource budget, which shows the need for end-to-end estimates.

To assess how tailoring circuits to low-degree drift affects feasibility within our EFTQC window, we compare Walsh–KvN with a block-encoding circuit (BE–KvN)[38] on the same wake input, transport generator, grid and interval. Each call to this block encoding costs only $O(K(\log_2 N_x)^2)$ gates for $N_x$ cells and drift degree $K$. The encoding, however, loads the cell centres as normalized amplitudes, whose normalization grows with refinement on a fixed interval. Restoring the physical drift scale and applying the discrete derivative give an $N_x^{1+K/2}$ factor in the call bound at fixed duration and tolerance (Supplementary Note 13). At the reference error probability, 813 accepted BE–KvN shots take 10.63 years on the minimum-qubit design and 317.3 days on the fastest. Both methods fit below one million physical qubits (Fig. 3b,c). At fixed hardware and effective per-call cost, normalization at the generator's 671-fold smaller spectral norm would give 17.4 h (Supplementary Note 13).

To match the grid of the LCHS–KvN benchmark[19], we run Walsh–KvN on 128 radial cells. Doubling the cells, with steps doubled (33 to 66) to hold Trotterization error, raises total task time for 1% field error from 1.62 to 5.09 h, with one extra logical qubit and less than 1% more physical qubits (Table 1; Supplementary Fig. S3b and Note 6). We compile the LCHS–KvN circuit for its own, lighter benchmark: ten steps of a drift polynomial in sin $x$, which simplifies its block encoding, without phase qubits. At 815 accepted shots, not at matched accuracy, its fastest design takes 35.5 years on 2.98 million physical qubits, or 36.6 years under a one-million-qubit cap (Methods; Supplementary Table S11). Its upwinding suppresses grid-scale oscillations. At fixed generator norm, evolution time and drift degree, its optimized selector[19] gives a gate bound polylogarithmic in cell count (Supplementary Note 13).

**Table 1 | Fastest estimated fault-tolerant designs and classical grid scaling.**

| Calculation | Grid cells | Logical qubits | Steps | Shots | T gates per attempt | Physical qubits | Total task time | Operation bound‡ |
|---|---|---|---|---|---|---|---|---|
| Walsh–KvN, radial | 64 | 9 | 33 | 815 | $3.77 \times 10^5$ | 559,953 | 1.62 h | $O(N_t n^3)$ |
| Walsh–KvN, refined radial | 128 | 10 | 66 | 809 | $1.19 \times 10^6$ | 561,553 | 5.09 h | $O(N_t n^3)$ |
| BE–KvN, matched wake | 64 | 25 | — | 813 | $1.35 \times 10^9$ | 953,563 | 317.3 d | $O(T N_x^{5/2} n^2)$ |
| Three-coordinate, late | 64×32×64 | 20 | 33 | 100 | $7.10 \times 10^6$ | 702,712 | 4.14 h | $O(N_t n^2)$ |
| Three-coordinate, growth | 256×32×256 | 24 | 264 | 437 | $1.02 \times 10^8$ | 809,048 | 11.87 d | $O(N_t n^2)$ |
| LCHS–KvN, sine drift† | 128 | 26 | 10 | 815 | $4.46 \times 10^{10}$ | 2,983,519 | 35.5 yr | $O(\lVert\mathbf{A}_{\mathrm{LCHS}}\rVert T \mathrm{poly}(n))$ |
| Classical Eulerian | $N_1 \times \cdots \times N_{N_c}$ | — | $N_t$ | — | — | — | — | $O(N_t N_1 \cdots N_{N_c})$ |

Quantum designs without a qubit cap at physical error probability $5 \times 10^{-4}$ and error budget $10^{-4}$ per attempt. Wake rows target 1% relative RMS field error against their respective continuum ensembles at the final time, requiring an effective bias below 1% to leave a positive sampling budget. The three-coordinate runs additionally require noiseless relative field error below 1% at all 67 diagnostic times; inputs, grids and readouts differ between the radial and three-coordinate benchmarks (Methods). Shots count accepted outputs, times include retries, and T counts include rotation synthesis.

† For the sine-drift circuit, we use 815 accepted shots to estimate total task time. The 1% field-error target applies to the wake circuits.

‡ $N_x = 2^n$ is the largest number of cells along one coordinate, and $n$ the corresponding register width. Under the assumptions of Supplementary Note 15, $N_t = O_\epsilon(N_x)$ steps suffice for coordinate-wise even–odd transport at fixed dimension $N_c$, domain, input, duration and state-error tolerance $\epsilon$. $T$ is the dimensionless evolution time and $\mathbf{A}_{\mathrm{LCHS}}$ the LCHS generator, whose bound assumes the optimized selector of ref. 19 (Supplementary Notes 3, 7 and 13). These bounds alone do not compare total computational costs at matched accuracy.

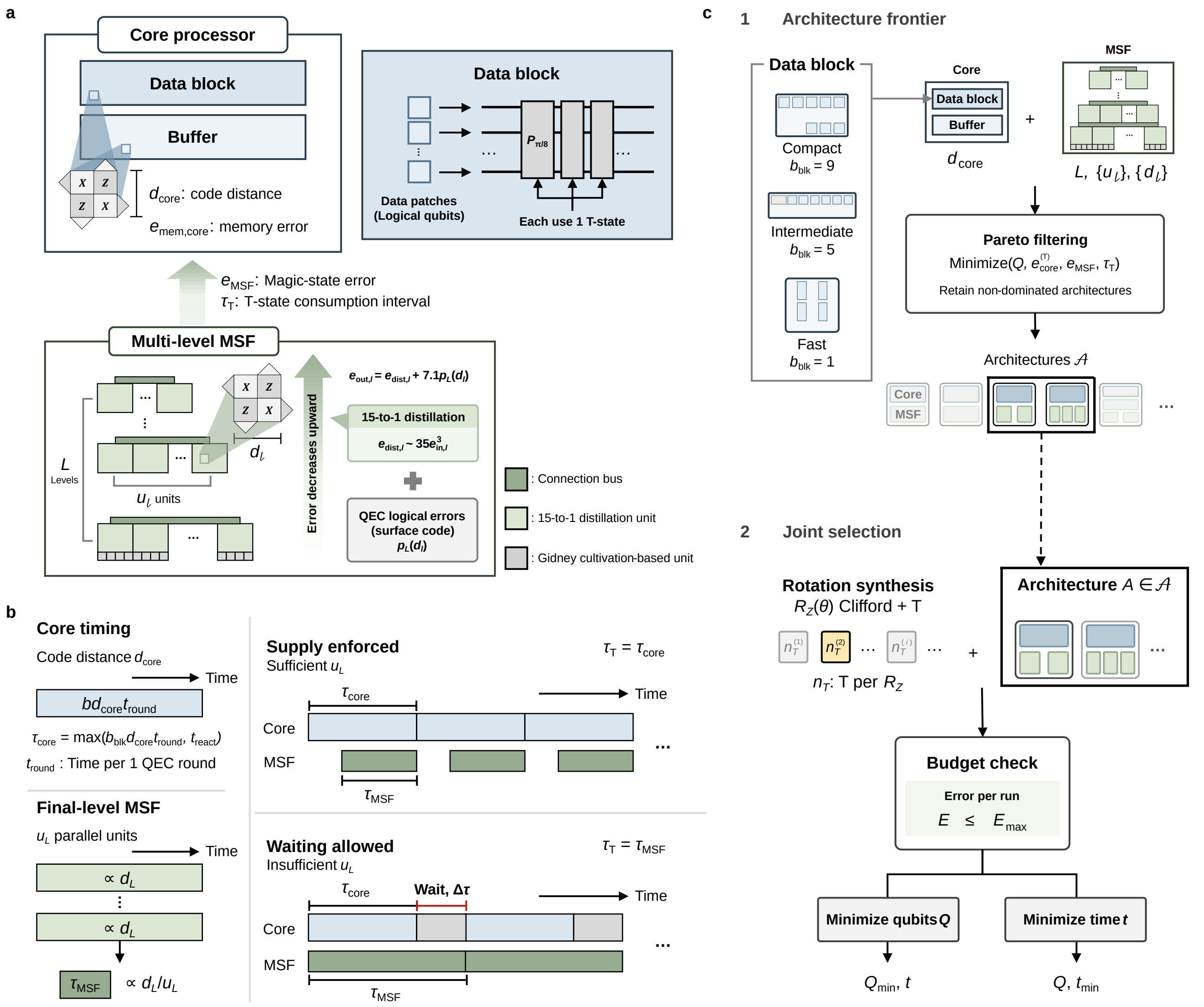

**Fig. 2 | Architecture, magic-state supply and joint selection. a**, Surface-code core and multilevel magic-state factory (MSF). Each Pauli π/8 rotation consumes one T state. Factory output accounts for cultivation, distillation and logical faults. **b**, Core and final-level factory timing. The core can consume one T state per core interval $\tau_{core}$, and the final-level factory supplies one per mean supply interval $\tau_{MSF}$. The data block schedules $b_{blk}$ logical cycles per T state (1 fast, 5 intermediate, 9 compact). The QEC round duration and classical reaction time are $t_{round}$ and $t_{react}$, respectively. With supply enforced, $\tau_T$ equals the nominal core interval $\tau_{core}$; with waiting allowed, insufficient final-level units set $\tau_T$ to the mean supply interval $\tau_{MSF}$. **c**, We construct the architecture frontier by Pareto filtering data-block layouts, code distances and MSF configurations using physical-qubit count $Q$, effective core error per T state $e^{(T)}_{core}$, factory-output error $e_{MSF}$ and T-state consumption interval $\tau_T$. Joint selection evaluates retained architectures and synthesis precisions for arbitrary rotations, minimizing qubit count or total task time when the error budget $E_{max}$ covers the estimated error per attempt ($E$ in c).

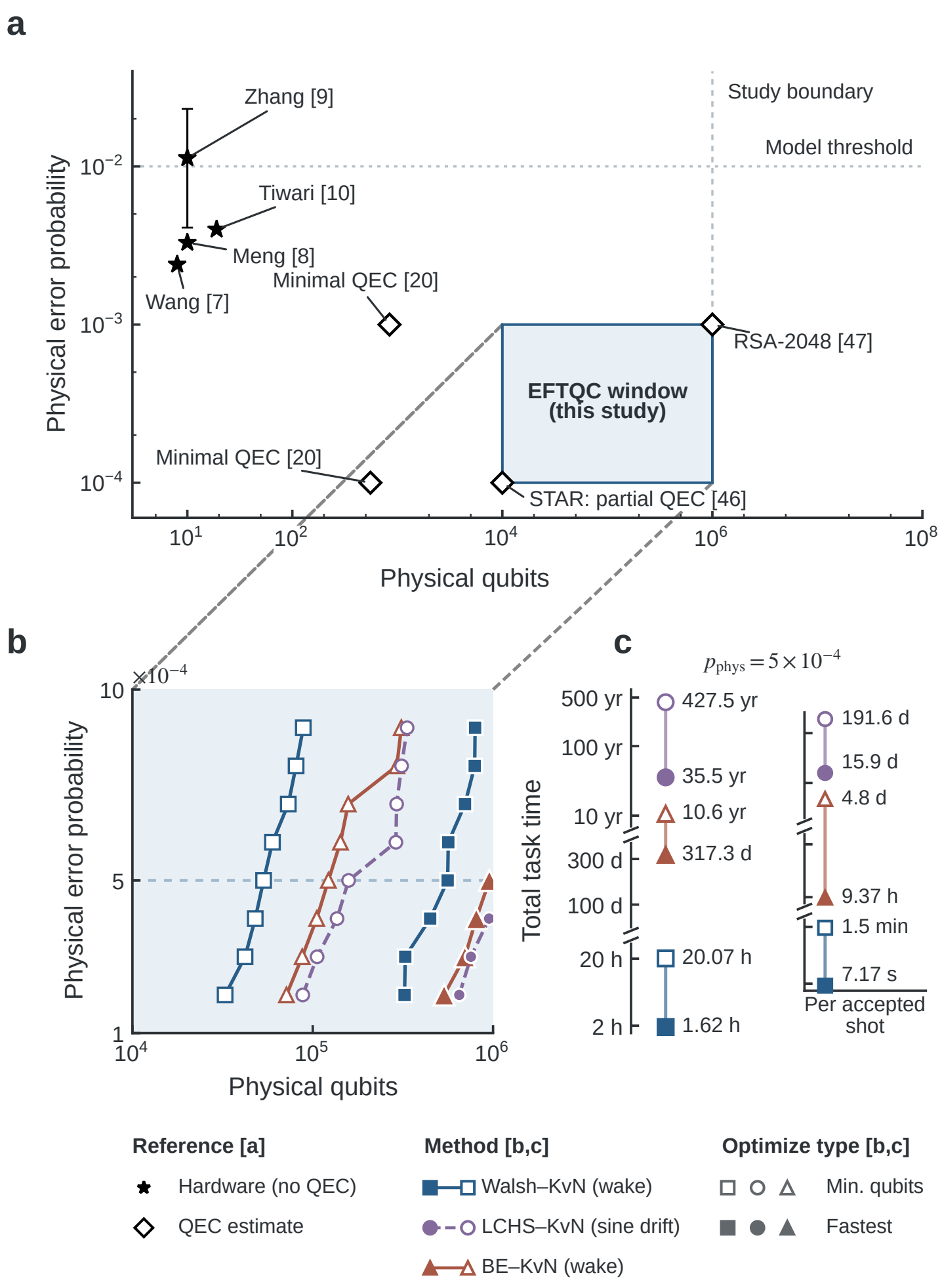

**Fig. 3 | Physical resources for nonlinear ensemble prediction. a**, Hardware experiments and estimates for error-corrected designs. The shaded EFTQC window spans $10^4$–$10^6$ physical qubits at physical error probabilities of $10^{-4}$–$10^{-3}$. The partially fault-tolerant STAR architecture[46] marks the lower-left corner. The upper-right corner rounds the 897,864-qubit RSA-2048 estimate[47] up to one million. The vertical line marks the study boundary ($10^6$ physical qubits) and the horizontal line marks the model threshold ($10^{-2}$). **b**, Minimum-qubit (open) and fastest (filled) designs at a per-attempt error budget of $10^{-4}$. **c**, Total task time at physical error probability $5 \times 10^{-4}$ and expected time per accepted shot (inset). For the same wake and 1% relative RMS field error, Walsh–KvN requires 815 shots and BE–KvN[38] requires 813 accepted shots. LCHS–KvN[19] uses 815 accepted shots for the sine drift. Its fastest design (2.98 million qubits) exceeds the axis limit in b. Axis breaks separate time scales. Supplementary Notes 10, 11 and 13 give the assumptions.

## Ensemble simulation of vortex shedding with Walsh–KvN

Walsh–KvN recovers alternating transverse-velocity lobes seen in the FEM solution, using 64 radial cells and 33 steps (Fig. 4a). That reconstruction rests on a design variable, the polynomial degree, which has a price. For each degree, we fit three of four consecutive time blocks and predict the fourth in turn (Fig. 4b). The cubic lowers mean error against the projected FEM field from 5.62% to 2.85% but raises the Walsh rotation bound almost sixfold (226 to 1,297 in total for one layer in each of three 64-cell coordinates). We adopt the 12-coefficient quadratic model.

Ideal circuit means follow the FEM spiral (Fig. 4c), which approaches $z \approx r^2$ near saturation (Fig. 4d). Figure 4e separates three sources of error. For one initial state, the three-coordinate model (3-mode) differs from FEM by 6.42% at the final time, including projection and dynamics. The radial model (Radial), which applies the shift closure and averages over phase, differs from 3-mode by 0.380% at the final time and by less than 0.60% throughout. For the ensemble, the field from ideal circuit means differs from the continuum Radial solution (the radial model solved without a spatial grid) by 0.107% after 33 steps. The 1% target applies to this last comparison and bounds the total of this bias, sampling noise and implementation error, which sets the 815 shots.

Following Deng et al.[35], we model drag and lift as polynomials in the modal coordinates fitted to FEM forces over times 10–60; the circuit estimates their averages without reconstructing the field. Within this calibration window, 815 additional ideal shots per time point give relative RMS errors against FEM of 0.21% for drag and 10.5% for lift (Fig. 4f; Methods). Drag is dominated by constant and radial terms, whereas lift depends on shedding phase; lift's RMS is also about one-eighth of drag's, so the same absolute error weighs more.

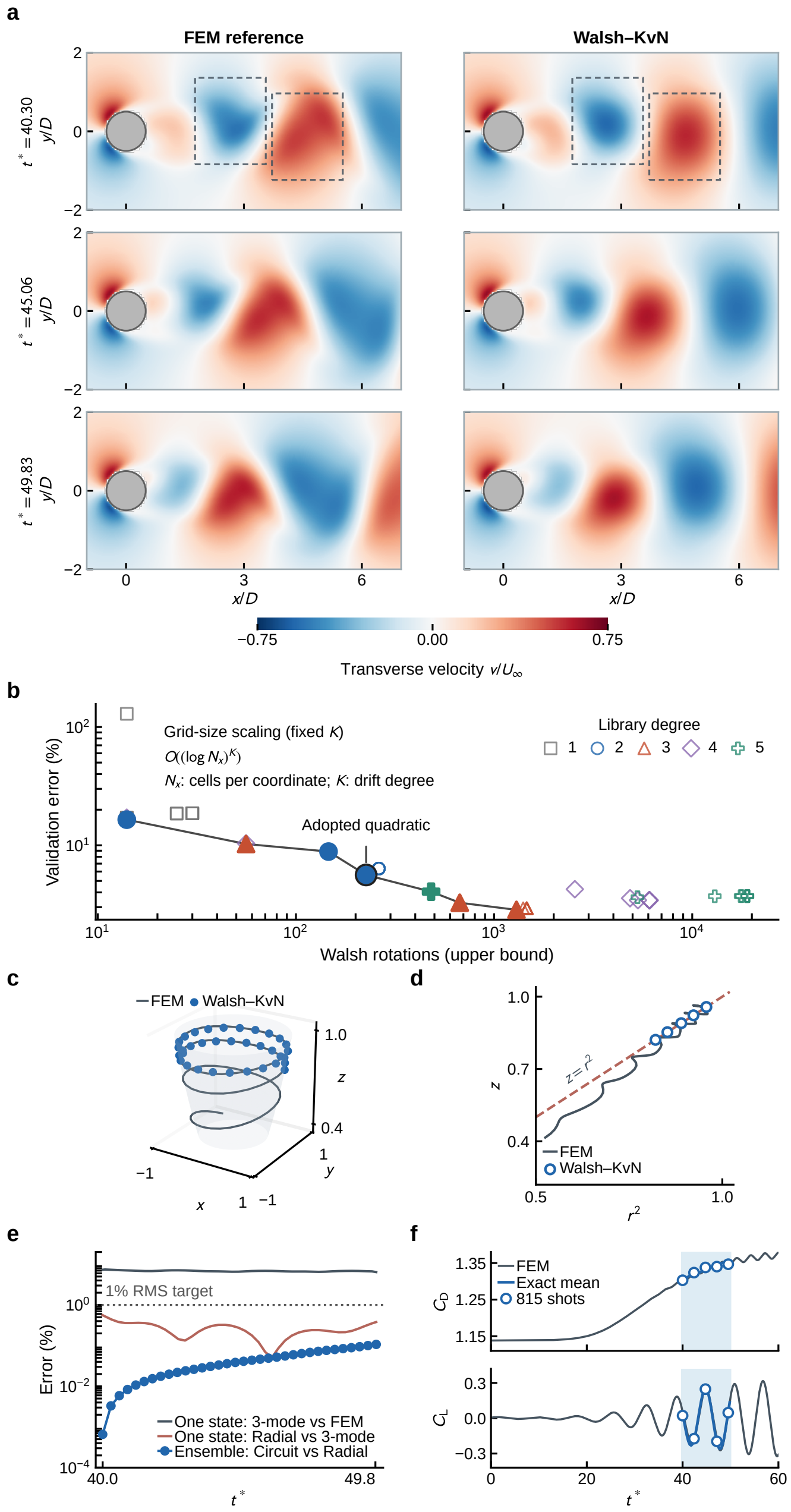

**Fig. 4 | Wake reconstruction and model selection. a**, Transverse velocity at Reynolds number 100. Rows show times and columns compare FEM and Walsh–KvN. **b**, Validation error against the projected FEM field and total rotation bound for one layer on each of the three modal coordinates. Filled symbols mark the Pareto frontier and the circle the selected quadratic fit. **c**,**d**, Modal trajectories and shift from ideal circuit means. **e**, For one initial state, grey and red give model differences divided by the FEM norm. For the ensemble, blue gives circuit bias relative to the continuum Radial field. The dotted 1% target includes sampling and implementation errors. **f**, Drag and lift in the shaded circuit interval, within the force calibration window (nondimensional times 10–60). Blue lines (exact mean) give ideal expectations. Open points share 815 additional ideal shots per time point across four measurement settings.

### Higher degrees and dimensions

Beyond the quadratic wake model (Fig. 4b), we ask how cost and accuracy change with polynomial degree and dimension (Fig. 5). Transport takes $O(N_c s_c^{K}(\log_2 N_x)^K)$ arbitrary rotations per step for $N_c$ transported coordinates of up to $N_x$ cells each and a drift of total degree $K$ whose components each depend on at most $s_c$ coordinates. Full coupling, $s_c = N_c$, gives $O(N_c^{K+1}(\log_2 N_x)^K)$. Per step, coupling enters polynomially and the grid only polylogarithmically, with the degree setting both exponents, provided the grid and boundary treatment keep the rotation angles polynomial (Supplementary Note 13).

To check whether accuracy holds across degrees, we test one-dimensional quadratic (Q2), cubic (C3) and quintic (Q5) drifts, each from inputs near a stable point, in the interior and near an unstable point, against matched classical trajectories (Fig. 5a,b). With 10,000 shots, all nine cases have relative RMS moment errors below 1%, combining noiseless bias and sampling variance. Errors are lower near stable points, consistent with contraction damping trajectory perturbations, and higher near unstable points, where repelling dynamics amplify them (Supplementary Note 8). Contraction also narrows the probability density and can demand finer grids: in a separate Q5 near-boundary test, insufficient resolution produces parasitic grid-scale oscillations in the centred discretization and a moment bias that additional shots cannot remove (Supplementary Fig. S4 and Note 8).

In two dimensions, a coupled cubic drift (2D C3) also stays below 1% (Fig. 5a); 674 shots meet the 1% moment-error target, and both the minimum-qubit and fastest designs fit our EFTQC window (Fig. 5c and Supplementary Note 10).

We drop the shift closure and propagate the fitted three-coordinate quadratic model, so that the shift evolves independently through growth and saturation (Fig. 5d). Radius and shift use second-order centred differences on finite grids, and Fourier modes of the angle keep the rotation angles polynomial for Walsh synthesis (Supplementary Note 7). At 1% RMS field error against each continuum ensemble, the late run, from $t^* = 40$, takes 33 steps on 64 × 32 × 64 cells and about 4 h, whereas the growth run, from $t^* = 29.8$, takes 264 steps on 256 × 32 × 256 cells and about 12 days; both end at $t^* = 49.829$. Both fastest designs fit our EFTQC window (Table 1, Fig. 5e); the extra coordinates cost mainly runtime.

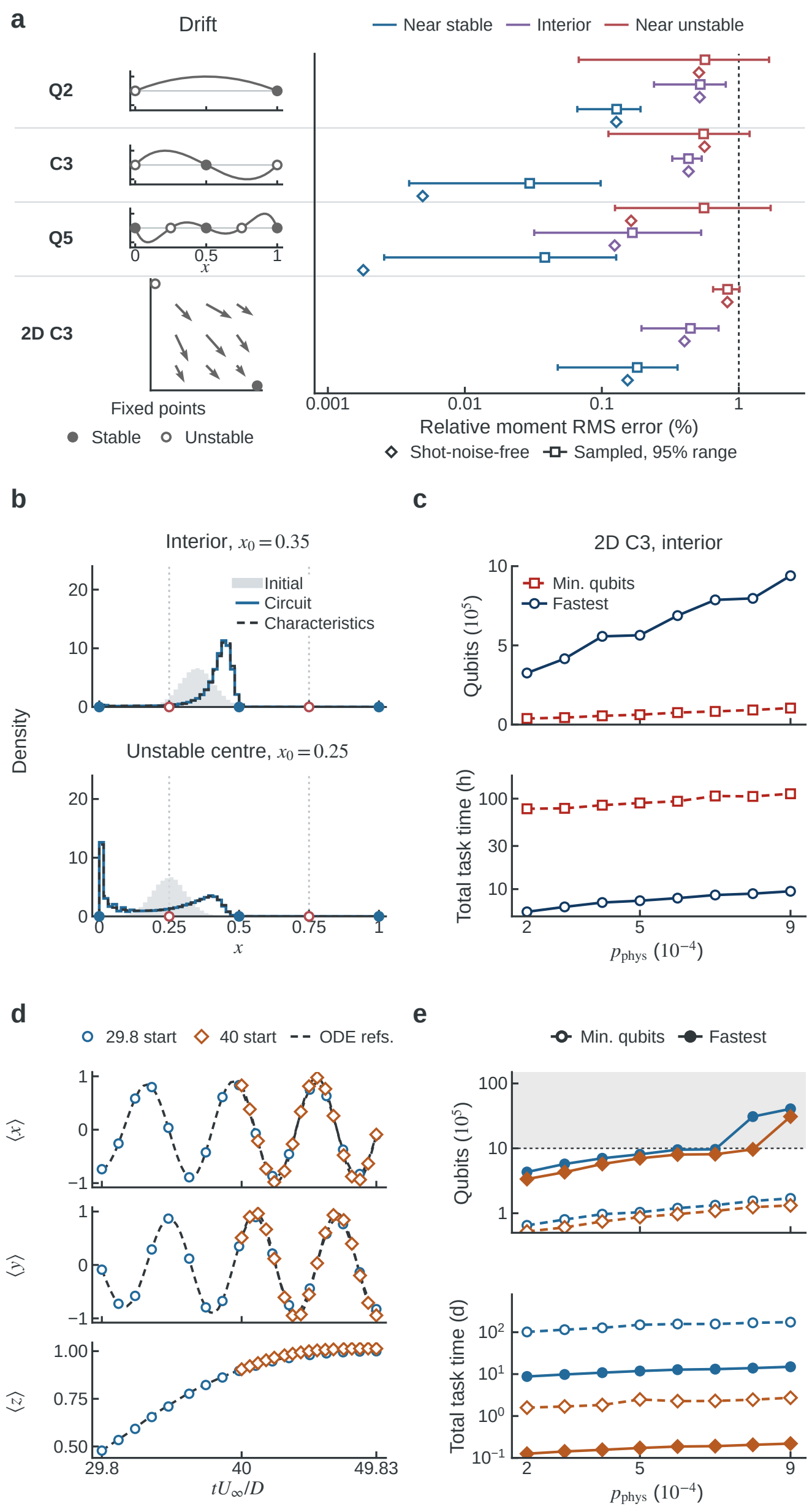

**Fig. 5 | Nonlinear benchmarks and three-coordinate wake simulation. a**, Quadratic (Q2), cubic (C3), quintic (Q5) and two-dimensional coupled cubic (2D C3) drifts. Filled/open fixed points are stable/unstable. Blue, purple and red denote near-stable, interior and near-unstable inputs. Diamonds show shot-noise-free relative moment errors; squares/bars show medians and 2.5th–97.5th percentiles of 5,000 estimates, each with 10,000 shots. **b**, Initial (grey) and final quintic densities from circuit (blue) and classical trajectories (characteristics, dashed). $x_0$ is the Gaussian centre before truncation to $[0,1]$. **c**, Qubits and total task time for coupled cubic input, using 674 shots without a qubit cap. **d**, Three-coordinate wake means from growth (blue, 29.8 start) and late (orange, 40 start) circuits versus continuum ensembles (ODE references, dashed); the time axis is $t^* = tU_\infty/D$, with $U_\infty$ the free-stream speed and $D$ the cylinder diameter. **e**, Physical resources without a qubit cap. Dashed/open and solid/filled designs minimize qubits and time. In c,e the RMS target is 1% and the error budget is $10^{-4}$. Grey marks qubits outside the EFTQC window.

### Hardware experiment

On current hardware, constant-speed circle rotation has been demonstrated on three IBM Quantum System One qubits using Koopman evolution[49]. We demonstrate nonlinear transport towards an attracting state on four IBM qubits using Walsh–KvN, under drift proportional to $r - r^3$, with 16 cells and four or six steps (Extended Data Fig. 1). Corrected outer-cell probability rises from 0.798 to 0.962 (steps 1–6). For mitigation, we fit an effective depolarizing model to reference circuits (Methods). At 16 cells the cubic drift uses all eight Walsh terms per layer, versus 26 of 32 at 64 cells, so no rotations are saved here. At step five, corrected probability is 3.7 bootstrap standard deviations below ideal (Supplementary Note 14). At step six the corrected mean radius, 0.998, exceeds the largest cell centre (31/32) because mitigation permits negative weights. These results motivate early fault-tolerant tests of rotation savings over longer evolutions with field reconstruction.

## Discussion

The end-to-end test is whether favourable asymptotic scaling survives the full cost of delivering a physical prediction at the required accuracy. We show feasibility for a reduced model of the cylinder wake: the estimated cost of matching its continuum solution to 1% stays within our EFTQC window.

For sparse KvN Hamiltonians, quantum simulation can offer an exponential gain over deterministic Eulerian Liouville solvers[5]. Realizing this gain for a full PDE also requires efficient ensemble preparation and propagation. For a scalar field $u$ on $M$ spatial points, $N_u$ value bins per point give ${N_u}^M$ joint field configurations[5,27]. Direct KvN embedding therefore needs $O(M\log_2 N_u)$ position qubits. While this linear dependence on physical grid size can offset the potential quantum gain, model reduction replaces the grid-point variables with $N_\mathrm{c}$ modal coordinates, requiring $O(N_\mathrm{c}\log_2 N_x)$ position qubits for $N_x$ cells per coordinate.

Combining SINDy-derived ROMs with Walsh synthesis[42] also keeps costs polynomial in dimension. As in total-degree Carleman embeddings[14], a library for $N_\mathrm{c}$ coordinates and total degree at most $K$ contains $\binom{N_\mathrm{c}+K}{K}$ monomials, including the constant. This count grows as $O(K^{N_\mathrm{c}})$ with degree at fixed dimension, but only as $O({N_\mathrm{c}}^K)$ with dimension at fixed low degree, and SINDy can select a sparse subset[1]. With each drift component depending on at most $s_\mathrm{c}$ coordinates, the Walsh–KvN rotation bound per step is linear in $N_\mathrm{c}$ at fixed $s_\mathrm{c}$, degree and grid resolution. In this regime KvN can offer a gain over deterministic Eulerian Liouville solvers on a joint grid of ${N_x}^{N_\mathrm{c}}$ configurations[5]. The drift degree thus governs how the library grows with additional coordinates and how the rotation bound grows with grid refinement.

Halving the sampling error requires four times as many independent shots, as in classical trajectory sampling. Amplitude estimation could reduce circuit calls quadratically[5] at the cost of deeper circuits. Benchmarking total costs against Monte Carlo and incorporating diffusion could be next steps.

## Methods

### Physical model and computational task

We construct three spatial modes to represent wake velocities computed by the finite-element method (FEM). Proper orthogonal decomposition of the antisymmetric velocity in snapshots from nondimensional times 20–80 gives the two leading shedding modes. We rescale their modal coordinates to equal mean squares, preserving their zero mean product, and choose the second coordinate's sign so that phase increases on average. We multiply each mode by the factor used to divide its coordinate, leaving the velocity field unchanged. The third mode represents the mean-flow shift[3]. We subtract the unstable steady flow from the velocity field averaged over two shedding cycles, remove the components along the shedding modes and normalize the remainder. Least-squares projection gives each snapshot's three modal coordinates (Supplementary Note 1)[1,36].

To test prediction across disturbances, we ran six FEM simulations (cases A–F) from the same unstable steady flow at Reynolds number 100, varying the amplitude, centre and width of the initial streamfunction perturbation. Case A supplies the modes and 292 training and validation snapshots from nondimensional times 10–80. We split these into four consecutive blocks of 73 snapshots each. In turn, we fit on three blocks and forecast the fourth from the modal coordinates of its first snapshot. We forecast cases B–F using the modes and dynamics fitted to case A.

Using SINDy, we fit each coordinate's time derivative to symmetry-allowed monomials, omitting the constant. For each maximum degree from 1 to 5 we test thirteen coefficient cutoffs (Fig. 4b). We divide each coordinate by its training root-mean-square value. We estimate each coordinate's time derivative with a fourth-order centred difference spanning five snapshots. All five snapshots must belong to the training set. We evaluate each monomial at these times, scale its column of values to unit length and fit by ridge regression with penalty weight $10^{-4}$. After undoing the column scaling, we discard coefficients below the cutoff in magnitude and refit.

We choose the polynomial degree by comparing validation error with the Walsh rotation-count bound. Validation error is the relative RMS velocity-field error on the block left out of each fit. We score predictions against the three-mode projection of the FEM field, exclude the initial snapshot and average the four block errors (Fig. 4b and its Source Data). For each degree and coefficient cutoff, we refit the model using all training snapshots and compute the rotation bound from the monomials retained by the cutoff. On a common Cartesian grid with $N_x = 64$ cells per coordinate, we sum the bounds for one layer of pair rotations along each of the three coordinates, before excluding pairs that cross the domain boundaries. The count excludes the decrement and increment, repeated layers, state preparation, rotation synthesis, shots and error correction. The quadratic model used in the simulations retains 12 of the 13 symmetry-allowed coefficients at cutoff 0.03.

To measure physical model error over a forecast window, we take the norm of the difference between the predicted and FEM velocity fields and divide by the FEM field norm, summing squared velocities over snapshots, sites and both components. The projection error is the smallest such error attainable with these three modes and the same steady flow, obtained by least-squares projection. Model validation uses the three-mode projection of the FEM field, including the steady background, as its reference. Quantum simulation uses the corresponding continuum reduced-model solution as its reference. Its 1% field-error target lies below the observed physical model errors. Supplementary Fig. S1 compares forecast and projection errors (a,b) and identifies the six disturbances (c).

## Radial Walsh–KvN transport and readout

To obtain the radial and phase equations, we fit the mean-flow shift to squared shedding amplitude through the origin, where both vanish[1]. We substitute this relation into the two shedding-coordinate equations, resolve their rates in the radial and angular directions and average over phase at fixed amplitude, giving (1a)–(1b) (Supplementary Note 1).

The two 64-cell wake circuits, Walsh–KvN and BE–KvN, use the same radial input. We choose a Gaussian of standard deviation 0.25 before truncation, truncate it to the unit interval and normalize it. Differences of its cumulative distribution at cell faces give the target cell probabilities. We choose the Gaussian centre so that the probability-weighted mean of the cell centres equals the initial projected FEM radius, 0.9017. A matrix product state (MPS) of bond dimension two approximates the square roots of the cell probabilities. To construct the continuum reference, we spread each cell's prepared probability uniformly within that cell, then evolve the same equations without a spatial grid. Each attempt executes the preparation circuit compiled offline (Supplementary Note 5).

Following ref. 12, transport moves amplitude between neighbouring radial cells in even and odd layers of two-state rotations. On the uniform grid, we discretize the spatial transport operator with second-order centred differences in a real skew-symmetric form. For cells in increasing index order, the shared-face block $\mathbf{A}_{\text{face}}$ and its exact two-state evolution are

$$\mathbf{A}_{\text{face}} = \frac{f_{\text{face}}}{2h}\begin{pmatrix} 0 & -1 \\ 1 & 0 \end{pmatrix}, \tag{5a}$$

$$e^{\delta t \mathbf{A}_{\text{face}}} = R_Y\left(\frac{f_{\text{face}}\delta t}{h}\right). \tag{5b}$$

Here $f_{\text{face}}$ is the drift at the face, $h$ the cell width and $\delta t$ the layer duration. For the step of duration $\delta$ in (6b), we use $\delta t = \delta/2$ in each even layer and $\delta t = \delta$ in the intervening odd layer. Zero drift at the endpoints removes the coupling between the last and first cells (Supplementary Note 2). In (5b), the rotation angle is proportional to the drift, so polynomial drift gives polynomial angles that admit sparse Walsh synthesis. By contrast, block encodings that write a normalized coefficient $g(x)$ into the $|1\rangle$ amplitude of $R_Y(\theta)|0\rangle$ require $\theta = 2\arcsin g(x)$, which generally loses this polynomial structure.

For $N_x = 2^n$ cells, $n - 1$ bits encode the pair number in the even layer and the lowest bit distinguishes its two cells. Pairs across odd faces are brought to this form by subtracting one from each cell number. Adding one after the rotations restores the numbering. We implement the decrement and its inverse, the increment, with incrementer A, a sequence of multi-controlled bit flips that returns its single work qubit to zero (Supplementary Fig. S2b and Note 3). Each binary parity function corresponds to a product of Pauli $Z$ operators. For each nonzero Walsh coefficient, controlled-NOT (CNOT) gates compute a bit parity, a target rotation applies the term and the inverse CNOT gates clear the parity (Supplementary Fig. S2c). A degree-$K$ term uses at most $2K$ CNOT gates. Equation (3) bounds the rotation count. Supplementary Notes 2, 3 and 13 give boundary conditions, gate counts and the coupled-coordinate bound.

To measure the radius-weighted phase moments in (4b), Hadamard gates prepare the two phase qubits $P_{\text{X}}$ and $P_{\text{Y}}$ in equal superpositions. Both acquire the same residual phase $\Phi$; measuring one in X and the other in Y allows both moments to be estimated from the same shots. A phase update proportional to the squared-radius operator, conditioned on each phase qubit, precedes and follows each radial step. For step duration $\delta$, we use

$$D(\delta) = \exp\left[i\gamma\delta\,\hat{r}^2(\hat{n}_{\text{X}} + \hat{n}_{\text{Y}})\right], \tag{6a}$$

$$U_f(\delta) = D(\delta/2)\left[U_r(\delta) \otimes I_{\text{ph}}\right]D(\delta/2). \tag{6b}$$

Here $U_f$ denotes the combined radius–phase step, $U_r$ transports radius, $\hat{r}$ is the radius operator and $I_{\text{ph}}$ is the phase-register identity. The occupation operator $\hat{n}_\alpha = (I - Z_\alpha)/2$ projects phase qubit $P_\alpha$ onto state one for $\alpha = \text{X}, \text{Y}$, with $Z_\alpha$ the Pauli Z operator on that qubit. Consecutive steps share two half updates, which combine as $D(\delta/2)D(\delta/2) = D(\delta)$. The phase qubits undergo no intermediate measurement or reset (Fig. 1d). The residual phase $\Phi = \theta - \theta_{\text{c}}$ is the trajectory-dependent part remaining after subtracting the common ensemble phase $\theta_{\text{c}}$, as defined in (4a)–(4e). Supplementary Note 4 gives the moment estimators and field reconstruction.

We separate spatial error from Trotterization error in the reconstructed mean velocity field. The matrix exponential of the grid generator propagates amplitudes without time splitting. Its difference from the continuum field for the same prepared density measures spatial error. The circuit field's difference from this exact-grid field measures Trotterization error. Both differences are normalized by the continuum field norm. Field bias denotes the noiseless circuit field's difference from the continuum reference; the relative RMS field error also includes sampling and implementation contributions (equations (7)–(8)).

For the radial Walsh–KvN benchmark in Table 1, we use $N_x = 64$ cells and $N_t = 33$ steps to match the late-time three-coordinate calculation over $t^* = 40$–$49.829$. At fixed $N_t$, we optimize the surface-code architecture, code distances, magic-state supply and rotation-synthesis precision.

We test every integer step count $N_t$ from 1 to 64, plus 66, 80, 96, 128, 132 and 256, keeping the input, grid, forecast interval, readout and continuum reference fixed. At each $N_t$, we evaluate the effective bias bound $b_{\text{eff}}$ and single-shot variance bound $v_{\text{eff}}$ for the relative RMS field-error target $\varepsilon = 0.01$. We require $b_{\text{eff}} < \varepsilon$, so the squared-error budget $\varepsilon^2 - b_{\text{eff}}{}^2$ available for sampling is positive, and use equation (8) to compute the shot count $S$. For each qualifying $N_t$, we reoptimize rotation-synthesis precision and the surface-code design. Supplementary Table S9 summarizes selected step counts; Supplementary Note 12 gives the full comparison, and Note 6 gives the grid-refinement tests.

Centred differences can produce grid-scale oscillations when the density is poorly resolved. We compare them with first-order upwind differences, which use the cell on the side from which flow arrives[19]. For the near-boundary input of the quintic test drift (Supplementary Note 8), upwinding damps these oscillations and improves density accuracy but increases moment bias. Such oscillations are absent in the resolved wake. For the wake comparison, both calculations use the same prepared input, domain, interval and radius–phase readout. We compute both grid-generator exponentials classically and normalize the upwind output because its evolution does not preserve the state norm. Supplementary Note 6 reports the wake field errors; Note 8 reports the quintic density and moment errors.

## Three-coordinate Walsh–KvN transport and readout

During wake growth and late-time shedding we evolve the three-coordinate quadratic model in radius $r$, angle $\theta$ and mean-flow shift $z$. The shedding coordinates obey $\frac{\mathrm{d}}{\mathrm{d}t^*}(x,y)^{\mathrm{T}} = (\mathbf{J}_0 + z\mathbf{J}_1)(x,y)^{\mathrm{T}}$, where $\mathbf{J}_0$ and $\mathbf{J}_1$ are constant $2 \times 2$ matrices fitted to the FEM modal trajectories. We evolve $z$ with its own quadratic differential equation, without imposing the radial closure $z \approx r^2$. Substituting $x = r\cos\theta$ and $y = r\sin\theta$ gives equations for $r$, $\theta$ and $z$ whose angular factors are $1, \cos(2\theta), \sin(2\theta)$ (Supplementary equations (S23a)–(S23c)). The double-angle factors couple Fourier mode $k$ to $k \pm 2$, where integer $k$ labels an angular Fourier mode.

We centre Cartesian Gaussian inputs on the corresponding classical model trajectories, with standard deviations 0.12, 0.12 and 0.09 in $x, y$ and $z$, respectively, and condition them on the retained radial and $z$ intervals. An MPS of bond dimension four approximates their square-root amplitudes. We choose the radial and $z$ limits so that the estimated probability of classical trajectories leaving the domain at any sampled time is below $10^{-6}$. We require the noiseless circuit mean field to differ from the continuum reference by less than 1% in relative field norm at all 67 diagnostic times. At the final time, the effective bias bound $b_{\text{eff}}$ must satisfy $b_{\text{eff}} < \varepsilon$, where $\varepsilon = 0.01$, leaving a positive sampling-error budget in equation (8). For each input, we vary the numbers of radial, angular and $z$ cells at fixed step count and stencil. Among the grids that pass both error tests, we select the one with the smallest product of required shots and estimated T gates per attempt (Supplementary Note 7).

After preparation, a forward Fourier transform puts the angle register in the Fourier-mode basis. We use second-order centred differences in radius and $z$ in both runs. Finite differences in $r$ or $z$ couple cell pairs, while the factors $\cos 2\theta$ and $\sin 2\theta$ couple Fourier modes $k$ and $k \pm 2$. We group these couplings into disjoint two-state blocks for terms changing only the cell or mode, and four-state blocks for terms changing both. The angles have total degree at most two in $r, k, z$. Each coordinate is affine in its binary digits, so expanding the angles produces products of at most two digits and hence Walsh parities involving at most two coordinate bits. An inverse Fourier transform restores the angular grid for measurement.

Two work qubits flag coupled pairs of basis states whose cell or mode number would wrap when incremented or decremented across a radial, $z$ or Fourier-mode boundary. These pairs are excluded from the rotations. A third work qubit supports arithmetic, and all work qubits are cleared after each layer of disjoint rotations. We check the compiled preparation gates against the target MPS and the compiled Fourier gates against the discrete Fourier transform. Starting from the state produced by the compiled preparation gates, we simulate propagation by applying each block's exact rotation matrix to the corresponding amplitudes. We obtain the continuum reference from probability-weighted classical trajectories of the conditioned Gaussian and check convergence by increasing the number of integration points. Bias relative to this reference includes the MPS preparation error (Supplementary Note 7).

### Field error and shot estimation for both wake models

For the radial circuit, each shot measures the radius and one Pauli observable on each phase qubit. At residual phase Φ, each phase qubit is in $|q_\Phi\rangle = (|0\rangle + e^{i\Phi}|1\rangle)/\sqrt{2}$. Measuring Pauli $X$ on one and Pauli $Y$ on the other gives expectations $\langle X\rangle_\Phi = \cos\Phi$, $\langle Y\rangle_\Phi = \sin\Phi$ (Supplementary Note 4). The vector of outcomes from one shot is $\boldsymbol{\xi} = (rs_{\mathrm{x}}, rs_{\mathrm{y}}, r^2)^{\mathrm{T}}$, where $r$ is the measured radius and $s_{\mathrm{x}}, s_{\mathrm{y}} \in \{-1, +1\}$ are the two signs. Averaging this vector gives the moments in (4b).

For the three-coordinate circuit, measuring the three position registers returns the shot vector $\boldsymbol{\xi} = (r\cos\theta, r\sin\theta, z)^{\mathrm{T}}$.

For either wake circuit, we separate field bias from the sampling variance of the reconstructed mean. Let $b$ be relative field bias, $\boldsymbol{\Sigma}$ the covariance of the shot vector $\boldsymbol{\xi}$, $\mathbf{A}$ the linear map from moments to field values, $\mathbf{W}$ the spatial weights, $\bar{\mathbf{u}}_{\mathrm{ref}}$ the continuum reference field and $S$ the number of independent shots. The mean squared relative field error, derived in Supplementary Note 9, is

$$\mathrm{E}[\varepsilon_{\mathrm{field}}^2] = b^2 + \frac{v}{S}, \tag{7a}$$

$$v = \frac{\mathrm{tr}(\mathbf{A}^{\mathrm{T}}\mathbf{W}\mathbf{A}\boldsymbol{\Sigma})}{\|\bar{\mathbf{u}}_{\mathrm{ref}}\|_{\mathbf{W}}^2}. \tag{7b}$$

Here $v$ is the variance that a single shot contributes to the field, in the weighted norm and relative to the squared norm of the reference. The square root of (7a) gives relative RMS error over independent repetitions of the $S$-shot estimate. We obtain bias from the ideal circuit mean and the continuum reference, and covariance from the corresponding ideal joint readout distribution. We weight all field sites equally, so $\mathbf{W}$ is the identity. Implementation error changes the measured distribution. From the per-attempt error budget $E_{\max}$, we bound its total variation distance after conditioning on success. The smallest and largest values of each observable and each pairwise product on the grid then bound shifts in the means and second moments, giving effective bounds $b_{\mathrm{eff}}, v_{\mathrm{eff}}$. For relative RMS target $\varepsilon$, the sufficient shot count is derived in Supplementary Note 10:

$$S = \max\{1, \lceil \frac{v_{\mathrm{eff}}}{\varepsilon^2 - b_{\mathrm{eff}}^2} \rceil\}, \qquad b_{\mathrm{eff}} < \varepsilon. \tag{8}$$

If the effective bias $b_{\mathrm{eff}}$ reaches the target $\varepsilon$, more shots cannot meet it under this bound.

We test other drifts by applying the same relative RMS criterion to the vector of moments. One-coordinate benchmarks use position and its square. Two-coordinate benchmarks use $x$, $y$, $x^2$, $y^2$ and $xy$. Supplementary Table S4 lists the inputs, and Supplementary Fig. S5 traces the four coupled mean paths. Source Data for Fig. 5a give reference moment norms and central 95% intervals at 10,000 shots. For the coupled cubic with input centred at (0.5, 0.5), equation (8) gives 674 shots for 1% five-moment error including implementation error (Fig. 5c, Supplementary Note 10). This circuit uses 32 × 32 cells, 64 steps and 11 logical qubits. All shot counts assume independent repetitions. Coherent amplitude estimation requires additional forward and inverse circuit calls and reversible evaluation of the observable[5,50].

## Drag and lift from the radial model

To reconstruct cylinder drag and lift, we fit polynomials of the modal coordinates that respect reflection symmetry[35] to FEM modal coordinates and cylinder surface forces over $10 \leq t^* \leq 60$. The times plotted in Fig. 4f fall within this calibration window. At each of 34 times, drag and lift share 815 shots sampled from the ideal circuit distributions, divided among the four pairs of X or Y measurement bases on the two phase qubits. Single-qubit readings weighted by the same shot's $r$ or $r^3$ give the first-harmonic moments for lift, and products of the two readings weighted by $r^2$ give the second-harmonic moments for drag. We rotate these moments by $\theta_c$ and $2\theta_c$, respectively, to recover the physical phase $\theta = \theta_c + \Phi$. Force RMS errors combine bias against FEM with analytic sampling variance, normalized by the corresponding FEM RMS force. Supplementary Note 4 gives the estimators. These additional simulated measurements total 27,710 shots and are excluded from the resource estimates in Table 1. Source Data for Fig. 4f give force coefficients, sampled forces, covariance and standard errors.

### Fault-tolerant resource estimation for all circuits

We model a surface-code processor in which the core executes the logical circuit and magic-state factories supply T states (Fig. 2a)[40,41]. A patch is a connected region of surface code that stores or manipulates logical information. Compact, intermediate and fast core data blocks use progressively more patches to consume T states more frequently[40]. One round of quantum error correction (QEC) takes 1 μs. At code distance $d$, a logical cycle lasts $d$ rounds. The 10 μs classical reaction time covers decoding a measurement and choosing the next operation.

Gidney et al.'s cultivation procedure prepares inputs for one or two levels of 15-to-1 distillation[41,51]. We estimate the accepted T-state error as twice the S-state error simulated with their public code, following their calibration at fault distance three[51] (Supplementary Note 11).

We allocate $2(d+1)^2$ physical qubits per standard patch[52]. Supplementary Note 11 gives the factory parameters and layouts.

Clifford basis changes map transport rotations to $R_Z$ rotations. The typical T count for Clifford+T synthesis grows logarithmically with the inverse of the allowed rotation error[39].

We measure the error of one circuit attempt by the normalized diamond distance[53], before conditioning on circuit success. The reference per-attempt error budget $E_{\text{max}}$ is $10^{-4}$. Core faults, faulty T states and rotation synthesis contribute to the error model. Let $N_{\text{T}}^{\text{native}}$ count native T and inverse T gates, $N_{R_Z}$ count arbitrary Z rotations and $n_{\text{T}}$ count T gates used to synthesize each rotation. The total T-state consumption per attempt is $N_{\text{T}} = N_{\text{T}}^{\text{native}} + n_{\text{T}} N_{R_Z}$. We evaluate

$$E_{\text{model}} = 1 - (1 - E_{\text{core}})(1 - e_{\text{MSF}})^{N_{\text{T}}} + N_{R_Z} 2^{-n_{\text{T}}/3} \tag{9}$$

and require $E_{\text{model}} \leq E_{\text{max}}$. Here $E_{\text{core}}$ is the probability of at least one core fault during the entire attempt, whereas $e_{\text{MSF}}$ is the error probability of each delivered T state. The synthesis contribution uses the typical per-rotation estimate $2^{-n_{\text{T}}/3}$; it is not a certified bound (Supplementary Note 10). At physical error probability $p_{\text{phys}}$ and code distance $d$, the logical fault probability per patch per QEC round is

$$p_{\mathrm{L}}(p_{\mathrm{phys}}, d) = 0.1\left(\frac{p_{\mathrm{phys}}}{p_{\mathrm{th}}}\right)^{(d+1)/2}, \tag{10}$$

with threshold $p_{\mathrm{th}} = 10^{-2}$ (refs. 40,41). We count memory faults on every core patch throughout each T-state consumption interval, including waits for decoding or supply. During gates, the data-block patches and the 18 fixed buffer patches assumed in our core layout also incur gate faults (Supplementary Note 11). For distillation Cliffords, we use the one-cycle memory-fault probability at the same distance[41]. Noise tailoring and error correction motivate treating $p_{\mathrm{phys}}$ as an effective stochastic error probability (Supplementary Note 10, equation (S39)). Its reference value is $5 \times 10^{-4}$ and the sweep spans $2 \times 10^{-4}$ to $9 \times 10^{-4}$.

We time the circuit by T-state consumption. We commute Clifford gates into the final measurement bases, leaving Pauli rotations that each consume one T state[40]. The core interval $\tau_{\mathrm{core}}$ is the longer of the reaction time and the data block's scheduled period of one (fast), five (intermediate) or nine (compact) logical cycles. The supply interval $\tau_{\mathrm{MSF}}$ is the inverse of the final level's mean output rate. The circuit's T-state consumption interval is $\tau_{\mathrm{T}} = \max(\tau_{\mathrm{core}}, \tau_{\mathrm{MSF}})$. For $S$ accepted shots and circuit success probability $P_{\mathrm{s}}$, the per-attempt and expected total task times are

$$t_{\mathrm{attempt}} = N_{\mathrm{T}}\tau_{\mathrm{T}}, \tag{11a}$$

$$t_{\mathrm{task}} = \frac{S}{P_{\mathrm{s}}} t_{\mathrm{attempt}}. \tag{11b}$$

Here $S$ counts accepted shots, that is, attempts that pass the circuit's success checks. We evaluate $P_{\mathrm{s}}$ for the ideal compiled circuit. Walsh–KvN has no postselection, so $P_{\mathrm{s}} = 1$; BE–KvN and LCHS–KvN have $P_{\mathrm{s}} < 1$. Total task time includes failed attempts. Each three-coordinate attempt includes preparation, propagation and one forward and one inverse quantum Fourier transform (QFT); both QFTs are included in the gate counts and resource estimates. Timing excludes classical fitting and compilation, factory start-up before the first output, additional reset and readout delays and classical data transfer. We use mean supply intervals (Supplementary equation (S60b)) without detailed chip routing or concurrent scheduling.

We compare two supply policies: waiting allowed lets the core pause for T states, whereas supply enforced requires the mean factory output rate to meet core demand. The default search allows waiting. Supplementary Fig. S9 compares minimum-qubit designs under the two policies, optimizing code distances and synthesis precision separately for each.

For a fixed number of propagation steps, we search for the fewest physical qubits and, separately, the shortest total task time within the per-attempt error budget (Fig. 2c). We vary the data block, code distances, distillation levels and numbers of units, testing final-level unit counts from one up to the smallest count at which mean supply meets core demand ($\tau_{\mathrm{MSF}} \le \tau_{\mathrm{core}}$). We discard Pareto-dominated architectures, pair each remaining one with candidate synthesis precisions and select, among the designs that meet the budget, the one with the fewest qubits and the one with the shortest total task time. Unless a cap is stated, the fastest-design search has no qubit cap. Supplementary Note 11 details the search; Supplementary Figs. S8 and S10 give the sensitivity to error budget, factory architecture and qubit limit (Supplementary Note 12).

## Comparison circuits

The quantum propagation gate bounds in Table 1 count gates per attempt before rotation synthesis and exclude shot repetitions. They hold at fixed drift degree and dimension and, where applicable, stencil order and angular factors in the drift $(1, \cos(2\theta), \sin(2\theta)$ for the three-coordinate model). The classical bound counts arithmetic operations for deterministic, explicit updates at fixed dimension $N_{\mathrm{c}}$ and stencil, with $N_j$ cells in coordinate $j$. The step-count estimate in footnote ‡ fixes the input, duration and state-error tolerance. It applies to coordinate-wise even–odd transport in any fixed dimension under Supplementary Note 15's assumptions. For the three-coordinate Fourier circuit, we test step counts by doubling the steps at fixed grid and by doubling the radial and $z$ cell counts together with the steps (Supplementary Note 7). Supplementary Note 13 gives the assumptions behind the LCHS bound.

The BE–KvN resource estimates are for our reconstruction of the block encoding of Guseynov et al.[38] for the same 64-cell wake generator, input, interval and observables as Walsh–KvN. Its coordinate loader, the subcircuit that writes the cell coordinates into amplitudes, divides them by a normalization factor $\beta_{\mathrm{R}}$. We restore that factor in the drift polynomial and block normalization $\alpha$ to encode the physical wake generator (Supplementary Note 13). An odd polynomial of the coordinate operator encodes the linear and cubic drift terms, and an even polynomial encodes the squared-radius coupling to the phase qubits. The Hamiltonian is the sum of radial transport and this coupling. Quantum singular value transformation approximates its evolution over the full interval with Jacobi–Anger polynomials, without time steps, followed by one round of oblivious amplitude amplification (Supplementary Table S12). We evaluate the amplified evolution polynomial at this Hamiltonian's eigenvalues, propagate the prepared state and reconstruct its mean field to measure bias against the continuum. With sampling variance and implementation error, this gives 813 accepted shots for 1% relative RMS field error. The BE–KvN and LCHS–KvN circuits each use one shared work qubit, initialized and returned to $|0\rangle$, to decompose each multi-controlled gate. Their widths and gate counts enter the same error model, timing model and search (equations (9), (10) and (11a)–(11b)).

Separately, we estimate resources for the LCHS–KvN sine-drift benchmark of Novikau and Joseph[19], whose drift is a cubic polynomial in the sine of the coordinate. Seven Hadamard gates prepare equal amplitudes on 128 cells for ten steps of size 0.1. We tally single-target multi-controlled gates from definitions and angles in QuCF[54], the authors' public circuit framework, and check the total against their saved simulation output. Weighting each decomposition by its multiplicity gives Clifford, native T and arbitrary Z-rotation counts. Multi-controlled X gates use the Khattar–Gidney construction[55]. For other controlled one-qubit gates, we use the Zindorf–Bose decomposition to implement each gate exactly, including the relative phase needed for interference between control branches[56]. Each architecture uses the smallest $n_{\mathrm{T}}$ meeting the budget, including the error from replacing source angles near multiples of $\pi/4$ by those exact values. We use 815 accepted shots, matching the radial Walsh–KvN benchmark's shot count, to set the sampling workload. This sine-drift calculation does not reproduce the wake task or enforce its 1% field-error target, so its total time is not a comparison at matched prediction accuracy. The saved amplitudes give success probability 0.8763 for timing. Applying this sine encoding to the wake would require approximating the drift composed with the inverse sine[19,57].

To reproduce the QuCF circuit, we use its original LCHS kernel, which assigns Lorentzian weights to the Hamiltonian evolutions (Supplementary Note 13). Later LCHS kernels[58–61] may reduce the physical resource requirements of the LCHS–KvN benchmark, but we do not quantify this reduction here.

## Radial transport experiment on IBM hardware

We test outward radial transport on IBM hardware with 16 cells prepared by a four-qubit MPS of bond dimension two (Extended Data Fig. 1). Four or six steps span nondimensional times 40–49.829. Step zero measures the prepared input. For each later step, a reference circuit sets the transport circuit's Walsh angles to zero. The four- and six-step datasets use 16,000 and 32,000 shots per circuit, respectively. Supplementary Note 14 describes the device and calibration. The Source Data for Extended Data Fig. 1 give the measured and corrected 16-cell results with their uncertainties. They also include a separate five-qubit, 32-cell control run over the first step of the four-step schedule.

To correct for noise at step $k$, we fit the measured reference distribution to a mixture of the ideal input and a uniform distribution (an effective depolarizing model) by least squares. The fitted weight of the ideal input defines the reference contrast $F_k$. It equals one when the measurements reproduce the ideal input and zero when all cells are equally likely. Assuming that the transport and reference circuits attenuate the measured observable by the same contrast, we add the transport–reference difference divided by $F_k$ to the ideal input observable (Supplementary equation (S86)). Corrected probabilities sum to one and can be negative. Dividing by $F_k$ amplifies shot noise, so we attach bootstrap error bars to the corrected values: 1,000 resamples of both circuits' counts, with the contrast refitted each time. The bars cover shot noise only. These hardware circuits act on the radial register alone, without the phase qubits.

### Use of generative AI

The authors used Claude (Anthropic) and ChatGPT (OpenAI) for assistance with language editing of the manuscript, code development and drafting the derivation of the step-count bound in Supplementary Note 15. The authors reviewed and verified all text, code, derivations and results, and take full responsibility for the content.

## Data availability

Numerical source data supporting the figures and tables, including Supplementary Data 1–10, are deposited in Zenodo at https://doi.org/10.5281/zenodo.24505927.

## Code availability

The code for simulation, resource estimation, analysis and figure reproduction is maintained at https://github.com/hsnoh9003-lab/walsh-kvn.

## Acknowledgements

This work was supported by Korea National Research Foundation (NRF) grant No. NRF-RS-2026-25612745, ICT R&D program of IITP-2023-2021-0-01810, RS-2023-00225385, and AFOSR grant FA2386-22-1-4052. The work was partly supported by the Cleveland Clinic Quantum Start-up Catalyzer Program and Amazon Web Services. Authors acknowledge the use of IBM Quantum services for this work through the Cleveland Clinic's Quantum Start-up Catalyzer Program. The views expressed are those of the authors, and do not reflect the official policy or position of IBM or the IBM Quantum team. This work was also supported by the National Quantum Laboratory at Maryland (QLab). We acknowledge QLab for providing quantum computing resources on IonQ devices during this project. The views expressed are those of the authors, and do not reflect the official policy or position of IonQ.

## Author contributions

H.N. devised the overall framework for early fault-tolerant flow prediction, combining low-degree models identified with SINDy, Walsh synthesis of Koopman–von Neumann quantum circuits and velocity-field reconstruction from joint radius–phase readout. H.N. and U.O. worked on resource estimation and quantum circuit development. H.N. and S.B. worked on reduced-order modelling and finite-element flow simulations. D.A. proposed investigating quantum algorithms for nonlinear dynamics and acquired funding. D.A. performed the IBM hardware experiments. H.N. wrote the manuscript. All authors discussed the results and commented on and revised the manuscript.

## Competing interests

D.A. is a co-founder of Singularity Quantum Inc. and holds an equity interest in the company. The other authors declare no competing interests.

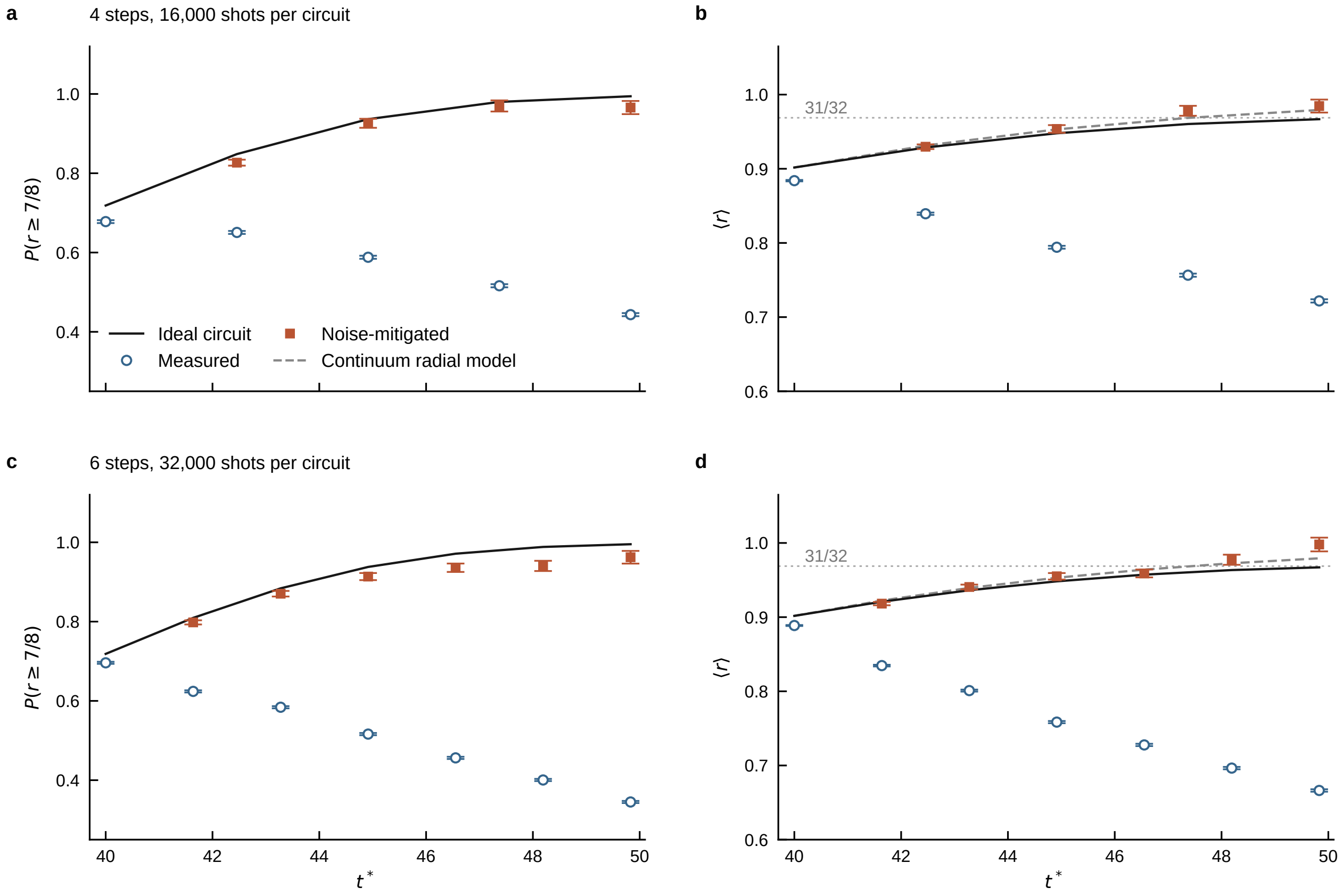

**Extended Data Fig. 1 | Nonlinear radial KvN transport on IBM hardware.** We test whether probability moves towards the saturated wake amplitude on four ibm_cleveland qubits. A bond-dimension-two matrix product state prepares 16 cells. Four steps (a,b) use 16,000 shots per circuit and six (c,d) 32,000. **a**,**c**, Probability in the two outermost cells ($r \geq 7/8$, centres 29/32 and 31/32). **b**,**d**, Mean radius. Dotted lines mark 31/32, the largest possible mean for nonnegative cell probabilities. Correction can give larger estimates. Solid lines show ideal circuits and dashed lines continuum means. Bars show one standard error from shot noise for measured circles and one bootstrap standard deviation for corrected (noise-mitigated) squares (1,000 resamples of both circuits' counts). Small bars lie inside the markers. See Supplementary Note 14.

# Supplementary Information

*Nonlinearity-preserving quantum simulation of flows in the early fault-tolerant regime*

Hansol Noh[1,*], Unam Oh[1], Sangkyu Baek[1], Doyeol (David) Ahn[1,2,*]

[1] Department of Electrical and Computer Engineering and Center for Quantum Information Processing, University of Seoul, Seoul 02504, Republic of Korea.

[2] Singularity Quantum Inc., Newport Beach, CA, USA.

*Correspondence to Hansol Noh (hs.noh9003@gmail.com) and Doyeol (David) Ahn (dahn@uos.ac.kr).

## Supplementary Note 1  Wake model and benchmark definition

Case A alone supplies the spatial modes and fitted coefficients of the wake model; five further cases, B–F, test it. The six cases differ in the signed amplitude, centre and width of the initial streamfunction perturbation to the same unstable steady wake (Reynolds number 100, Fig. S1c). Two modes describe vortex shedding, the periodic release of vortices behind the cylinder. The third describes the change in mean flow. At each time, three modal coordinates weight these patterns. Adding their sum to the unstable steady-state solution $\mathbf{u}_{\mathrm{st}}$ approximates the velocity field[1–3].

$$\mathbf{u}(t^*) \simeq \mathbf{u}_{\mathrm{st}} + \boldsymbol{\varphi}_1 a_1(t^*) + \boldsymbol{\varphi}_2 a_2(t^*) + \boldsymbol{\varphi}_3 a_3(t^*). \tag{S1}$$

To isolate shedding, we reflect 250 case A snapshots, velocity fields saved at nondimensional times 20–80, across the centreline, $(\mathcal{R}\mathbf{u})(x,y,t^*) = \left(u(x,-y,t^*), -v(x,-y,t^*)\right)^{\mathrm{T}}$. Proper orthogonal decomposition (POD) extracts the two dominant patterns of the antisymmetric field $\mathbf{u}_{\mathrm{a}} = (\mathbf{u} - \mathcal{R}\mathbf{u})/2$. In the $(a_1, a_2)$ plane, distance from the origin measures shedding amplitude and angle gives phase. POD gives $\langle a_1 a_2 \rangle = 0$, where brackets denote the average over these snapshots. For a common scale, we divide each modal coordinate by $\sqrt{2}$ times its snapshot root-mean-square (RMS), giving $\langle a_1^2 \rangle = \langle a_2^2 \rangle = 1/2$, so that the mean squared shedding amplitude $\langle a_1^2 + a_2^2 \rangle$ is one. We multiply each mode by the same factor, leaving the velocity field unchanged. We choose the sign of $a_2$ so phase advances on average.

Let $\mathbf{u}_{\mathrm{time}}$ be the velocity averaged over two shedding cycles (times 75.55–87.33). Subtracting the unstable steady-state solution $\mathbf{u}_{\mathrm{st}}$ gives the mean change. We remove its shedding components and normalize the remainder to obtain the mean-flow shift mode[1].

$$\boldsymbol{d} = \mathbf{u}_{\mathrm{time}} - \mathbf{u}_{\mathrm{st}}, \tag{S2a}$$
$$\boldsymbol{d}_\perp = \left(\mathbf{I} - \boldsymbol{\Psi}_0 \boldsymbol{\Psi}_0^{\mathrm{T}}\right)\boldsymbol{d}, \tag{S2b}$$
$$\boldsymbol{\varphi}_3 = \boldsymbol{d}_\perp / \|\boldsymbol{d}_\perp\|_2. \tag{S2c}$$

In (S2b), $\boldsymbol{\Psi}_0$ is an orthonormal basis for $\boldsymbol{\Psi} = [\boldsymbol{\varphi}_1, \boldsymbol{\varphi}_2]$. Since the rescaled modes $\boldsymbol{\Psi}_{\mathbf{a}} = (\boldsymbol{\varphi}_1, \boldsymbol{\varphi}_2, \boldsymbol{\varphi}_3)$ are orthogonal but not of unit norm, (S3b) uses their inner products, collected in $\mathbf{G}$, to find $\mathbf{a} = (a_1, a_2, a_3)^{\mathrm{T}}$ by minimizing the sum of squared velocity differences over field sites and both components. Equation (S3c) gives the reduced-order model (ROM) reconstruction $\mathbf{u}_{\mathrm{ROM}}$.

$$\mathbf{G} = \boldsymbol{\Psi}_{\mathbf{a}}^{\mathrm{T}} \boldsymbol{\Psi}_{\mathbf{a}}, \tag{S3a}$$
$$\mathbf{G}\mathbf{a}(t^*) = \boldsymbol{\Psi}_{\mathbf{a}}^{\mathrm{T}} \left[\mathbf{u}(t^*) - \mathbf{u}_{\mathrm{st}}\right], \tag{S3b}$$
$$\mathbf{u}_{\mathrm{ROM}}(t^*) = \mathbf{u}_{\mathrm{st}} + \boldsymbol{\Psi}_{\mathbf{a}} \mathbf{a}(t^*). \tag{S3c}$$

### Choosing the polynomial dynamics and fitting the mean-flow shift

We choose the polynomial degree by weighing field accuracy against circuit cost. For each maximum degree from 1 to 5, sparse identification of nonlinear dynamics (SINDy) fits each coordinate's time derivative by linear regression on candidate monomials such as $a_1 a_3$. We split 292 case A snapshots (times 10–80) into four consecutive blocks. We fit on three blocks and forecast the fourth from its first state, testing each block in turn. We divide each coordinate by its training root-mean-square value so the three have comparable scales. To estimate a derivative at one snapshot, we use a centred difference spanning five consecutive snapshots, two before and two after the central one. Its error is fourth order in the sampling interval. We use a training derivative only when all five snapshots belong to the training set.

Let $\mathbf{v}_k$ be the vector of values of candidate term $k$ at the training snapshots used for derivatives. For example, for $a_1 a_3$ each entry is the product of the two scaled coordinates at that snapshot. The corresponding entry of $\mathbf{y}$ is the estimated time derivative of one scaled coordinate at the same snapshot. The two vectors have one entry per usable training time. We fit the weights $b_k$ by minimizing

$$\left\| \mathbf{y} - \sum_k b_k \frac{\mathbf{v}_k}{\| \mathbf{v}_k \|_2} \right\|_2^2 + 10^{-4} \sum_k b_k^2. \tag{S3d}$$

The first term measures error in the predicted derivatives. The second penalizes large weights. Scaling each column to unit length puts all terms on the same penalty scale. We then remove terms whose coefficient magnitude $|b_k|/\| \mathbf{v}_k \|_2$ falls below the cutoff and refit the rest.

Against the three-mode projection of the simulated velocity field, the mean relative RMS velocity-field error across the four test blocks is 5.62% for the quadratic fit (cutoff 0.03) and 2.85% for the cubic (cutoff 0.01). We choose the quadratic because the cubic raises the rotation bound nearly sixfold. On a Cartesian grid with 64 cells per coordinate, summing one layer of pair rotations along each of the three coordinates before excluding pairs that cross the domain boundaries gives a bound of 226 Walsh rotations for the quadratic and 1,297 for the cubic (Note 3). In the equation for coordinate $i \in \{1,2,3\}$, a monomial with powers $(p_1, p_2, p_3)$ can produce Walsh subsets with at most $p_j$ bits from coordinate $j \in \{1,2,3\}$; coordinate $i$ supplies its $n - 1 = 5$ pair bits and the others $n = 6$ bits. Counting the distinct subsets over the retained monomials, with overlaps counted once, gives $84 + 84 + 58 = 226$ for the quadratic and $424 + 379 + 494 = 1{,}297$ for the cubic. The quadratic retains 12 coefficients, including four of the five terms in the shift equation (S4b).

In original units, the shedding coordinates $\mathbf{a}_\perp = (a_1, a_2)^\mathrm{T}$ and the shift obey

$$\frac{\mathrm{d}\mathbf{a}_\perp}{\mathrm{d}t^*} = (\mathbf{L} + a_3 \mathbf{B}_{\mathrm{dyn}}) \mathbf{a}_\perp, \tag{S4a}$$

$$\frac{\mathrm{d}a_3}{\mathrm{d}t^*} = \lambda_3 a_3 + q_{11} a_1{}^2 + q_{12} a_1 a_2 + q_{22} a_2{}^2 + q_{33} a_3{}^2. \tag{S4b}$$

In (S4a), $\mathbf{L}$ governs shedding when the shift is zero, and a nonzero shift modifies it by $a_3 \mathbf{B}_{\mathrm{dyn}}$. In (S4b), $\lambda_3$ is the fitted coefficient of the linear shift term. The SINDy-fitted coefficients $q_{11}, q_{22}, q_{33}$ multiply the squares of the three coordinates in order, and $q_{12}$ multiplies their cross term $a_1 a_2$. We round to four significant figures.

$$\mathbf{L} = \begin{pmatrix} 0.06772 & -0.7906 \\ 0.7156 & 0.1172 \end{pmatrix}, \tag{S4c}$$

$$\mathbf{B}_{\mathrm{dyn}} = \begin{pmatrix} -0.002720 & -0.009648 \\ 0.01253 & -0.004077 \end{pmatrix}. \tag{S4d}$$

Thresholding (S4b) gives $q_{12} = 0$ (scaled coefficient −0.019, whose magnitude is below the 0.03 cutoff). We model the shift as $a_3 \approx c_0 + c_2 \varrho^2$, where $c_0$ is its value at zero shedding amplitude. Training snapshot $i$ pairs squared shedding amplitude $\varrho_i^2 = a_{1,i}^2 + a_{2,i}^2$ with shift $a_{3,i}$. Both vanish at steady state, so $c_0 = 0^2$. Linear regression minimizes $\sum_i \left(a_{3,i} - c_2 \varrho_i^2\right)^2$, giving

$$c_2 \sum_i \left(\varrho_i^2\right)^2 = \sum_i \varrho_i^2 a_{3,i} \quad \Rightarrow \quad c_2 = \frac{\sum_i \varrho_i^2 a_{3,i}}{\sum_i \varrho_i^4}. \tag{S5}$$

For $\varrho > 0$, the shedding coordinates $a_1 = \varrho\cos\theta$ and $a_2 = \varrho\sin\theta$ give the radial direction $(\cos\theta, \sin\theta)$. Project the fitted rates in (S4a) onto this direction for amplitude growth and onto the tangent $(-\sin\theta, \cos\theta)$ for speed around the circle.

$$\dot{\varrho} = \dot{a}_1\cos\theta + \dot{a}_2\sin\theta, \qquad \text{(S6a)}$$
$$\varrho\dot{\theta} = -\dot{a}_1\sin\theta + \dot{a}_2\cos\theta. \qquad \text{(S6b)}$$

Substitute (S4a) and $a_3 = c_2\varrho^2$ into (S6a)–(S6b), then average uniformly over one turn at fixed amplitude (brackets with subscript θ). With $J = L + a_3 B_{\text{dyn}}$, the radial rate is $\varrho[J_{11}\cos^2\theta + (J_{12} + J_{21})\sin\theta\cos\theta + J_{22}\sin^2\theta]$. Averaging replaces $\cos^2\theta$ and $\sin^2\theta$ by $1/2$ and removes the mixed term, leaving $\varrho\,\mathrm{tr}(J)/2$. The angular rate likewise averages to $(J_{21} - J_{12})/2$. Here $L_{ij}$ and $B_{ij}$ denote the entries in row $i$, column $j$ of $\mathbf{L}$ and $\mathbf{B}_{\text{dyn}}$, respectively.

$$\langle\dot{\varrho}\rangle_\theta = \mu\varrho - \beta\varrho^3, \qquad \text{(S6c)}$$
$$\langle\dot{\theta}\rangle_\theta = \omega + \gamma_{\text{raw}}\varrho^2, \qquad \text{(S6d)}$$
$$\mu = \frac{L_{11} + L_{22}}{2}, \qquad \text{(S6e)}$$
$$\beta = -c_2\frac{B_{11} + B_{22}}{2}, \qquad \text{(S6f)}$$
$$\omega = \frac{L_{21} - L_{12}}{2}, \qquad \text{(S6g)}$$
$$\gamma_{\text{raw}} = c_2\frac{B_{21} - B_{12}}{2}. \qquad \text{(S6h)}$$

For a nonzero intercept, replace $L$ by $L + c_0 B_{\text{dyn}}$. With saturation radius $R = (\mu/\beta)^{1/2}$, where the averaged radial rate (S6c) vanishes, set $r = \varrho/R$, $x = a_1/R$, $y = a_2/R$ and $z = (a_3 - c_0)/(c_2R^2)$. Then $r^2 = x^2 + y^2$ and $z \approx r^2$. Time $\tau = \mu(t^* - 40)$ starts at zero at time 40, where the radial benchmark and the late three-coordinate run (Note 7) begin, near saturation. The growth run of the same circuit begins earlier, at time 29.8.

$$\frac{\mathrm{d}r}{\mathrm{d}t^*} = \mu(r - r^3), \qquad \text{(S7a)}$$
$$\frac{\mathrm{d}r}{\mathrm{d}\tau} = r - r^3, \qquad \text{(S7b)}$$
$$\gamma = \gamma_{\text{raw}}R^2. \qquad \text{(S7c)}$$

$$\frac{\mathrm{d}\theta}{\mathrm{d}t^*} = \omega + \gamma r^2, \qquad \text{(S8a)}$$
$$\frac{\mathrm{d}\theta}{\mathrm{d}\tau} = \frac{\omega}{\mu} + \frac{\gamma}{\mu}r^2. \qquad \text{(S8b)}$$

The factor $R^2$ in (S7c) comes from $\varrho = Rr$.

We split the phase into a known part and a residual, $\theta = \theta_c + \Phi$. The known part advances at a constant rate, $\theta_c = \theta_0 + \omega(t^* - 40)$, with $\theta_0 = \mathrm{atan2}(a_2, a_1)$ interpolated to $t^* = 40$. The circuit accumulates only the residual $\Phi$, which starts at zero and grows at $\mathrm{d}\Phi/\mathrm{d}t^* = \gamma r^2$.

**Table S1 | Radial wake parameters.** The Source Data for Table S1 give their full precision.

| Parameter | Value | Parameter | Value |
|---|---|---|---|
| $\mu$ | 0.092482 | $R$ | 1.1023 |
| $\omega$ | 0.75309 | $\gamma$ | 0.30173 |
| $\theta_0$ | 0.36385 | $c_2$ | 22.396 |

The coefficients $\mu$ and $\gamma$ obtained from (S6e)–(S7c) set the rotation angles of radial transport (S7a) and of the phase updates (S8a). The circuit estimates three ensemble averages, or moments, $\mathbf{m} = (m_1, m_2, m_3)^T =$ $(\langle r\cos\Phi\rangle, \langle r\sin\Phi\rangle, \langle r^2\rangle)^T$. To reconstruct velocity from the stored mode columns $\mathbf{\Psi}_\mathrm{m}=(\boldsymbol{\varphi}_1,\boldsymbol{\varphi}_2,c_2\boldsymbol{\varphi}_3)$, $\mathbf{M}$ rotates the first two moments by the known phase $\theta_\mathrm{c}$ and rescales them by $R$. The third scales by $R^2$ because it measures squared radius. The field $\overline{\mathbf{u}}$ in (S9b) is the ensemble mean. For the fitted zero-intercept model, the spatial modes in Fig. 1e are $\boldsymbol{\varphi}_x = R\boldsymbol{\varphi}_1$, $\quad\boldsymbol{\varphi}_y = R\boldsymbol{\varphi}_2$, $\quad\boldsymbol{\varphi}_z = c_2R^2\boldsymbol{\varphi}_3$.

$$\mathbf{M}=\begin{pmatrix} R\cos\theta_\mathrm{c} & -R\sin\theta_\mathrm{c} & 0 \\ R\sin\theta_\mathrm{c} & R\cos\theta_\mathrm{c} & 0 \\ 0 & 0 & R^2 \end{pmatrix}, \tag{S9a}$$

$$\overline{\mathbf{u}}=\mathbf{u}_\mathrm{st}+\mathbf{\Psi}_\mathrm{m}\mathbf{M}\mathbf{m}. \tag{S9b}$$

The benchmark's 1% field-error target is measured against a continuum reference: the same radial model and initial distribution, evolved without a spatial grid.

## Testing forecasts against finite-element velocity fields

Projection measures how closely three modes represent each velocity snapshot computed by the finite-element method (FEM). The radial model predicts the modal coordinates from the first snapshot of each forecast window (40.08–49.68 and 20.16–90). Over the longer window, case D, the weakest disturbance, gives the largest gap between forecast and projection (19.6% versus 7.3%; Fig. S1b). Supplementary Data 1 give the short-window errors in Fig. S1a.

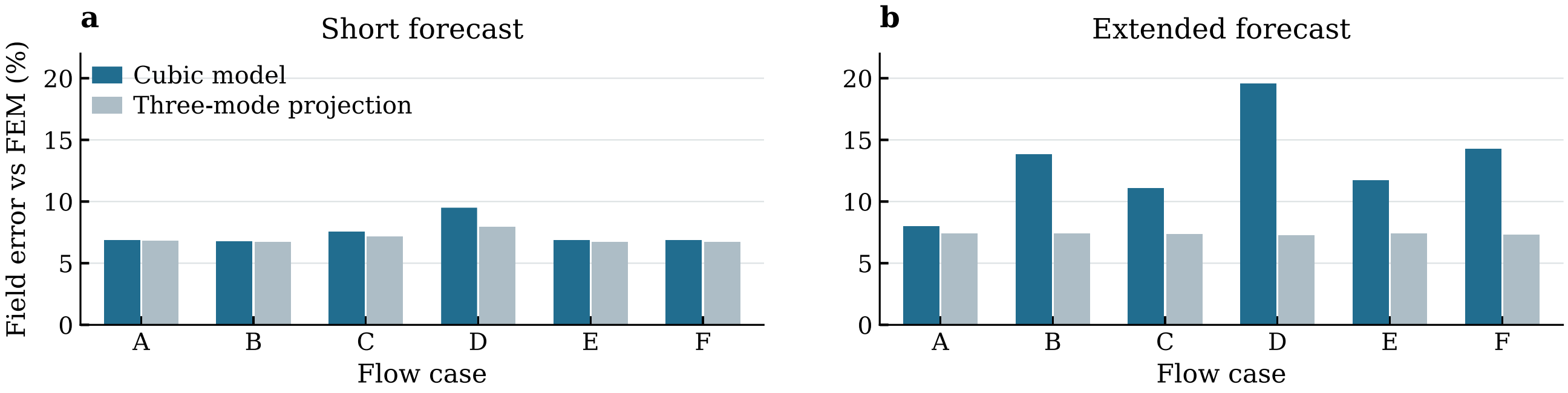

**c**

$$\delta\psi(x,y) = A_\psi\, C(x)\exp\left[-\frac{(x-x_c)^2+(y-y_c)^2}{w^2}\right]$$

| **Initial disturbance** | **A** | **B** | **C** | **D** | **E** | **F** |
|---|---|---|---|---|---|---|
| Amplitude, $A_\psi$ | +0.020 | +0.035 | −0.015 | +0.010 | −0.030 | +0.050 |
| Downstream of cylinder centre, $x_c$ (diameters) | 2.0 | 2.7 | 1.8 | 2.3 | 2.5 | 3.0 |
| Height above wake centreline, $y_c$ (diameters) | 0.00 | +0.25 | −0.30 | −0.20 | +0.15 | 0.00 |
| Width, $w$ (diameters) | 0.70 | 0.80 | 0.60 | 0.65 | 0.75 | 0.90 |

**Supplementary Figure S1 | Wake forecast errors against FEM. a**, **b**, Radial forecasts (dark) and three-mode projections (light) over 40.08–49.68 and 20.16–90. Case D, the weakest disturbance, gives the largest extended-window gap (19.6% versus 7.3%). **c**, We measure $x$ downstream from the cylinder centre in diameters. To preserve the velocity at the surface of the cylinder, the Gaussian streamfunction multiplier $C(x)$ is zero through $x = 0.7$ and one from $x = 1.2$. It rises smoothly 0.2–0.7 diameters behind the rear surface of the cylinder.

## Register size with and without model reduction

Model reduction changes register size. Direct encoding uses $N_\mathrm{f}=sN_\mathrm{s}$ coordinates ($s$ components at $N_\mathrm{s}$ sites), with $N_\mathbf{u}$ cells each. Encoding $N_\mathrm{c}$ modal coordinates with $N_r$ cells each gives (S10a)–(S10c). Here $\mathcal{N}_\mathrm{full}$ counts full-field configurations, and $q_\mathrm{full}$ and $q_\mathrm{ROM}$ count the position qubits for the full and reduced fields. With 64 cells per coordinate, the three-coordinate model uses 18 position qubits. Direct encoding would use six per field coordinate.

$$\mathcal{N}_\mathrm{full}=N_\mathbf{u}^{N_\mathrm{f}}, \tag{S10a}$$

$$q_\mathrm{full}=N_\mathrm{f}\lceil\log_2 N_\mathbf{u}\rceil, \tag{S10b}$$

$$q_\mathrm{ROM}=N_\mathrm{c}\lceil\log_2 N_r\rceil. \tag{S10c}$$

## Supplementary Note 2 Probability transport on the radial grid

An ensemble can concentrate while conserving total probability. To reproduce this on a circuit, we use Koopman–von Neumann (KvN) evolution. The drift $F$ gives each member's velocity, the continuity equation $\partial_t \rho = -\nabla\cdot(\mathbf{F}\rho)$ transports the ensemble density, and the corresponding amplitude ($\rho=\psi^*\psi$) obeys $\partial\psi/\partial t = L\psi$, with

$$\mathrm{L}\psi = -\mathbf{F}\cdot\nabla\psi - 1/2\,(\nabla\cdot\mathbf{F})\psi. \tag{S11}$$

The half-divergence term gives $\partial_t|\psi|^2 = -F\cdot\nabla|\psi|^2 - (\nabla\cdot F)|\psi|^2$. The right-hand side equals $-\nabla\cdot(F|\psi|^2)$, so the probability density obeys the continuity equation.

To find the probability flow across the boundary of the domain $D$, integrate by parts using outward unit normal $\mathbf{n}$,

$$\langle f, Lg\rangle + \langle Lf, g\rangle = -\int_{\partial D} f^*\, g\, \mathbf{F}\cdot\mathbf{n}\,\mathrm{d}S. \tag{S12}$$

For $f = g = \psi$ this is the net probability inflow per unit time. Zero flux makes $H = i\mathrm{L}$ symmetric on its domain[4,5] and preserves total probability.

On a grid of cells of width $h$, a centred difference uses the drift $F_{j+1/2}$ at the face shared by two cells. Equal and opposite couplings rotate the two amplitudes and conserve their combined probability. The shared-face value $F_{j+1/2} = f_{\text{face}}$ is the one used in Methods equations (5a)–(5b). The discretized generator matrix $\mathbf{\Omega}$ has entries

$$\Omega_{j+1,j} = \frac{F_{j+1/2}}{2h}, \tag{S13a}$$

$$\Omega_{j,j+1} = -\frac{F_{j+1/2}}{2h}. \tag{S13b}$$

Row $j$ reads $(\Omega\psi)_j = \left[F_{j-1/2}\psi_{j-1} - F_{j+1/2}\psi_{j+1}\right]/(2h)$. For smooth $F$ and $\psi$, expanding about the cell centre gives $-F\psi' - F'\psi/2 + O(h^2)$, so $\Omega$ approximates $L$ in (S11) to second order, including the half-divergence term. Opposite signs make $\Omega$ real and skew-symmetric, so $e^{t\Omega}$ preserves the norm. Zero drift at the endpoints removes the coupling between the last and first cells.

Even pairs (0,1), (2,3), … rotate independently, as do odd pairs such as (1,2) and (3,4). We write $\Omega = \Omega_\mathrm{e} + \Omega_\mathrm{o}$ for the even-pair and odd-pair couplings. Each face uses an $R_Y$ rotation through $F_{j+1/2}\delta/h$ over duration $\delta$, since the coupling $F_{j+1/2}/(2h)$ turns the pair by $F_{j+1/2}\delta/(2h)$ and $R_Y(\vartheta)$ turns it by $\vartheta/2$. An even half step, odd full step and even half step give[6]

$$U_r(\delta) = e^{\delta\Omega_\mathrm{e}/2} e^{\delta\Omega_\mathrm{o}} e^{\delta\Omega_\mathrm{e}/2} = e^{\delta\Omega} + O(\delta^3). \tag{S14}$$

The exact quadratic term $\delta^2(\Omega_\mathrm{e} + \Omega_\mathrm{o})^2/2$ contains both mixed products ($\Omega_\mathrm{e}\Omega_\mathrm{o}, \Omega_\mathrm{o}\Omega_\mathrm{e}$), each with coefficient $\delta^2/2$. The even half steps reproduce both, so the expansions agree through $\delta^2$. The one-step error is therefore $O(\delta^3)$. At fixed time $T$, summing this error over $T/\delta$ steps gives $(T/\delta)\,O(\delta^3) = O(\delta^2)$. Exact-grid evolution applies the matrix exponential of the grid generator for the full duration. The field difference introduced by splitting each step into layers is the Trotterization error. On 64 cells, doubling the steps from 33 to 66 cuts it fourfold, from 0.0230% to 0.00573% (Fig. S3c).

## Supplementary Note 3 Walsh synthesis and common-target alignment

The Walsh-synthesized KvN circuit (Walsh–KvN) sets each pair's rotation from the position of the face shared by its two cells. The grid has $N_x = 2^n$ cells of width $h = 1/N_x$. The $n$ register qubits store a cell number in binary. We number the pairs in each layer from zero. In an even pair, the lowest bit distinguishes the two cells.

The other $m = n - 1$ bits give its pair number, written $b = (b_0, \dots, b_{m-1})$, with $b_i$ equal to 0 or 1 and $b_0$ the least significant pair bit.

**Walsh synthesis of the pair rotations**

The Walsh transform expresses each rotation angle as a function of the bits of the pair number. The shared face is at $x(b) = x_0 + \sum_i w_i b_i$. For the first pair its coordinate $x_0$ is $h$ in the even layer and $2h$ in the odd layer. Successive faces in either layer are $2h$ apart, so the distance $w_i$ contributed when bit $b_i$ is 1 equals $2h \times 2^i$. The rotation angle $\vartheta$ is proportional to the drift at the shared face, so for a drift of degree $K$ the angle has degree $K$ in $x$. Because $b_i^a = b_i$ for every positive integer $a$, each term contains at most $K$ distinct bits. Writing $b_i = (1 - z_i)/2$ with $z_i = (-1)^{b_i}$ expresses the angle as a sum of Walsh terms. Each term multiplies a coefficient $\hat{\vartheta}_S$ by a sign for a subset $S$ of at most $K$ bit positions. The sign is +1 if an even number of those bits equal 1 and $-1$ if an odd number do[7].

$$\vartheta(b) = \sum_{|S| \le K} \hat{\vartheta}_S \prod_{i \in S} (-1)^{b_i}, \tag{S15a}$$

$$N_\mathrm{W} \le \sum_{j=0}^{\min(K,m)} \binom{m}{j}. \tag{S15b}$$

The binary parity basis corresponds to products of Pauli $Z$ operators through $(\prod_{i \in S} Z_i)|b\rangle = (\prod_{i \in S} z_i)|b\rangle$, where $|b\rangle$ stores the pair bits and $Z_i$ acts on bit $i$. Each nonzero Walsh coefficient $\hat{\vartheta}_S$ therefore requires one target rotation. In Fig. S2c, controlled-NOT gates compute the selected parity on the target. Applying a Y rotation and reversing those gates gives the rotation angle the sign of that parity. $N_\mathrm{W}$ counts these rotations in one layer. The sum in (S15b) counts subsets by their size $j$. A cubic on five pair bits allows $1+5+10+10=26$ subsets, giving at most 78 rotations in a three-layer step.

## Incrementers for aligning odd pairs

To align an odd pair with the lowest bit, decrement both cell numbers. For pair number $q$ on $N_x = 2^n$ cells, this sends $(2q+1, 2q+2)$ to $(2q, 2q+1)$ (Fig. S2b). The even pairs already have this form. The increment after the odd-layer rotations restores the original cell numbers.

Incrementer A uses multi-controlled bit flips[8]. Incrementer B uses Gidney's split-register construction[9] with a Cuccaro adder[10]. We run either circuit in reverse for the decrement and forward for the increment. Both use exact Toffoli gates and return their single work qubit to zero.

While B needs fewer Toffoli gates on larger registers, we use A for the 64-cell wake because each increment or decrement on its six-qubit radial register takes 16 Toffoli gates with A and 28 with B. The count grows quadratically with register width for A and linearly for B. At twelve qubits, A takes 100 Toffoli gates per application and B takes 88. For the general transport bound, we use B. Its decrement and increment add $O(n)$ gates per step, preserving the $O((\log_2 N_x)^K)$ transport gate bound before rotation synthesis for every fixed drift degree $K \ge 1$. With A, the cost of these two operations would dominate for $K = 1$ (Note 13).

In reflected Gray code[6], neighbouring cell numbers differ in one bit. Rotating that bit avoids the decrement and increment used for alignment, but the target depends on the pair. On an $n$-bit grid, both even half steps target bit 0 and use at most $W_K(n-1)$ rotations each. Here $W_K(m)$ counts the subsets of size at most $K$ among $m$ free bits for drift degree $K$ (S15b). Odd pairs sharing target bit $j$ form a group, with $j=1,\ldots,n-1$. In each group the lower $j$ bits have a fixed pattern, leaving $n-j-1$ free bits and at most $W_K(n-j-1)$ Walsh terms.

Rotating bit $j$ without checking the lower bits would also mix cells outside the group. Let $\phi$ be the rotation angle of one Walsh term for the given upper bits. We first rotate the target by $+\phi/2$. If the lower bits match the group's pattern, we flip the target, rotate it by $-\phi/2$ and flip it back. Exchanging the two target states reverses the intervening $Y$ rotation, so the second half also contributes $+\phi/2$ and the net angle is $\phi$. For any other pattern, the two rotations act without the flips and cancel, leaving those cells unchanged. This gives two rotations per Walsh term in the odd-layer sum in (S16a). With alignment, all three layers use bit 0, giving (S16b).

$$R_{\text{Gray}} \leq 2W_K(n-1)+2\sum_{j=1}^{n-1} W_K\,(n-j-1)=O\left(n^{K+1}\right), \tag{S16a}$$

$$R_{\text{aligned}} \leq 3W_K(n-1)=O(n^K). \tag{S16b}$$

At fixed $K$, each group contributes $O(n^K)$ rotations. The $n-1$ odd target groups give the extra power of $n$ in (S16a). For cubic drift on 64 cells, $W_3(5)=26$ and the five odd-group Walsh bounds sum to 30. The per-step rotation bounds are $2\times 26+2\times 30=112$ for Gray addressing and $3\times 26=78$ for common-target alignment.

Reusing a computed bit parity saves controlled-NOT gates. We order the retained Walsh terms to reduce changes in the parity stored on the target. Between two terms, we apply one controlled-NOT gate for each bit that changes, rather than rebuilding the parity for every rotation (the "reused" and "rebuilt" columns of Table S2). After the last rotation, controlled-NOT gates clear the remaining parity from the target qubit[7].

One attempt uses 4,803 rotations before synthesis: 19 for preparation, 33 × 78 for transport and 34 × 65 for phase (Table S2). It runs preparation, all steps and the final measurement once. The two phase qubits accumulate the radius-dependent phase (Note 4) in half steps on either side of each transport step. In 33 steps, adjacent phase half steps merge at 32 internal boundaries. Two endpoint half steps bring the total to 34 layers. For six radius bits, the $r^2$ phase has 22 Walsh terms (1 + 6 + 15 subsets of at most two bits). Since $\hat{n}_X+\hat{n}_Y=I-(Z_X+Z_Y)/2$, each Walsh term of $r^2$ appears alone and multiplied by $Z_X$ or $Z_Y$, giving three rotations. For the constant term the first is a global phase, leaving two. Thus one phase layer takes $2+3\times 21=65$ rotations. Preparation uses $19=4n-5$ rotations (Note 5).

**Table S2 | Operations in one 64-cell wake attempt with incrementer A.** Counts are per use before rotation synthesis. The total combines all uses. $R_Z$ counts arbitrary-angle rotations about any axis.

| Operation | Uses | $R_Z$ per use | Native T per use | CNOT per use, rebuilt | CNOT per use, reused |
|---|---|---|---|---|---|
| Prepare radial state | 1 | 19 | 0 | 9 | 9 |
| Transport radial amplitudes | 33 | 78 | 224 | 524 | 284 |
| Apply one phase layer | 34 | 65 | 0 | 174 | 94 |
| Total per attempt | — | 4,803 | 7,392 | 23,217 | 12,577 |

Each transport step uses one decrement and one increment, giving 32 Toffoli gates with incrementer A, or 224 T gates at seven T gates per exact Toffoli.

With incrementer A, parity reuse reduces the circuit total from 26,268 to 15,628 Cliffords; this is the nine-qubit design with a total task time of 1.62 h in main-text Table 1. The circuit with incrementer B and parity reuse needs 12,936 native T gates and 25,066 Cliffords. Native T counts include the fixed T and inverse T gates already present before rotation synthesis. Arbitrary-angle synthesis adds T gates and readout adds single-qubit Cliffords. We checked the circuits using either incrementer against direct cell-pair rotations on all 256 basis inputs (one step) and on a complex superposition (33 steps), with the work qubit initialized to zero.

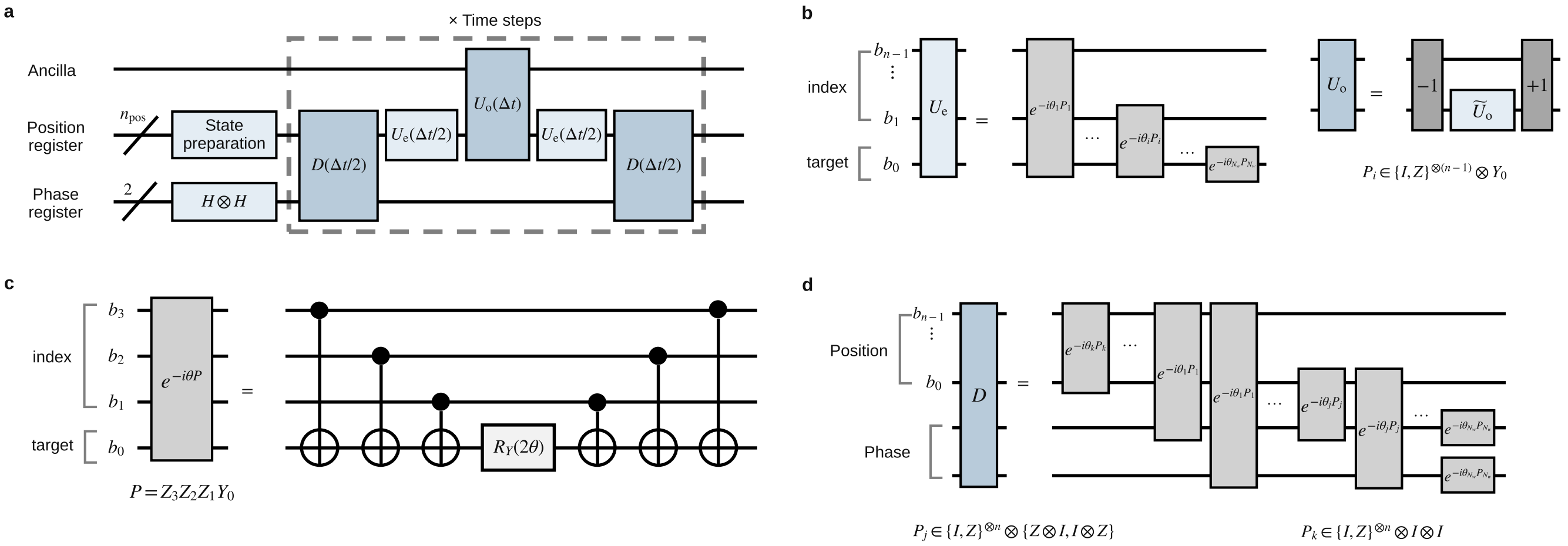

**Supplementary Figure S2 | Circuit for transport between neighbouring cells and phase accumulation. a**, Prepare the ensemble, evolve it and read radius and phase (X and Y). **b**, Relabel odd pairs to rotate the same bit. **c**, Implement one Walsh term by computing the parity of the selected bits, applying the target rotation and undoing the parity computation. **d**, Position-only and position–phase operators.

## Supplementary Note 4 Joint radius–phase readout and field reconstruction

After the last step, we measure the radius and both phase qubits in each shot. These measurements give the means of $r\cos\Phi$, $r\sin\Phi$ and $r^2$ needed to reconstruct the field in (S9b). Angle brackets denote ensemble moments, as in main-text equation (4b); $\mathbb{E}$ denotes expectation over measurement or sampling outcomes. Along a trajectory of radius $r$, residual phase grows at $d\Phi/dt^*=\gamma r^2$ with fitted coefficient $\gamma$. From equal superposition at $t^*$=40, each qubit gains a factor $e^{i\Phi}$ on its state-1 amplitude,

$$\Phi(t^*) = \gamma \int_{40}^{t^*} r(s)^2\, ds, \tag{S17a}$$

$$|q_\Phi\rangle = \frac{|0\rangle + e^{i\Phi}|1\rangle}{\sqrt{2}}. \tag{S17b}$$

Each qubit carries the same phase. A single $X$ or $Y$ measurement returns a sign $s = \pm1$. Its mean gives the cosine or sine, respectively. The sign eigenstates are

$$|s\rangle_X = (|0\rangle + s|1\rangle)/\sqrt{2}, \tag{S17c}$$

$$|s\rangle_Y = (|0\rangle + is|1\rangle)/\sqrt{2}. \tag{S17d}$$

Squaring the overlap gives each probability. The $X$ cross term selects the real part of $e^{i\Phi}$. Conjugating the $Y$ eigenstate contributes $-i$, selecting the imaginary part.

$$\Pr_X(s \mid \Phi) = \frac{|1 + se^{i\Phi}|^2}{4} = \frac{1 + s\cos\Phi}{2}, \tag{S17e}$$

$$\Pr_Y(s \mid \Phi) = \frac{|1 - ise^{i\Phi}|^2}{4} = \frac{1 + s\sin\Phi}{2}. \tag{S17f}$$

Subtracting the probability of $-1$ from that of $+1$ gives the mean sign at fixed phase,

$$\mathbb{E}[s_x \mid \Phi] = \cos\Phi, \tag{S17g}$$

$$\mathbb{E}[s_y \mid \Phi] = \sin\Phi. \tag{S17h}$$

A Hadamard reads $X$, and an inverse phase gate then a Hadamard reads $Y$. Pairing each sign with the same shot's radius preserves their correlations,

$$\mathbb{E}[rs_x] = \langle r\cos\Phi\rangle, \tag{S17i}$$

$$\mathbb{E}[rs_y] = \langle r\sin\Phi\rangle. \tag{S17j}$$

Transport can entangle radius with phase because the radius sets the angle of each subsequent phase rotation. Basis changes before measurement read X on phase qubit $P_X$ and Y on phase qubit $P_Y$. The resulting $4 \times 4$ basis map $\mathbf{U}_{read}$ acts on the four phase-branch amplitudes $\psi_k(j)$ in cell $j$ at radius $r_j$, where $k$ labels the branch. The squared magnitude of the transformed amplitude for sign pair $i$ gives its joint probability $p_{ij}$ with that cell. Each shot yields $\xi_{ij} = (r_j s_x, r_j s_y, r_j^2)^T$. Its mean $\boldsymbol{m}$ gives the three moments used to reconstruct the field in (S9b).

$$p_{ij}=\left|\sum_{k=0}^{3}\mathbf{U}_{\mathrm{read}\,ik}\psi_k(j)\right|^2, \tag{S17k}$$

$$\mathbf{m}=\sum_{ij}p_{ij}\boldsymbol{\xi}_{ij}. \tag{S17l}$$

For the three stored modes, restoring $\theta_c$ and summing their contributions costs $O(N_s)$ operations for $N_s$ field sites.

## Drag and lift observables

We infer cylinder forces from the wake coordinates using polynomials fitted to finite-element drag and lift (times 10–60). Reflection symmetry allows $1, z, x^2, xy, y^2, z^2$ for drag and $x, y, xz, yz$ for lift. Forces are scaled by $\rho U_\infty^2 D/2$ for density $\rho$, inflow speed $U_\infty$ and cylinder diameter $D$ (main-text ref. 35).
Writing the fitted coefficients as $c$ with subscripts naming their monomials, the force polynomials are

$$C_{\mathrm{D}} = c_1 + c_z z + c_{xx}x^2 + c_{xy}xy + c_{yy}y^2 + c_{zz}z^2, \tag{S17m}$$

$$C_{\mathrm{L}} = c_x x + c_y y + c_{xz}xz + c_{yz}yz. \tag{S17n}$$

With $x = r\cos\theta,\ y = r\sin\theta,\ z = r^2$, their ensemble means are

$$\begin{aligned}\langle C_{\mathrm{D}}\rangle \ &= c_1 + \left[c_z + \frac{c_{xx}+c_{yy}}{2}\right]\langle r^2\rangle + c_{zz}\langle r^4\rangle \\ &+\frac{c_{xx}-c_{yy}}{2}\langle r^2\cos 2\theta\rangle + \frac{c_{xy}}{2}\langle r^2\sin 2\theta\rangle,\end{aligned} \tag{S17o}$$

$$\langle C_{\mathrm{L}}\rangle = \langle (c_x r + c_{xz}r^3)\cos\theta\rangle + \langle (c_y r + c_{yz}r^3)\sin\theta\rangle. \tag{S17p}$$

The full phase is $\theta = \theta_c + \Phi$, so $2\theta = 2\theta_c + 2\Phi$. We rotate the measured first harmonic by $\theta_c$ for lift and the second by $2\theta_c$ for drag. Along a trajectory with residual phase $\Phi$, the mean X reading is $\cos\Phi$ and the mean Y reading is $\sin\Phi$ (S17g,h).
At each of 34 saved times (initially and after each of the 33 steps), we draw 815 simulated measurement shots from the ideal circuit distributions and use them jointly for drag and lift. These samples are additional to the velocity-field readout and are excluded from the resource estimates in main-text Table 1. We use 204 shots each for XX, XY and YX and 203 for YY. The first letter labels $P_{\mathrm{Y}}$ and the second $P_{\mathrm{X}}$. Each shot gives a radius $r$ and a reading of $+1$ or $-1$ from each phase qubit. We estimate each radial moment by averaging the four basis-pair means. The $z^2 = r^4$ term in drag requires the fourth moment.
For drag, $x^2 - y^2 = r^2\cos(2\theta)$ and $2xy = r^2\sin(2\theta)$. Before restoring $2\theta_c$, we estimate the residual-phase moments $\langle r^2\cos(2\Phi)\rangle$ and $\langle r^2\sin(2\Phi)\rangle$. A shot contributes $+r^2$ if the two qubit readings agree and $-r^2$ if they differ. Along a trajectory with residual phase $\Phi$, each phase qubit carries a copy of the state in (S17b). The product of the two readings therefore has mean $\cos^2\Phi$ for XX, $\sin^2\Phi$ for YY and $\sin\Phi\cos\Phi$ for XY or YX. We average the radius-weighted products within each basis pair. The XX mean minus the YY mean gives the cosine moment because $\cos^2\Phi - \sin^2\Phi = \cos(2\Phi)$. The XY mean plus the YX mean gives the sine moment because $2\sin\Phi\cos\Phi = \sin(2\Phi)$.
For lift, the same shots give the first-harmonic moments $\langle r\cos\Phi\rangle$ and $\langle r\sin\Phi\rangle$ for $x$ and $y$ before restoring $\theta_c$. The $xz$ and $yz$ terms require $\langle r^3\cos\Phi\rangle$ and $\langle r^3\sin\Phi\rangle$ because $z = r^2$. We weight each X or Y reading by the corresponding power of the radius from the same shot. As in the velocity-field readout (S17i,j), we use $X$ from $P_{\mathrm{X}}$ and $Y$ from $P_{\mathrm{Y}}$. XX and YX thus give two estimates of the cosine component, and YX and YY give two of the sine component. We average the two estimates of each component.
For the standard errors in Fig. 4f, we compute the single-shot drag and lift covariance matrix within each basis pair and divide by that pair's shot count. Each basis-pair mean has weight one quarter, so the covariance of the force estimate is one sixteenth of the sum of the four matrices. Its diagonal entries give the squared standard errors.

## Supplementary Note 5 Initial ensembles and state preparation

For the radial benchmark input near saturation at time 40, we keep the tail of a Gaussian centred beyond the grid (1.4597, standard deviation 0.25). Normalizing its probabilities on [0, 1] gives the projected FEM mean radius, 0.90170, on 64 cells. Amplitudes are square roots of cell probabilities. The 64- and 128-cell grids use these same probabilities in the continuum comparison (Fig. S3b,c and Supplementary Data 2).

For refinement (Fig. S3d,e), we truncate a Gaussian (centre 0.35, standard deviation 0.12) to [0, 1]. The coupled-cubic test (Note 6) adds an independent Gaussian (centre 0.65, standard deviation 0.10). Table S4 lists other initial ensembles.

A matrix product state (MPS) stores $N_x=2^n$ cell amplitudes in one tensor for each of $n$ qubits. The bond dimension $\chi$ limits how many independent components link the two sides of a cut. At fixed precision, preparation costs $O(n\chi^2)$ gates[11]. A layer of bond dimension two takes $4n-5$ one-qubit rotations and $2n-3$ controlled-NOT gates: three rotations and one controlled-NOT gate for the first two qubits, then four rotations and two controlled-NOT gates for each added qubit.

Stacking $L_{\mathrm{MPS}}$ layers of bond dimension two can represent a state of higher bond dimension at cost $O(L_{\mathrm{MPS}}n)$. A fixed-depth alternative (ref. 12) also has gate count linear in $n$. Fidelity is the squared overlap with the target state. In the grid-refinement tests, each coordinate input reached a fidelity of at least 0.9999 using one to seven layers (Supplementary Data 3). For inputs that factorize over coordinates, the total fidelity is the product of the per-coordinate fidelities. Finding the MPS tensors costs $O(N_x\chi^2)$ classical operations, done once beforehand. The resulting circuit loads the signed amplitudes at the start of every attempt. Here signed amplitudes are the real preparation amplitudes, retaining any negative entries.

## Supplementary Note 6 Spatial and temporal refinement

We distinguish spatial and Trotterization errors by varying cells and steps separately at fixed input and duration. To refine the grid without changing the input, we split each cell into $k$ equal parts. Amplitude $\psi_j$ on width $h_0=1/64$ gives density $\rho_0=|\psi_j|^2/h_0$. Amplitude $\psi_j/\sqrt{k}$ in each of $k$ subdivisions preserves density and sign. An extra qubit in $|+\rangle$ halves each original cell's probability at 128 cells.

For the continuum reference, the DOP853 Runge–Kutta method integrates classical radius and phase trajectories from 32 Gauss–Legendre points per cell, weighted by cell probability and quadrature weight. With 16 points, each moment changes by less than $2 \times 10^{-15}$.

On the wake, refining the grid at a fixed step count lowers the spatial error but raises the Trotterization error; refining cells and steps together lowers the total bias. Spatial error compares the exact-grid field with the continuum field for the same input, and total bias compares the noiseless circuit mean with the continuum field. All errors use that field's norm. Doubling the cells at a fixed 33 steps lowers the spatial error (0.0870% to 0.0274%) but raises the Trotterization error (0.0230% to 0.0922%), raising total bias from 0.1072% to 0.1173%. The reason is that, for a layer of duration $\Delta t$ and face speed $F$, a trajectory spans $|F|\Delta t/h$ cells of width $h$. This ratio is the magnitude of the rotation angle, so doubling the cells at a fixed step count doubles the angle at the same face speed. Doubling steps too restores the ratio, giving 0.0228% Trotterization error and 0.0489% total bias (rings, Fig. S3b and Supplementary Data 2). Across all six combinations of 64 or 128 cells with 33, 66 or 132 steps, the Trotterization error is proportional to the square of $N_x/N_t$ to within 1% (Supplementary Data 2), consistent with the fixed-input bound below.

The radial benchmark retains 33 steps to match the late three-coordinate run over the same interval; it is not the minimum step count required for the 1% field-error target. Among the step counts tested on this grid, six is the first that meets the target and eight gives the shortest total task time (Note 12, Table S9).

We also test the spatial derivative using first-order upwinding, which estimates the derivative using the neighbour from which flow arrives. The input, duration and readout are those of the centred-difference run. We omit couplings across the domain boundary and normalize the output, because upwind evolution does not preserve the state norm. Noiseless field bias decreases from 1.105% at 64 cells to 0.572% at 128 cells and falls below the 64-cell centred result at 1,024.

## Step count at fixed input and accuracy

For coordinate-wise even–odd transport, let $\epsilon_{\text{Trot}}$ denote the Strang splitting error in the Euclidean norm of a normalized state, relative to exact evolution on the same grid. Assume drift components with bounded derivatives through order two and, in each direction, periodic boundaries or zero normal drift on the boundary faces. For duration $T$, step count $N_t$, step size $\delta=T/N_t$, and smallest cell width $h_{\text{min}}$, a general operator-norm estimate gives $\epsilon_{\text{Trot}}=O\left(T\delta^2 h_{\text{min}}^{-3}\right)$. For inputs whose two-cell difference quotients (S89) have Euclidean norms bounded uniformly in the grid, cancellation between neighbouring amplitudes in the double commutators improves this to $\epsilon_{\text{Trot}}\leq C(\delta/h_{\text{min}})^2$, with a grid-independent constant $C$, for $\delta/h_{\text{min}}$ below a fixed constant. A fixed piecewise-constant input is handled by smooth approximation, giving $N_t=O_\epsilon(N_x)$ at fixed dimension, domain, input, duration and state-error tolerance $\epsilon$, where $N_x$ is the largest number of cells along one coordinate. With the cubic Walsh bound, the radial circuit then needs $O_\epsilon(N_x(\log N_x)^3)$ transport rotations before rotation synthesis. The proof covers coordinate-wise spatial even–odd splitting, including the coupled model in (S19a)–(S19b), and radial transport–phase splitting at fixed growth rate $\mu$ and frequency-shift coefficient $\gamma$ (equation (S8b)). For the three-coordinate circuit, two errors lie outside this proof: time splitting between layers of disjoint Fourier-mode rotation blocks and truncation to finitely many modes. The refinement results in Note 7 are numerical checks for this circuit. Note 15 gives the proof.

## Testing spatial and Trotterization errors beyond the wake

Refinement tests beyond the wake compare three scalar drifts and a coupled cubic. The scalar drifts on $[0,1]$ share peak speed 0.25 after rescaling, preserving equilibria and stability (Table S3).

$$\frac{\mathrm{d}x}{\mathrm{d}t}=aP(x), \tag{S18a}$$

$$a=\frac{1}{4\max\limits_{x\in[0,1]}|P(x)|}. \tag{S18b}$$

**Table S3 | Scalar test drifts and factors that set their peak speed to 0.25.**

| Label | $P(x)$ | Degree | $a$ |
|---|---|---|---|
| Q2 | $x(1-x)$ | 2 | 1 |
| C3 | $x(1-x)(1/2-x)$ | 3 | 5.1962 |
| Q5 | $x(1-x)(x-1/4)(x-1/2)(x-3/4)$ | 5 | 70.496 |

To test motion in interacting coordinates, we use a cubic on $[0,1]^2$ in which each velocity depends on the other coordinate,

$$\frac{\mathrm{d}x}{\mathrm{d}t}=\frac{x(1-x)(1+2y)}{3}, \tag{S19a}$$

$$\frac{\mathrm{d}y}{\mathrm{d}t}=\frac{y(1-y)(-2+x)}{2}. \tag{S19b}$$

Zero normal drift prevents boundary crossings. The coupled circuit applies even and odd cell-pair layers along $x$ and along $y$. It applies the first three layers for a half step, the fourth for a full step and the first three in reverse for another half step, giving seven layers per step. The time-refinement test in Fig. S3e uses 16 cells per axis; Fig. 5 and Fig. S5 use 32.

We measure means $\langle x\rangle$, $\langle x^2\rangle$ and, in two dimensions, $\langle y\rangle$, $\langle y^2\rangle$ and the cross moment $\langle xy\rangle$. Equation (S20) takes the root mean square over $p$ moment errors, with $p = 2$ for scalar drifts and $p = 5$ for the coupled cubic. Relative error divides the error vector's length by the reference vector's length.

$$e_{\mathrm{RMS}}=\left[\frac{1}{p}\sum_{j=1}^{p}\left(m_j-m_j^{\mathrm{ref}}\right)^2\right]^{1/2}. \tag{S20}$$

We identify sources of moment error by changing one part of the calculation at a time. Classical trajectories start from the target Gaussian ($\mathbf{m}_{\mathrm{A}}$), its exact cell probabilities ($\mathbf{m}_{\mathrm{C}}$) or the prepared probabilities ($\mathbf{m}_{\mathrm{P}}$), spreading cell mass uniformly in the latter two cases. The prepared amplitudes initialize both exact-grid evolution by matrix exponential ($\mathbf{m}_{\mathrm{H}}$) and the circuit ($\mathbf{m}_{\mathrm{Q}}$).

$$\mathbf{m}_{\mathrm{Q}}-\mathbf{m}_{\mathrm{A}}=\left(\mathbf{m}_{\mathrm{Q}}-\mathbf{m}_{\mathrm{H}}\right)+\left(\mathbf{m}_{\mathrm{H}}-\mathbf{m}_{\mathrm{P}}\right)+\left(\mathbf{m}_{\mathrm{P}}-\mathbf{m}_{\mathrm{C}}\right)+\left(\mathbf{m}_{\mathrm{C}}-\mathbf{m}_{\mathrm{A}}\right). \tag{S21}$$

In (S21) the four differences measure Trotterization, spatial discretization, state preparation and representing the continuous input by uniform density within each cell, in order (Supplementary Data 4). Open symbols in Fig. S3d show spatial error. Filled symbols include both input approximations. The four differences are signed, so they can partly cancel in the total.

To isolate Trotterization error, Fig. S3e fixes input $\psi_0$ and grid. After $N_t$ steps to $T=1$ we compare the circuit component $\psi_{N_t}(T)$ with work qubits zero to exact evolution under the grid generator $\mathbf{\Omega}$ on $N_x$ cells. A common phase leaves measurement probabilities unchanged. We choose $\alpha\in\mathbb{R}$ to minimize the distance between states, retaining the component's norm to detect any numerical leakage into nonzero work-qubit states.

$$\epsilon_{\mathrm{Trot}}(N_t)=\min_{\alpha\in\mathbb{R}}\left\|\psi_{N_t}(T)-e^{i\alpha}e^{T\Omega}\psi_0\right\|_2. \tag{S22}$$

At fixed duration, the distance (S22) scales approximately as the inverse square of the step count for all four drifts. Supplementary Data 5 list 64-step gate counts including preparation.

For resolved scalar and coupled inputs, the Trotterization errors in the moments remain approximately unchanged when grid and step counts are doubled together. In this test we vary grid and step count together while fixing the drift, domain, duration and prepared input. Uniform subdivision preserves each original cell's probability and amplitude phase. Scalar grids use 64, 128 and 256 cells, and coupled grids use 32, 64 and 128 per coordinate; steps rise from 64 to 128 and 256. Moment errors are measured against exact evolution on each grid. Boundary inputs are tested for Q5 and the coupled cubic only. For the inputs of Table S4, the maximum Trotterization errors in the moments across the interior, near-stable and near-unstable inputs remain 0.122–0.128% (Q2), 0.141–0.146% (C3) and 0.0180–0.0186% (Q5) as cells and steps double together. Across all four coupled inputs, including the boundary input, the range is 0.0291–0.0314%. The poorly resolved quintic boundary input departs from this pattern between 64 and 128 cells. The numerical refinement records are in Supplementary Data 9.

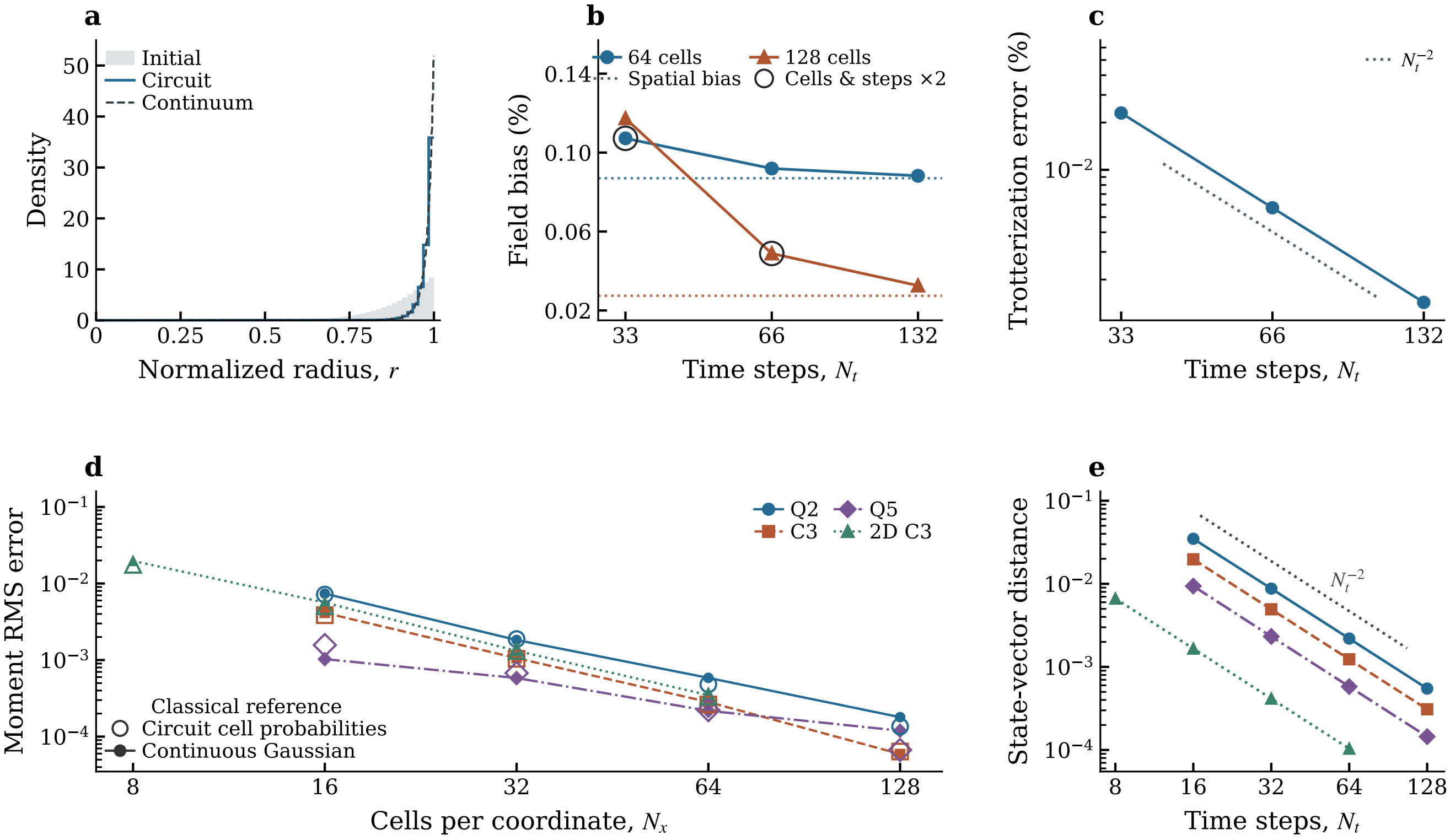

**Supplementary Figure S3 | Separating spatial and Trotterization errors. a**, Initial and final wake densities on 64 cells. **b**, Field bias against the continuum. Dotted lines use exact-grid evolution. Rings mark 64 cells with 33 steps and 128 with 66. **c**, Trotterization error relative to exact evolution on 64 cells. **d**, Moment errors with grid refinement for quadratic (Q2), cubic (C3), quintic (Q5) and coupled cubic (2D C3) drift. Exact-grid moments are compared with trajectories from prepared probabilities spread uniformly within each cell (open) or the target truncated Gaussian (filled). **e**, Trotterization error on 64 scalar or 16 × 16 coupled cells. We measure the distance between states after removing a common phase difference invisible to measurements. Dotted guides in c,e fall fourfold when step count doubles.

## Supplementary Note 7 Three-coordinate Fourier–Walsh wake circuit

During growth the mean-flow shift $z$ can lag behind the oscillation, so we evolve it alongside radius and angle. For $x = r\cos\theta$ and $y = r\sin\theta$, let $J(z)$ be the two-by-two matrix that advances $(x, y)$ in (S4a) after rescaling. Its entries $J_{ij}$ are affine in the shift. Substituting this matrix into the radial and tangent projections (S6a)–(S6b) gives the angle-dependent rates that the radial model averages in (S6c)–(S6d).

Write $A_\mathrm{p} = (J_{11} + J_{22})/2$, $B_\mathrm{p} = (J_{11} - J_{22})/2$, $C_\mathrm{p} = (J_{12} + J_{21})/2$ and $D_\mathrm{p} = (J_{21} - J_{12})/2$. In the shift equation, $q$ collects terms linear and quadratic in the shift, and $G, H, I$ are the fitted weights of the constant, cosine and sine terms multiplying squared radius. Thresholding sets $q_{12} = 0$ in (S4b), so $I = 0$ for the fitted model. For $r > 0$, the three rates are

$$\frac{\mathrm{d}r}{\mathrm{d}t^*} = r\big(A_\mathrm{p} + B_\mathrm{p}\cos 2\theta + C_\mathrm{p}\sin 2\theta\big), \tag{S23a}$$

$$\frac{\mathrm{d}\theta}{\mathrm{d}t^*} = D_\mathrm{p} + C_\mathrm{p}\cos 2\theta - B_\mathrm{p}\sin 2\theta, \tag{S23b}$$

$$\frac{\mathrm{d}z}{\mathrm{d}t^*} = q(z) + r^2(G + H\cos 2\theta + I\sin 2\theta). \tag{S23c}$$

The physical polar volume element is $r\,dr\,d\theta\,dz$. We integrate the probability density with this Jacobian and load the square roots of the cell probabilities. In the continuum this corresponds to the amplitude $\tilde{\psi} = \sqrt{r}\,\psi$, normalized in the flat measure $dr\,d\theta\,dz$, where $\psi$ is normalized in the physical polar measure. For coordinate rates $\mathbf{f} = (f_r, f_\theta, f_z) = \big(\dot{r}, \dot{\theta}, \dot{z}\big)$, the generator acting on $\tilde{\psi}$ is $\tilde{H} = -i[\mathbf{f}\cdot\nabla + (\partial_r f_r + \partial_\theta f_\theta + \partial_z f_z)/2]$, with $\nabla=$ $(\partial_r, \partial_\theta, \partial_z)$. We expand the angle dependence in Fourier modes $e^{ik\theta}$ with integer index $k$. Multiplying by $\cos(2\theta)$ or $\sin(2\theta)$ couples mode $k$ to $k \pm 2$. The angular derivative multiplies mode $k$ by $ik$. In the symmetrized generator the derivative acts on either side of the angular multiplier, so a transition from $k$ to $k'$ carries the average $(k + k')/2$.

Writing $\tilde{\psi} = \sum_{k\in\mathbb{Z}} \psi_k\,(r, z)e^{ik\theta}$, $\beta = (B_\mathrm{p} - iC_\mathrm{p})/2$ and $\eta = (H - iI)/2$, mode $k$ obeys

$$\begin{aligned}\frac{\partial \psi_k}{\partial t^*} &= -\left[A_\mathrm{p}\left(r\,\partial_r + \frac{1}{2}\right) + ikD_\mathrm{p} + (q + Gr^2)\,\partial_z + \frac{q'}{2}\right]\psi_k \\ &\quad + \left[\beta\left(k - 1 - r\,\partial_r - \frac{1}{2}\right) - \eta r^2\,\partial_z\right]\psi_{k-2} \\ &\quad + \left[\beta^*\left(-k - 1 - r\,\partial_r - \frac{1}{2}\right) - \eta^* r^2\,\partial_z\right]\psi_{k+2}.\end{aligned} \tag{S23d}$$

Here $q' = dq/dz$, and the asterisks on $\beta^*$ and $\eta^*$ denote complex conjugation.

One Fourier transform puts the angle register in this basis. A spatial pair at fixed mode or a mode pair at fixed position gives a two-state block. For a spatial pair $j, j+1$ in radius or $z$ and a mode pair $k, k+2$, the four-state block consists of $|j,k\rangle$, $|j+1,k\rangle$, $|j,k+2\rangle$ and $|j+1,k+2\rangle$, with the remaining coordinate fixed. Its two cross-pairs are $(|j,k\rangle, |j+1,k+2\rangle)$ and $(|j,k+2\rangle, |j+1,k\rangle)$. In the mode basis, $\cos 2\theta$ becomes $(S_+ + S_-)/2$ and $\sin 2\theta$ becomes $(S_+ - S_-)/(2i)$, where $S_\pm$ raise or lower $k$ by two. Multiplying the real skew-symmetric spatial coupling by the first gives real antisymmetric entries with the same sign on both cross-pairs. Multiplying by the second gives imaginary entries of opposite sign. The cosine term applies $R_Y(2\alpha)$ to each cross-pair; the sine term applies $R_X(2\alpha)$ to the first and $R_X(-2\alpha)$ to the second. Here $\alpha$ is the real coupling coefficient multiplied by the layer duration. With the cell-pair bit $a = 0,1$ labelling $j, j+1$ and the mode-pair bit $b = 0,1$ labelling $k, k+2$, the cosine layer is $\exp(-i\alpha Y_a X_b)$ and the sine layer is $\exp(+i\alpha Y_a Y_b)$, where $X$ and $Y$ are Pauli operators.

The mode-only terms of (S23d) act at fixed position: $-ikD_\mathrm{p}$ gives a diagonal phase. On each mode pair $(k, k+2)$, the $\cos 2\theta$ part of (S23b) gives $R_X\left((k+1)C_\mathrm{p}\Delta t\right)$, and the $\sin 2\theta$ part gives $R_Y\left((k+1)B_\mathrm{p}\Delta t\right)$, for a layer of duration $\Delta t$ in $t^*$.

Let $k_\mathrm{min}$ be the lowest retained Fourier mode. One mode-pair layer contains $(k_\mathrm{min} + 4s + e, k_\mathrm{min} + 4s + e + 2)$ for integers $s \geq 0$ and $e \in \{0,1\}$ whenever both modes are retained. The other layer shifts both entries by two. Together these layers cover the retained $k \leftrightarrow k+2$ couplings, and no mode belongs to two pairs within one layer. Combining a spatial pair layer with a mode-pair layer gives disjoint four-state blocks.

Since $J(z)$ is affine in $z$, the radial couplings in (S23a) contain terms proportional to $r$ and $rz$. For the angular couplings from (S23b), the Fourier factor $(k + k')/2$ equals $k$ or $k \pm 1$, so $kz$ is the only possible quadratic term. In (S23c), $q(z)$ is at most quadratic and $G, H, I$ are constants multiplying $r^2$. The block rotation angles are proportional to these couplings and thus have total degree at most two in $r, k, z$. Since these variables are affine in their register bits, each Walsh term involves at most two coordinate bits. An inverse transform restores angles for measurement. The gate inventory includes both endpoint quantum Fourier transforms once per attempt, not once per propagation step.

Two work qubits flag increments or decrements that would wrap across spatial or Fourier boundaries (e.g. 7 to 0 on eight cells). We skip flagged rotations and clear the flags using the original spatial or Fourier indices. Arithmetic uses a third zero-initialized qubit.

For register width $n$, a fixed finite-difference stencil (the set of neighbouring points used) and fixed bond dimension, rotations and index arithmetic cost $O(N_t n^2)$ gates over $N_t$ steps. Fourier transforms add $O(n^2)$ and preparation $O(n)$. Finer grids may need more steps.

## Input preparation and field readout

To compare growth and near-saturation costs, we start the growth and late runs at 29.8 and 40; both end at 49.829. These illustrative windows were chosen to show transient growth and the approach to saturation. Both runs use second-order centred spatial differences in radius and $z$. Each Gaussian centres on the corresponding classical three-coordinate model state at the starting time, with independent Cartesian modal standard deviations (0.12, 0.12, 0.09). We integrate over polar cells with their radial area factor and normalize, conditioning each Gaussian on the retained radial and $z$ intervals.

We set the radial and $z$ limits from probability-weighted classical trajectories over each forecast interval, with limits rounded to multiples of 0.05. To check the domain, we evolve quadrature nodes representing the conditioned Gaussian and sum the weights of trajectories that lie outside it at any sampled time. This estimated escaped probability is below $10^{-6}$. Growth uses $r \in [0,1.45]$, $z \in [0,1.50]$; the late run uses $r \in [0,1.65]$, $z \in [0.30,1.70]$.

At bond dimension four both inputs reach 0.9999 fidelity. We simulate the preparation gates and compare their output amplitudes with the MPS. We check the Fourier gates against the discrete Fourier transform, then advance the gate-prepared state using the transport blocks.

We compare the circuit and continuum means at 67 equally spaced diagnostic times, including both endpoints, independently of the step count. If a diagnostic time falls between full steps, a partial step acts on a copy of the state without changing the propagation. Shots return $r \cos \theta$, $r \sin \theta$ and $z$. The reference starts from the target Gaussian, so the bias includes the MPS preparation error. The continuum reference uses the DOP853 Runge–Kutta integrator with relative and absolute tolerances of $2 \times 10^{-10}$ and $2 \times 10^{-12}$. We compare references using $64 \times 128 \times 64$ and $48 \times 96 \times 48$ Gauss–Legendre points and require the maximum relative field difference to be below $10^{-6}$.

Growth uses 264 steps and 437 shots on $256 \times 32 \times 256$ cells (11.87 days), the late run 33 steps and 100 shots on $64 \times 32 \times 64$ (4.14 h), at physical error probability $5 \times 10^{-4}$. A design combines the hardware architecture with a rotation-synthesis precision (Note 11). The fastest design for each input fits under one million qubits.

We select the grid first and the step count second. We accept a grid only if its noiseless field error stays below 1% at all 67 diagnostic times and its final-time bias leaves room for sampling error within the 1% target (Note 10). Among the grids that pass both error tests, we select the one with the smallest product of the shot count and the T gates per attempt. We first assign half of the per-attempt error budget to rotation synthesis, then reoptimize the synthesis precision for the selected grid. After fixing these grids, we select the step count by the complete task time, reoptimizing synthesis and hardware for each feasible candidate. The radial and three-coordinate benchmarks use different inputs, grids and readouts, so their resource comparison does not isolate the effect of the radial closure $z \approx r^2$.

Time-step checks keep the grids, domains, signed MPS inputs and total durations fixed. We test 8, 16, 33, 66, 132, 264 and 528 steps for the late run and 66, 132, 264, 528, 1,056 and 2,112 for growth (Supplementary Data 9). The first tested counts satisfying both accuracy conditions are 16 and 264. The late run at 16 steps requires 342 shots and takes 6.63 h at physical error probability $5 \times 10^{-4}$, so 33 steps with 100 shots is faster. Across all eight physical error probabilities, 33 and 264 steps minimize the total task time among the tested feasible counts on these grids. Doubling these selected counts changes the final mean fields by 0.0777% and 0.1366%, respectively, relative to the corresponding continuum field norms; the largest changes over 67 diagnostic times are 0.0777% and 0.2015%.

We also double the radial and shift cells and the baseline step count, keeping 32 angular modes. At each refinement level, we compare the baseline count with twice that count on the same grid. For the late run, the final-time field difference changes from 0.0777% on 64 cells per refined axis (33 versus 66 steps) to 0.0596% on 128 cells (66 versus 132 steps). For growth, it changes from 0.1366% on 256 cells (264 versus 528 steps) to 0.1557% on 512 cells (528 versus 1,056 steps). The two refinement levels use different grid generators, so these differences need not decrease under joint refinement. On each growth grid separately, doubling the steps again reduces them to 0.0325% and 0.0376%, respectively, consistent with second-order time convergence. These differences diagnose time convergence (Supplementary Data 9).

## Supplementary Note 8 Ensemble transport near equilibria and boundaries

Moving the Gaussian centre tests transport near boundaries and points that attract or repel trajectories (stable or unstable equilibria, Table S4). Figures S4 and S5 track scalar density and two-dimensional mean position. The near-boundary quintic input is the hardest case, with its error dominated by spatial discretization.

**Table S4 | Gaussian centres before truncation to [0, 1].** Scalar widths are 0.06 (boundary 0.10). Coupled widths are 0.10 (near-unstable 0.04). The Source Data for Table S4 give preparation fidelities and moment errors against both references. Scalar cases use 64 cells and 64 steps; coupled cases in Fig. 5 and Fig. S5 use 32 × 32 cells and 64 steps. All evolve to time 1.

| Input | Q2 | C3 | Q5 | Coupled cubic |
|---|---|---|---|---|
| Interior | 0.35 | 0.35 | 0.35 | (0.5, 0.5) |
| Near boundary | 0.02 | 0.02 | 0.02 | (0.02, 0.5) |
| Near stable | 0.94 | 0.5 | 0.5 | (0.85, 0.15) |
| Near unstable | 0.06 | 0.06 | 0.25 | (0.02, 0.98) |

For the quintic input centred near the boundary, classical trajectories carry 93% of prepared mass into the first cell, versus 57% on the 64-cell centred grid (Fig. S4). Circuit and exact-grid moments nearly agree, tracing the discrepancy to spatial discretization.

Doubling cells and steps at fixed input and duration reduces noiseless moment bias to 0.61% at 2,048 cells. On the 2,048-cell grid, the total variation distance between the circuit probabilities and the continuum probabilities for the same prepared input is still 6.27%; this distance is half the sum of their absolute differences. Summing those probabilities into the original 64 spatial bins reduces the distance to 0.051% because neighbouring excesses and deficits cancel. At the $10^{-4}$ error budget, each moment can shift by at most $10^{-4}$ times its range. Dividing the worst-case shifted moment error by the small reference-moment norm in (S44), with the weight (S43a), raises the relative bias bound from 0.61% to 1.71%.

We test the derivative's role by repeating this input with normalized first-order upwinding. It damps 64-cell oscillations, improving density accuracy but increasing moment bias. The centred density is more accurate from 256 cells. Comparable oscillations are absent in the resolved wake.

In the coupled test (Fig. S5), the circuit means travel a shorter distance to the right and downward than the classical means. Both the circuit and the exact-grid means deviate most from the classical means near the unstable point. On 32 × 32 cells, five-moment bias stays below 1% for all four prepared inputs. A step-count check at fixed input and duration tests 8, 16, 32, 64, 128 and 256 steps. With the $10^{-4}$ per-attempt error budget, 64 is the smallest tested common count that leaves sampling room within the 1% target for all four inputs. The near-boundary bias bound falls from 1.029% at 32 steps to 0.946% at 64.

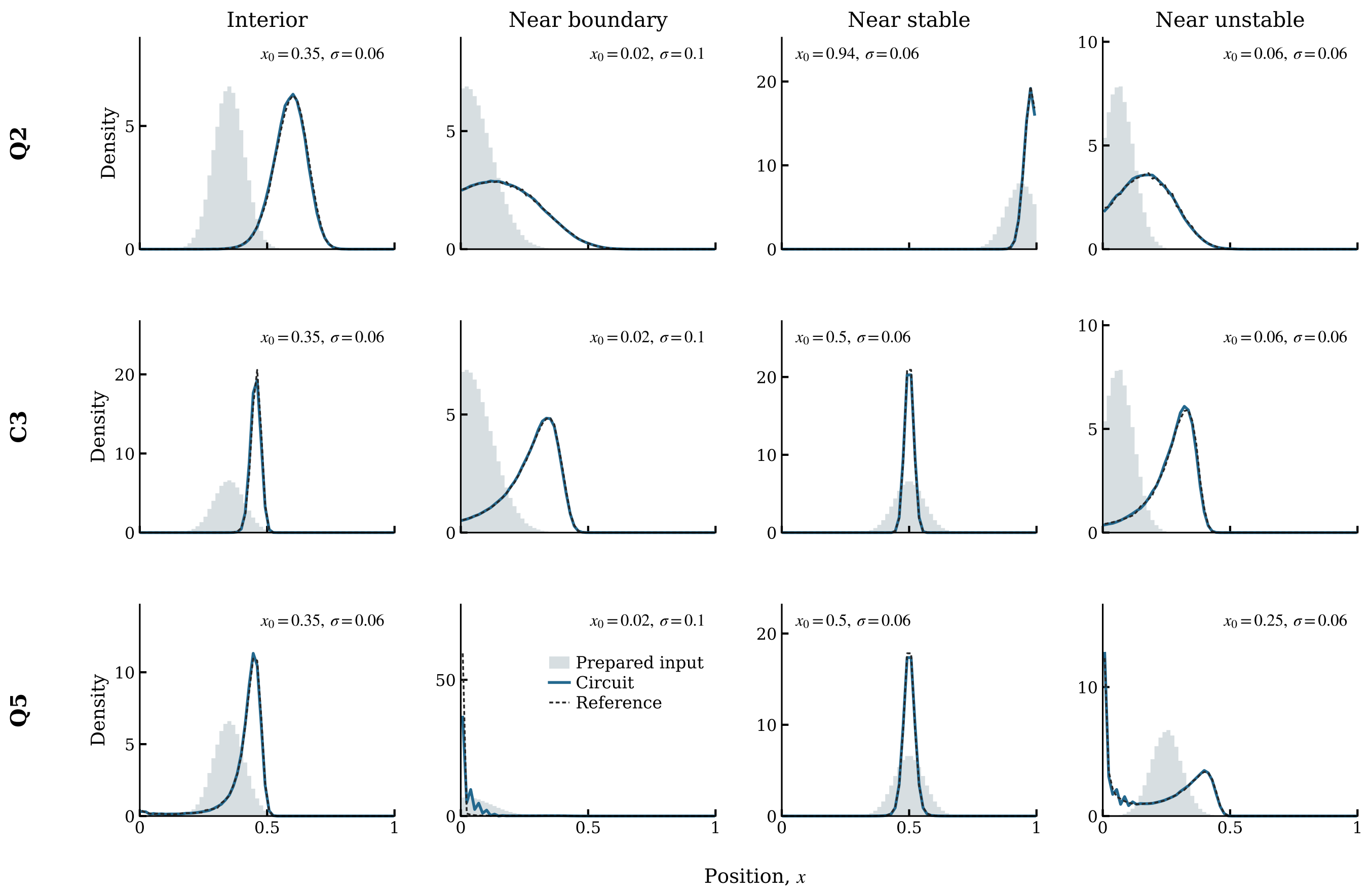

**Supplementary Figure S4 | Effect of starting position on scalar density.** Rows vary drift and columns vary input (Table S4). Grey shows prepared density, blue the circuit after 64 steps to time 1 on 64 cells, and dashed black trajectories from the same cell probabilities.

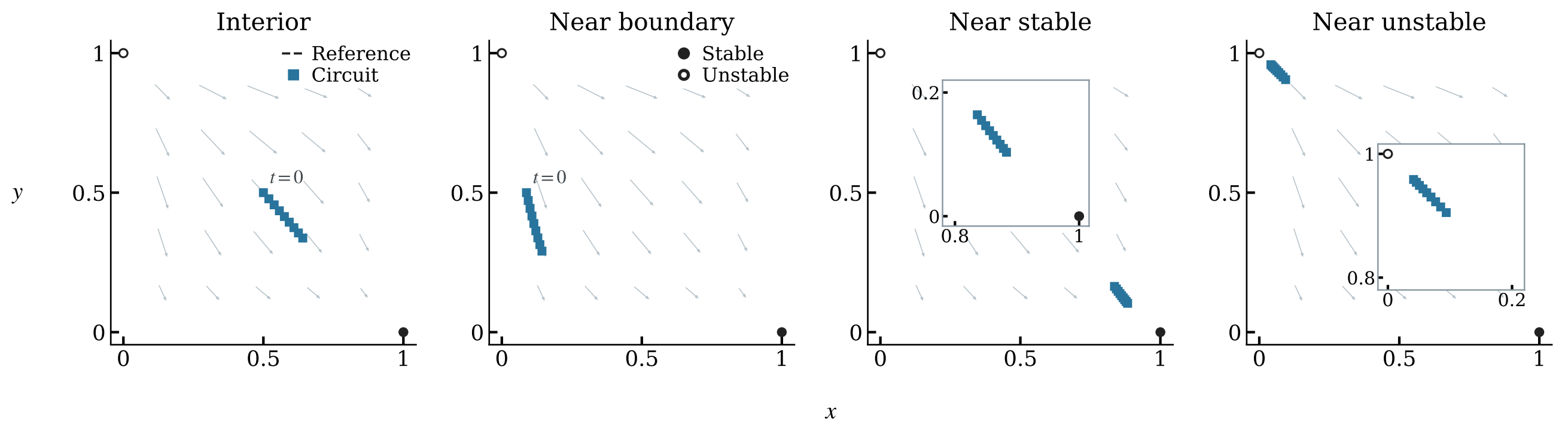

**Supplementary Figure S5 | Effect of starting position on the mean path under coupled cubic drift.** Paths track mean $x$ and $y$ from time 0 to 1 for inputs placed in the interior, near a boundary, near the stable point (1, 0) and near the unstable point (0, 1) (Table S4). Squares mark every eighth circuit step (64 steps, 32 × 32 cells). Dashed curves average classical trajectories from the same cell probabilities. Arrows give local drift. Insets enlarge the last two paths. Filled and open circles mark stable and unstable points.

## Supplementary Note 9 Sampling errors and complete prediction cost

This note relates the shot count to the field error and gives the complete cost of a prediction. The joint probabilities give the single-shot covariance $\mathbf{\Sigma}=\sum_{ij} p_{ij}\boldsymbol{\xi}_{ij}\boldsymbol{\xi}_{ij}^{\mathrm{T}}-\mathbf{m}\mathbf{m}^{\mathrm{T}}$, which describes how the three readings in (S17l) vary together. Supplementary Data 6 give the covariance for the final wake readout.

For $S$ independent shots, the sample mean $\widehat{\mathbf{m}}=S^{-1}\sum_s\boldsymbol{\xi}_s$ has covariance $\mathrm{Cov}(\widehat{\mathbf{m}})=\mathbf{\Sigma}/S$. The map $\mathbf{A}=\mathbf{\Psi}_{\mathrm{m}}\mathbf{M}$ gives the field estimate $\widehat{\mathbf{u}}=\mathbf{u}_{\mathrm{st}}+\mathbf{A}\widehat{\mathbf{m}}$. Its bias $\mathbf{e}=\mathbf{u}_{\mathrm{st}}+\mathbf{A}\mathbf{m}-\overline{\mathbf{u}}_{\mathrm{ref}}$ relative to the continuum field for the same initial ensemble and evolution persists as the shot count increases. The matrix $\mathbf{W}$ weights the field sites, so a real field vector $\mathbf{x}$ has squared norm $\|\mathbf{x}\|_{\mathbf{W}}{}^2=\mathbf{x}^T\mathbf{W}\mathbf{x}$. Expanding the square, the cross term vanishes because the sample mean is unbiased, and $\mathbb{E}[(\widehat{\mathbf{m}}-\mathbf{m})^T\mathbf{A}^T\mathbf{W}\mathbf{A}(\widehat{\mathbf{m}}-\mathbf{m})]=\mathrm{tr}(\mathbf{A}^T\mathbf{W}\mathbf{A}\mathbf{\Sigma})/S$. The mean squared field error is

$$\mathrm{E}\|\widehat{\mathbf{u}}-\overline{\mathbf{u}}_{\mathrm{ref}}\|_W^2=\|\mathbf{e}\|_W^2+\mathrm{tr}\left(\mathbf{A}^{\mathrm{T}}\mathbf{W}\mathbf{A}\mathbf{\Sigma}\right)/S. \tag{S24}$$

With equal site weights, $\mathbf{W}=I$. The trace uses the full covariance $\mathbf{\Sigma}$, including correlations between the three readings. Normalizing by the squared reference norm gives

$$b^2=\|\mathbf{e}\|_W^2/\|\overline{\mathbf{u}}_{\mathrm{ref}}\|_W^2, \tag{S25a}$$

$$v=\frac{\mathrm{tr}\left(\mathbf{A}^{\mathrm{T}}\mathbf{W}\mathbf{A}\mathbf{\Sigma}\right)}{\|\overline{\mathbf{u}}_{\mathrm{ref}}\|_W^2}. \tag{S25b}$$

Relative RMS error is the square root of $b^2+\frac{v}{S}$. Here $b$ is relative bias, $v$ is single-shot field variance divided by the squared reference norm and $S$ counts independent shots. Across 3,000 estimates (1,000 shots each), RMS scatter about the circuit mean was 0.892% of that norm (0.896% predicted, Supplementary Data 7).

Grouping shots by radius shows why phase noise persists as radius concentrates,

$$\mathbf{\Sigma}=\mathrm{Cov}_r[\mathrm{E}(\boldsymbol{\xi}|r)]+\mathrm{E}_r[\mathrm{Cov}(\boldsymbol{\xi}|r)]. \tag{S26}$$

The first term measures variation between radii, the second how the phase signs fluctuate at a given radius. Radius and phase are also correlated in the mean. Replacing $\langle r\cos\Phi\rangle$ and $\langle r\sin\Phi\rangle$ by $\langle r\rangle\langle\cos\Phi\rangle$ and $\langle r\rangle\langle\sin\Phi\rangle$, respectively, while retaining $\langle r^2\rangle$, raises the final-time field bias from 0.107% to 0.254%, which is why each sign is paired with the radius from the same shot (S17i,j).

### Complete prediction cost

Collecting $S$ accepted shots with success probability $P_{\mathrm{s}}$ takes $S/P_{\mathrm{s}}$ attempts on average. Each uses $G_{\mathrm{prep}}$ preparation rotations, $G_{\mathrm{step}}$ per step and $G_{\mathrm{read}}$ for readout. Loading arbitrary amplitudes exactly can cost linearly in their number[13,14], which is why $G_{\mathrm{prep}}$ uses the matrix product state (MPS) preparation of Note 5. The expected total is

$$C_Q\sim\frac{S}{P_{\mathrm{s}}}\left[G_{\mathrm{prep}}+N_tG_{\mathrm{step}}+G_{\mathrm{read}}\right]. \tag{S27}$$

The radial circuit has no postselection. On $N_x=2^n$ cells, angle degree $K\geq 2$ and MPS bond dimension at most $\chi$ give

$$C_{\mathrm{total}}=C_{\mathrm{offline}}+O(S[n\chi^2+N_t\,n^K]). \tag{S28}$$

$C_{\mathrm{offline}}$ covers input construction, compilation and validation. Finer grids may require more steps, larger bond dimension and more shots.

For target $\varepsilon$, upper bounds on bias $b_{\max}<\varepsilon$, variance $v_{\max}$ and attempt cost $G_{\max}$, with success at least $p_{\min}>0$, give

$$S_{\max}=\max\left(1,\left\lceil\frac{v_{\max}}{\varepsilon^2-b_{\max}^2}\right\rceil\right), \tag{S29a}$$

$$\mathrm{E}\left[C_Q\right]\leq\frac{S_{\max}}{p_{\min}}G_{\max}. \tag{S29b}$$

A deterministic Eulerian Liouville solver advances the density in every cell of the joint grid. With $N_{\mathrm{c}}$ coordinates, $N_x$ cells per coordinate and $N_t$ steps, its arithmetic cost for an explicit fixed-stencil update is

$$C_{\mathrm{Eu}}=O\left(N_tN_x^{N_{\mathrm{c}}}\right). \tag{S30}$$

Classical trajectory sampling estimates an observable by averaging independent trajectories. With cost $C_{\mathrm{traj}}$ per trajectory, observable variance $\sigma_O^2$ and mean-square tolerance $\varepsilon_O^2$, its cost is

$$C_{\mathrm{MC}}=O\left(C_{\mathrm{traj}}\sigma_O^2/\varepsilon_O^2\right). \tag{S31}$$

Independent quantum shots have the same inverse-square tolerance scaling. Equation (S30) counts cells in the reduced coordinates. End-to-end estimates for drag[15] and lattice-Boltzmann dynamics[16] also count the spatial grid used to resolve the physical flow.

## Supplementary Note 10 From circuit error to the required shot count

We seek enough accepted shots for 1% relative RMS field or moment error. The error budget $E_{\max} = 10^{-4}$ limits the modelled error in implementing one complete circuit attempt, before conditioning on success.

Finer approximations of $\mathrm{R}_Z(\theta)$ take more T gates. Ross and Selinger estimate the count $n_\mathrm{T}$ at tolerance $\varepsilon_\mathrm{R}$, the largest allowed deviation over normalized input states[17],

$$n_\mathrm{T} = 3\log_2(1/\varepsilon_\mathrm{R}) + O[\log(\log(1/\varepsilon_\mathrm{R}))]. \tag{S32}$$

With $\varepsilon_\mathrm{R} \approx 2^{-n_\mathrm{T}/3}$, the T count per attempt for $N_\mathrm{T}^\mathrm{native}$ native T or T† gates and $N_{R_Z}$ rotations is

$$N_\mathrm{T} = N_\mathrm{T}^\mathrm{native} + n_\mathrm{T} N_{R_Z}. \tag{S33}$$

Before conditioning on the circuit's success flag, the normalized diamond distance[18,19] between the ideal operation $\mathrm{U}_\mathrm{id}(\rho) = U\rho U^\dagger$ and its implementation $\mathrm{E}_\mathrm{impl}$ bounds measurement probability changes even on inputs entangled with a reference.

$$D_\diamond(\mathrm{E},\mathrm{U}) \equiv \frac{1}{2} \parallel \mathrm{E} - \mathrm{U} \parallel_\diamond. \tag{S34}$$

For input $\rho_0$, the total variation distance between ideal and implemented distributions $P$ and $Q$ is the largest probability change over events $B$, including success,

$$\begin{aligned} d_\mathrm{TV}(P,Q) &= \frac{1}{2}\sum_y |P(y) - Q(y)| = \sup_B |P(B) - Q(B)| \\ &\le \frac{1}{2} \parallel \mathrm{E}_\mathrm{impl}(\rho_0) - \mathrm{U}_\mathrm{id}(\rho_0) \parallel_1 \\ &\le D_\diamond(\mathrm{E}_\mathrm{impl}, \mathrm{U}_\mathrm{id}). \end{aligned} \tag{S35}$$

In the fault-free channel $\mathrm{U}_\mathrm{syn}$, the only error relative to $\mathrm{U}_\mathrm{id}$ comes from approximating the target rotations. For rotation $j$, $U_j$ is the target gate and $\widetilde{U}_j$ its synthesized approximation. If each rotation satisfies $\parallel U_j - \widetilde{U}_j \parallel_\infty \le \varepsilon_\mathrm{R}$, summing these bounds limits the error of the full circuit[18,19].

$$D_\diamond(\mathrm{U}_\mathrm{syn}, \mathrm{U}_\mathrm{id}) \le \min\{1, N_{R_Z}\varepsilon_\mathrm{R}\}. \tag{S36}$$

With the core fault probability $E_\mathrm{core}$ and an independent error $e_\mathrm{MSF}$ on each of the $N_\mathrm{T}$ T states delivered by the magic-state factory (MSF), both defined in Note 11, the probability that at least one fault occurs is

$$p_\mathrm{stoch} = 1 - (1 - E_\mathrm{core})(1 - e_\mathrm{MSF})^{N_\mathrm{T}}. \tag{S37}$$

The implemented channel is $\mathrm{E}_\mathrm{impl} = (1 - p_\mathrm{stoch})\mathrm{U}_\mathrm{syn} + p_\mathrm{stoch}\mathrm{E}_\mathrm{fault}$, with $\mathrm{E}_\mathrm{fault}$ conditional on a fault. Normalized diamond distance is at most one, so $D_\diamond(\mathrm{E}_\mathrm{impl}, \mathrm{U}_\mathrm{syn}) \le p_\mathrm{stoch}$. The triangle inequality and (S36) give

$$\begin{aligned} d_\mathrm{TV}(P,Q) &\le D_\diamond(\mathrm{E}_\mathrm{impl}, \mathrm{U}_\mathrm{id}) \\ &\le \min\{1, D_\diamond(\mathrm{E}_\mathrm{impl}, \mathrm{U}_\mathrm{syn}) + D_\diamond(\mathrm{U}_\mathrm{syn}, \mathrm{U}_\mathrm{id})\} \\ &\le \min\{1, p_\mathrm{stoch} + N_{R_Z}\varepsilon_\mathrm{R}\}. \end{aligned} \tag{S38}$$

Substituting the typical synthesis estimate $\varepsilon_\mathrm{R} \approx 2^{-n_\mathrm{T}/3}$ gives

$$E_\mathrm{model} = 1 - (1 - E_\mathrm{core})(1 - e_\mathrm{MSF})^{N_\mathrm{T}} + N_{R_Z} 2^{-n_\mathrm{T}/3}. \tag{S39}$$

The design search (Note 11) requires $E_\mathrm{model} \le E_\mathrm{max}$. More T gates per rotation lower the synthesis term but increase $N_\mathrm{T}$, exposing the core and the supplied T states to more faults. The search selects the T count per rotation $n_\mathrm{T}$ that balances these effects within the error budget.

Randomized compiling[20,21] and error correction[22,23] may reduce coherent errors, motivating the stochastic approximation in (S39). Following Litinski[24], we write the Clifford+T circuit as multiqubit Pauli π/8 rotations and absorb the remaining Clifford operations into the final measurement bases. Each rotation consumes one T magic state through joint lattice-surgery measurements, giving $N_\mathrm{T}$ opportunities for a T-state fault.

### Effect of postselection on prediction error

Keeping only successful shots can magnify probability errors. For success event $A$, ideal and implemented success probabilities are $p_A{=}P(A){=}P_{\mathrm{s}}$ and $q_A{=}Q(A)$. Conditioning on $A$ gives $P_A(B){=}P(B{\cap}A)/P_{\mathrm{s}}$ and $Q_A(B){=}Q(B{\cap}A)/q_A$, with $|p_A{-}q_A|{\le}E_{\max}$ from (S35) and (S38).

Changes in numerator and normalization each contribute at most $\frac{E_{\max}}{q_A}$. Since $q_A \ge P_{\mathrm{s}} - E_{\max} > 0$,

$$d_{\mathrm{TV}}(P_A, Q_A){\le}\frac{2E_{\max}}{P_{\mathrm{s}}{-}E_{\max}}, \qquad P_{\mathrm{s}}{>}E_{\max}. \tag{S40}$$

After conditioning on success, we write $\delta_{\mathrm{TV}}$ for the bound on the total variation distance between the ideal and implemented accepted-shot distributions $P_\star$ and $Q_\star$. With postselection, $\delta_{\mathrm{TV}}$ is the bound in (S40), capped at one because a total variation distance cannot exceed one. The radial circuit has no postselection, so $\delta_{\mathrm{TV}} = E_{\max}$. The block-encoding KvN circuit (BE–KvN, Note 13) uses the lower bound on success for accuracy and $S/P_{\mathrm{s}}$ attempts at nominal success $P_{\mathrm{s}}$ for timing.

For $\xi_i{\in}[a_i,b_i]$ of range $R_i{=}b_i{-}a_i$, mean $m_i{=}\mathrm{E}_{P_\star}\xi_i$ shifts by at most the range times the total variation distance. A 0/1 indicator spans one unit, a ±1 sign two. Applying the same bound to $\xi_i\xi_j$ of range $R_{ij}$ gives

$$e_i{=}\mathrm{E}_{Q_\star}\xi_i{-}m_i, \tag{S41a}$$

$$|e_i|{\le}R_i\delta_{\mathrm{TV}}, \tag{S41b}$$

$$d_{ij}{=}\mathrm{E}_{Q_\star}(\xi_i\xi_j){-}\mathrm{E}_{P_\star}(\xi_i\xi_j), \tag{S41c}$$

$$\left|d_{ij}\right|{\le}R_{ij}\delta_{\mathrm{TV}}. \tag{S41d}$$

Covariance subtracts the product of means from the mean product. Bounding both changes gives entrywise bounds for $\mathbf{\Sigma}_P$,$\mathbf{\Sigma}_Q$,

$$(\mathbf{\Sigma}_Q{-}\mathbf{\Sigma}_P)_{ij}{=}d_{ij}{-}m_ie_j{-}m_je_i{-}e_ie_j, \tag{S42a}$$

$$\left|(\mathbf{\Sigma}_Q{-}\mathbf{\Sigma}_P)_{ij}\right|{\le}(R_{ij}{+}|m_i|R_j{+}\left|m_j\right|R_i)\delta_{\mathrm{TV}}{+}R_iR_j\delta_{\mathrm{TV}}^2. \tag{S42b}$$

For $p$ moments, $\hat{\mathbf{m}}$, $\mathbf{m}$ and $\mathbf{m}_{\mathrm{ref}}$ are the sample, ideal-circuit and continuum means, respectively. The squared error $(\hat{\mathbf{m}} - \mathbf{m}_{\mathrm{ref}})^{\mathrm{T}}\mathbf{H}(\hat{\mathbf{m}} - \mathbf{m}_{\mathrm{ref}})$ uses one of two weights,

$$\mathbf{H}_{\mathrm{mom}} = \frac{\mathbf{I}_p}{\| \mathbf{m}_{\mathrm{ref}} \|_2^2}, \tag{S43a}$$

$$\mathbf{H}_{\mathrm{field}} = \frac{\mathbf{A}^{\mathrm{T}}\mathbf{W}\mathbf{A}}{\| \bar{\mathbf{u}}_{\mathrm{ref}} \|_{\mathbf{W}}^2}. \tag{S43b}$$

The identity $\mathbf{I}_p$ weights the $p$ moments equally ($p = 2$ for the scalar drifts, 5 for the coupled cubic). For the wake, $\mathbf{A}$ maps moments to field values, $\mathbf{W}$ weights sites and $\bar{\mathbf{u}}_{\mathrm{ref}}$ sets the reference magnitude.

Implementation can shift bias $\Delta\mathbf{m}_0 = \mathbf{m} - \mathbf{m}_{\mathrm{ref}}$ by the ranges $\mathbf{R} = (R_1, \ldots, R_p)^{\mathrm{T}}$ in (S41b), scaled by total variation distance. The squared bias is a convex function of the mean shifts, so its maximum over their allowed ranges is attained at a corner. Equation (S44) tests both signs $s_i$=±1 per component (eight corners for three readings, ⊙ denotes componentwise multiplication). Equation (S45) adds the covariance bound $\Gamma_{ij}$ (the right-hand side of (S42b)) to the ideal variance contribution from $\mathbf{\Sigma}_P$.

$$b_{\mathrm{eff}}^2(\delta_{\mathrm{TV}}) = \max_{\mathbf{s}\in\{-1,1\}^p}(\Delta\mathbf{m}_0 + \delta_{\mathrm{TV}}\mathbf{R} \odot \mathbf{s})^{\mathrm{T}}\mathbf{H}(\Delta\mathbf{m}_0 + \delta_{\mathrm{TV}}\mathbf{R} \odot \mathbf{s}), \tag{S44}$$

$$v_{\mathrm{eff}}(\delta_{\mathrm{TV}}) = \mathrm{tr}(\mathbf{H}\mathbf{\Sigma}_P) + \sum_{i,j}\left|H_{ij}\right| \Gamma_{ij}(\delta_{\mathrm{TV}}). \tag{S45}$$

At $S$ independent accepted shots, $b_{\mathrm{eff}}^2 + v_{\mathrm{eff}}/S$ bounds expected squared error. The remaining squared-error budget is $\varepsilon^2 - b_{\mathrm{eff}}{}^2$. When positive, dividing the variance bound $v_{\mathrm{eff}}$ by this budget and rounding up gives

$$S_{\mathrm{req}} = \max\left\{1, \left\lceil\frac{v_{\mathrm{eff}}(\delta_{\mathrm{TV}})}{\varepsilon^2 - b_{\mathrm{eff}}^2(\delta_{\mathrm{TV}})}\right\rceil\right\}, \qquad b_{\mathrm{eff}}(\delta_{\mathrm{TV}}) < \varepsilon. \tag{S46}$$

The 64-cell Walsh–KvN wake in Fig. 3c uses 815 shots. With bias bound $b_{\mathrm{eff}} = 0.00115157$ and single-shot variance bound $v_{\mathrm{eff}} = 0.0803515$, the relative RMS field-error bound falls from 1.00019% at 814 shots to 0.99958% at 815.

The comparison of four drifts in Fig. 5a uses 10,000 shots per estimate. For each interior, near-stable or near-unstable input, 5,000 independent estimates give the median and central 95% interval. All twelve inputs have expected relative RMS moment error below 1%.

For the interior coupled cubic on 32 × 32 cells, (S46) gives 674 shots for 1% relative RMS error across the five moments. The minimum-qubit and fastest designs in Fig. 5c both use fewer than one million physical qubits.

## Supplementary Note 11 Architecture for resource estimation

We estimate the physical-qubit count and total task time using Litinski's core processor layouts and the magic-state factory (MSF) architecture of Silva et al.[24,25] The core consumes T states supplied by the factory through a buffer.

**Table S5 | Hardware parameters.** QEC, quantum error correction.

| Parameter | Value | Ref. | Parameter | Value | Ref. |
|---|---|---|---|---|---|
| QEC round $t_{\mathrm{round}}$ | 1 µs | 26 | Threshold $p_{\mathrm{th}}$ | $10^{-2}$ | 24,25 |
| Classical reaction time $t_{\mathrm{react}}$ | 10 µs | 25,27,28 | Logical-error prefactor | 0.1 | 24,25 |
| Preparation rounds $r_{\mathrm{prep}}$ | 23 rounds | 27,29 | Qubits per patch $Q_{\mathrm{surface}}(d)$ | $2(d+1)^2$ | 26 |
| Preparation cycles $c_{\mathrm{prep}}(d_1)$ | $\left\lceil \frac{r_{\mathrm{prep}}}{d_1} \right\rceil$ | 27,29 | Physical error probability $p_{\mathrm{phys}}$ | $2 \times 10^{-4}$ to $9 \times 10^{-4}$<br>$5 \times 10^{-4}$ (reference) | This study |

Code distance sets patch size and protection. The core processor uses $d_{\mathrm{core}}$. Factory level l of $L$ has $u_l$ units at distance $d_l$. Each logical cycle at distance $d$ takes $d$ correction rounds. Gidney et al.'s 13 circuit rounds[29] plus the classical reaction time for decoding a measurement and choosing the next operation give $r_{\mathrm{prep}} = 13 + (10\,\mu\mathrm{s})/t_{\mathrm{round}} = 23$.

For factory and connector footprints, we reserve $d + 1$ lattice spacings along each patch axis: $d$ for the patch and one for the gap to its neighbour. Logical error probabilities and cycle times use the code distance $d$ itself.

### Errors in the magic-state factory

The factory uses one or two distillation levels (Fig. 2a)[25]. Preparation units grow an encoded T state and check it for faults during Gidney et al.'s 13-round cultivation circuit[29]. Their public code gives the qubit count and the fraction $P_{\mathrm{prep}}$ retained after error-detection and decoder-confidence checks. Replacing T gates with S gates allows efficient stabilizer simulation. For accepted states we estimate the T-state error $e_{\mathrm{prep}}$ as twice the simulated S-state error, following their calibration at fault distance three (Supplementary Data 10). This is the fault distance of the cultivation stage; the escape stage subsequently expands the state to a larger code distance. Trial counts range from 0.8 to 20 billion per escape distance, with 10 billion at each distance for the reference physical error probability. These prepared T states feed the first level. Each level uses 15-to-1 distillation[30]. Equations (S47) and (S48) give the output error and batch acceptance probability[25,31].

$$e_{\mathrm{out,l}} \simeq 35 e_{\mathrm{in,l}}^3 + 7.1 e_{\mathrm{Cliff,l}} \simeq 35 e_{\mathrm{in,l}}^3 + 7.1 e_{\mathrm{mem}}(d_l), \tag{S47}$$

$$P_l \simeq 1 - 15 e_{\mathrm{in,l}} - 356 e_{\mathrm{Cliff},l}. \tag{S48}$$

Equation (S47) uses the one-cycle memory error at code distance $d_l$ for Clifford faults[25]. All 15-to-1 units in the Silva model[25] use (S48). After $L$ distillation levels, level $L+1$ is the core processor. For the logical error per patch per error-correction round in (S49), we use Litinski's approximation for circuit-level noise[24]. Equation (S50) accumulates this error over a $d$-round cycle.

$$p_{\mathrm{L}}(p_{\mathrm{phys}}, d) = 0.1 \left(\frac{p_{\mathrm{phys}}}{p_{\mathrm{th}}}\right)^{(d+1)/2}, \tag{S49}$$

$$e_{\mathrm{mem}}(d)=1-\{1-p_{\mathrm{L}}(p_{\mathrm{phys}},d)\}^{d}. \tag{S50}$$

Independent distillation ($e_{\mathrm{out,l}}$), routing ($e_{\mathrm{path,l}}$) and patch enlargement ($e_{\mathrm{grow,l}}$) errors give the next level's input error,

$$e_{\mathrm{in,l+1}}=1-(1-e_{\mathrm{out,l}})(1-e_{\mathrm{path,l}})(1-e_{\mathrm{grow,l}}). \tag{S51}$$

Patch enlargement takes about one cycle at the original code distance. For $d_{\mathrm{l+1}}\geq d_l$,

$$e_{\mathrm{grow,l}}=\begin{cases} e_{\mathrm{mem}}(d_l), & d_{\mathrm{l+1}}>d_l, \\ 0, & d_{\mathrm{l+1}}=d_l. \end{cases} \tag{S52}$$

The bridges and the shared bus occupy routing space (Fig. S6). For rare faults, $e_{\mathrm{path},l}$ is proportional to the mean output-to-buffer distance in (S53). Adjacent outputs are separated by $s_l$ patch widths measured at the source code distance, with $s_1 = 11$ at level 1 and $s_l = 10$ for $l > 1$. The Litinski layouts compared below use (S73)–(S77).

$$\overline{N}_{path,l} = 1 + \frac{s_l}{u_l}\sum_{j=0}^{u_l-1} \left|j - \frac{u_l-1}{2}\right|, \tag{S53}$$

$$e_{\mathrm{path,l}}=1-\{1-p_{\mathrm{L}}(p_{\mathrm{phys}},d_l)\}^{\overline{N}_{\mathrm{path,l}}d_l}=1-\{1-e_{\mathrm{mem}}(d_l)\}^{\overline{N}_{\mathrm{path,l}}}. \tag{S54}$$

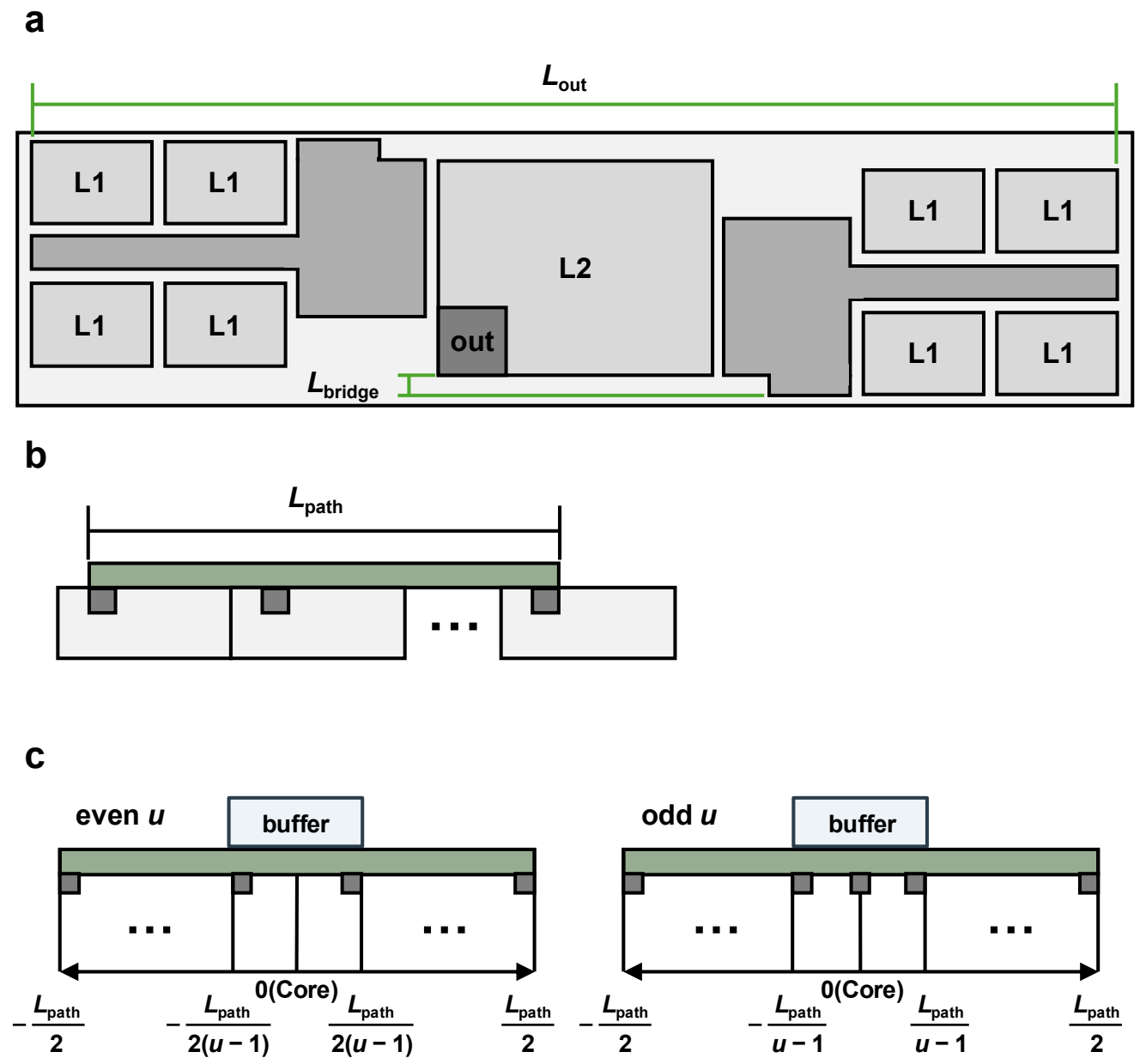

**Supplementary Figure S6 | Factory routing area and routing error.** **a**, $L_{\mathrm{out}}$ is the length of the side of a distillation unit where its output patch is exposed. A bridge connects each output patch to the shared bus. **b**, $L_{\mathrm{path}}$ is the length of the shared bus connecting all output patches. **c**, The routing error is estimated from the mean distance an output state travels to the core buffer or to the input of the next distillation level. Lengths are in lattice-site spacings.

From initial preparation error $e_{\mathrm{in,1}}=e_{\mathrm{prep}}$ through transfer to the core distance, the delivered-state error is

$$e_{\mathrm{MSF}}=1-(1-e_{\mathrm{out},L})(1-e_{\mathrm{path},L})(1-e_{\mathrm{grow},L}). \tag{S55}$$

### Factory supply and waiting

Each 15-to-1 distillation unit consumes 15 input magic states per batch. The batch takes $O_l$ logical cycles ($O_1$=13 at the first level and $O_l$=15 at higher levels) and delivers one output state with acceptance probability $P_l$[25]. For $u_l$ parallel units, the batch duration $\tau_l$, mean output supply rate $D_l$ and input consumption rate $C_l$ are

$$\tau_l = O_l d_l t_{\mathrm{round}}, \tag{S56}$$

$$D_l = \frac{u_l P_l}{\tau_l}, \tag{S57}$$

$$C_l = \frac{15 u_l}{\tau_l}. \tag{S58}$$

The fast, intermediate and compact data blocks in the core processor schedule $b_{\mathrm{blk}}$ = 1, 5 and 9 logical cycles per T state, respectively (Table S6). If the classical reaction time $t_{\mathrm{react}}$ exceeds the scheduled interval, it limits how quickly the core can consume states. The core interval $\tau_{\mathrm{core}}$ and consumption rate $C_{\mathrm{core}}$ are

$$\tau_{\mathrm{core}} = \max(b_{blk} d_{\mathrm{core}} t_{\mathrm{round}}, t_{\mathrm{react}}), \tag{S59a}$$

$$C_{\mathrm{core}} = \tau_{\mathrm{core}}^{-1}. \tag{S59b}$$

Within the factory, we size each level to meet the next level's input demand ($D_{\mathrm{l}-1} \geq C_l$ for adjacent distillation levels). With supply enforced, mean factory supply must meet core demand ($D_L \geq C_{\mathrm{core}}$). With waiting allowed, the factory has one final-level unit even if its output is slower than core consumption. Under either policy, the T-state consumption interval $\tau_{\mathrm{T}}$ is the longer of the core interval and the factory's mean delivery interval, $\tau_{\mathrm{MSF}}$.

$$\tau_{\mathrm{T}} = \max(\tau_{\mathrm{core}}, \tau_{\mathrm{MSF}}), \tag{S60a}$$

$$\tau_{\mathrm{MSF}} = \frac{1}{D_L}. \tag{S60b}$$

An attempt consumes $N_{\mathrm{T}}$ T states and succeeds with probability $P_{\mathrm{s}}$. If one attempt takes $t_{\mathrm{attempt}}$, the mean time per accepted shot is $t_{\mathrm{attempt}}/P_{\mathrm{s}}$. The expected total task time $t_{\mathrm{task}}$ for $S$ accepted shots is therefore

$$t_{\mathrm{attempt}} = N_{\mathrm{T}} \tau_{\mathrm{T}}, \tag{S61a}$$

$$t_{\mathrm{task}} = \frac{S}{P_{\mathrm{s}}} t_{\mathrm{attempt}}. \tag{S61b}$$

### Physical qubits for factory units, storage and connections

For $u_l$ parallel units, the full shared bus occupies $N_{\mathrm{path},l}$ patches (ref. 25, equations (15)–(16)),

$$N_{\mathrm{path},l} = \begin{cases} 1 + 11(u_1 - 1), & l = 1, \\ 1 + 10(u_l - 1), & l > 1. \end{cases} \tag{S62}$$

In the post-corrected form of auto-corrected T-state injection, the correction qubit is stored until earlier measurement outcomes determine its measurement basis[24,25]. We size storage for one correction qubit per logical cycle of duration $t_{\mathrm{round}} d_l$. Holding each qubit for reaction time $t_{\mathrm{react}}$ requires $\lceil t_{\mathrm{react}}/(t_{\mathrm{round}} d_l) \rceil$ storage patches. One bus patch per two storage patches gives the factor 1.5 in the patch-area estimate (Fig. S7).

$$N_{\mathrm{corr+bus},l} = 1.5 \left\lceil \frac{t_{\mathrm{react}}}{t_{\mathrm{round}} d_l} \right\rceil. \tag{S63a}$$

Each distillation unit has 29 fixed patches at level 1 and 28 at higher levels[25]. For the level patch counts $n_l$ in (S65) and (S66), we add one shared bus per level (S62) and correction storage with its access patches per unit (S63a). The total factory qubit count $Q_{\mathrm{MSF}}$ also includes $n_0$ effective preparation units.

$$Q_{\mathrm{MSF}} = n_0 Q_{\mathrm{prep,unit}}(d_1) + \sum_{\mathrm{l}=1}^{L} n_l \, Q_{\mathrm{surface}}(d_l), \tag{S63b}$$

$$n_0 = u_1 \frac{15}{13}\left(\frac{c_{\text{prep}}}{P_{\text{prep}}} + 1\right), \tag{S64}$$

$$n_1 = u_1\left[29 + N_{\text{corr+bus},1}\right] + N_{\text{path},1}, \tag{S65}$$

$$n_l = u_l\left[28 + N_{\text{corr+bus},l}\right] + N_{\text{path},l}, \qquad l \geq 2. \tag{S66}$$

The factor $15/13$ in (S64) is the input consumption rate per first-level unit. Each batch uses 15 prepared states in 13 first-level logical cycles. Here $P_{\text{prep}}$ is the preparation success probability and $c_{\text{prep}}$ is the preparation time in first-level logical cycles. Repeated attempts take $c_{\text{prep}}/P_{\text{prep}}$ cycles on average per accepted state. The $+1$ in the steady-state estimate allows one further cycle for the successful unit to transfer that state to a storage patch while other units prepare the next inputs[25]. Multiplying the total occupied time per accepted state by the required input rate $15u_1/13$ gives the effective number of preparation units $n_0$. A factory or connector patch at code distance $d$ uses $Q_{\text{surface}}(d) = 2(d+1)^2$ physical qubits[26]. For a preparation unit, we use the larger of the cost of one such patch at distance $d_1$ and the qubit count $Q_{\text{Gidney}}(d_1)$ recorded for Gidney et al.'s cultivation circuit at that distance.

$$Q_{\text{prep,unit}}(d_1) = \max\{Q_{\text{surface}}(d_1), Q_{\text{Gidney}}(d_1)\}. \tag{S67}$$

For a core data block holding $n_{\text{data}}$ logical qubits, Table S6 gives the patch count for each layout. We round each count up to the next integer when needed[24].

**Table S6 | Core data-block models.**

| Data block | Fast | Intermediate | Compact |
|---|---|---|---|
| Patch count $N_{\text{blk}}(n_{\text{data}})$ | $\lceil 2n_{\text{data}} + \sqrt{8n_{\text{data}}} + 1 \rceil$ | $2n_{\text{data}}+4$ | $\lceil 1.5n_{\text{data}}+3 \rceil$ |
| Core cycles per T state $b_{\text{blk}}$ | 1 | 5 | 9 |

The core buffer uses the same correction-storage allocation as a factory level and adds 18 fixed patches. Including the data block gives the core patch count $N_{\text{core}}$ and physical-qubit count $Q_{\text{core}}$.

$$N_{\text{core}} = N_{\text{blk}}(n_{\text{data}}) + 18 + 1.5\left\lceil \frac{t_{\text{react}}}{t_{\text{round}} d_{\text{core}}} \right\rceil, \tag{S68}$$

$$Q_{\text{core}} = N_{\text{core}} Q_{\text{surface}}(d_{\text{core}}). \tag{S69}$$

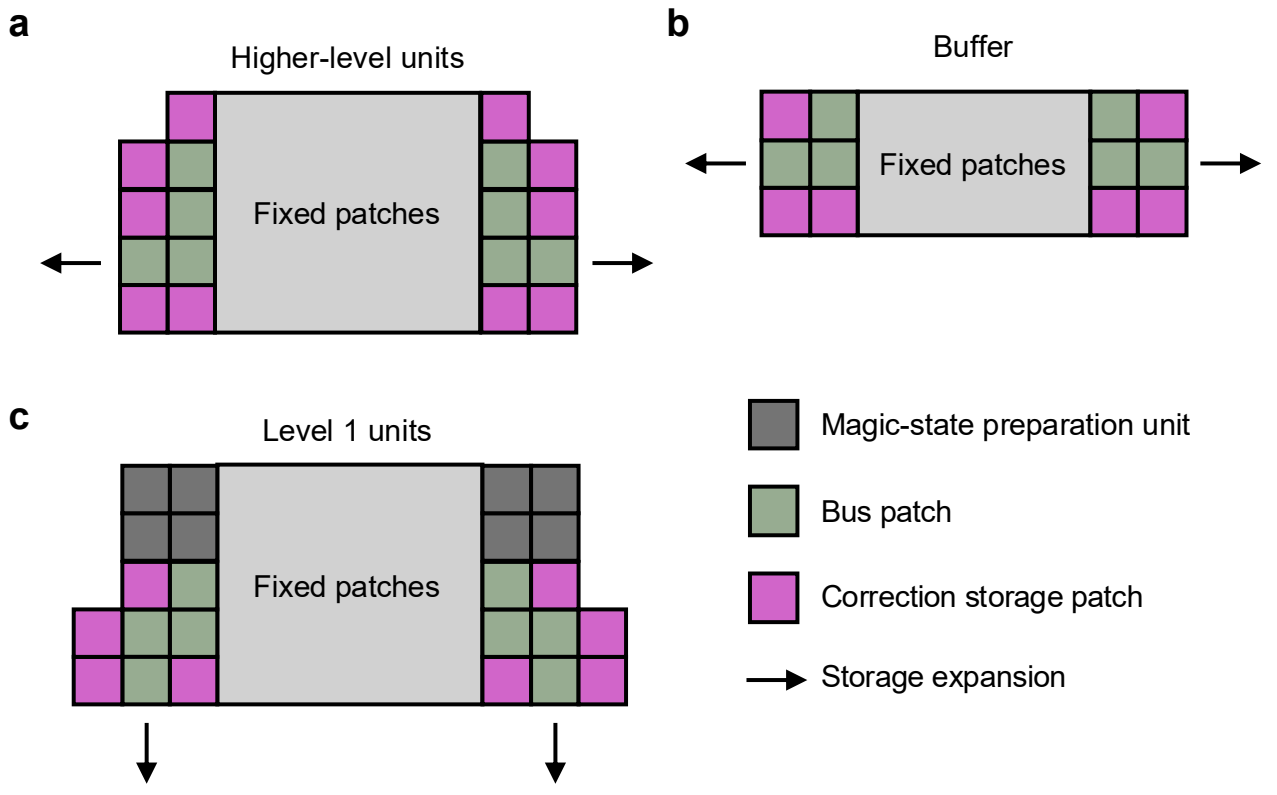

**Supplementary Figure S7 | Storage of correction ancillas pending measurement instructions.** Magenta marks correction-storage patches and green marks bus patches at a later distillation level (**a**), in the core buffer (**b**) and at level one (**c**). Arrows mark added storage. Each selected Silva design needed one correction-storage patch per distillation unit and one in the core buffer, so the patches fit within the layout without lengthening the shared bus (Source Data for Fig. S7).

One patch occupied for one logical cycle contributes one patch-cycle of memory exposure. During gates, every core patch contributes to the memory exposure $V_{\text{idle,base}}$. We also count operating faults on the data-block patches and the 18 fixed buffer patches in the active exposure $V_{\text{act}}$, each at the one-cycle memory-fault probability.

$$V_{\text{act}}=b_{blk}N_{\text{T}}[N_{\text{blk}}(n_{\text{data}})+18], \quad \text{(S70a)}$$
$$V_{\text{idle,base}}=b_{blk}N_{\text{T}}N_{\text{core}}. \quad \text{(S70b)}$$

If the interval $\tau_{\text{T}}$ between T-state consumptions exceeds the scheduled gate time $b_{blk}d_{\text{core}}t_{\text{round}}$, every core patch waits for the extra time $\Delta\tau$. This increases the core memory error.

$$\Delta\tau=\max(0,\tau_{\text{T}}-b_{blk}d_{\text{core}}t_{\text{round}}), \quad \text{(S71)}$$

$$E_{\text{core}}=1-\{1-e_{\text{mem}}(d_{\text{core}})\}^{V_{\text{act}}+V_{\text{idle,base}}}\{1-p_{\text{L}}(p_{\text{phys}},d_{\text{core}})\}^{N_{\text{T}}N_{\text{core}}\Delta\tau/t_{\text{round}}}. \quad \text{(S72)}$$

## Searching for the minimum-qubit and fastest designs

We use a two-stage grid search to find the minimum-qubit and fastest designs for a fixed circuit (Fig. 2c). First, we compare architectures by physical-qubit count $Q$, T-state consumption interval $\tau_{\text{T}}$, core fault probability per T gate $e_{\text{core}}^{(T)}$ and error per delivered T state $e_{\text{MSF}}$. Here $E_{\text{core}}=1-\left(1-e_{\text{core}}^{(T)}\right)^{N_{\text{T}}}$ follows from (S72). We discard an architecture if another is at least as good on all four measures and better on at least one (Pareto pruning). We then pair each remaining architecture with each candidate rotation-synthesis precision. Among the pairs that meet the error budget, we select the design with the fewest physical qubits and the one with the shortest total task time. For each optimization, ties in the primary objective are resolved by comparing the remaining criteria in a fixed order. For the fastest design, we additionally treat execution times within a relative tolerance of $10^{-6}$ of the minimum as tied, preventing negligible runtime gains from favouring substantially larger physical-qubit counts. The initial grid search tests odd core distances 13–49 and 45–200 T gates per rotation. The uncapped time search, which drops the one-million-qubit cap, extends the core-distance range as needed while retaining the 45–200 synthesis range. In the uncapped time search, we add the native T gates to the fewest synthesis T gates compatible with the error budget. The product of this sum and the core interval bounds the attempt time from below. We skip larger core distances for that data block once the bound exceeds the best feasible attempt time.

## Comparing magic-state factory architectures

We compare the Silva magic-state factory with two Litinski designs to test how the factory architecture changes the qubit requirement for the 33-step wake circuit[24,32]. Each alternative uses its own distillation layout and internal error model. Table S7 gives representative protocols, and its Source Data list the candidates.

**Table S7 | Distillation units at the reference physical error probability.** The first and last rows give the alternatives selected in Fig. S8b. Distance triples list X, Z and measurement code distances. Times include retries and are means per accepted batch. Dashes indicate that no separate batch acceptance probability is tabulated for two-level protocols.

| Protocol / code distances | Levels | Factory-output error | Mean rounds | Batch acceptance probability |
|---|---|---|---|---|
| 116-to-12 (output distance 19) | 1 | $2.58 \times 10^{-12}$ | 1,006.74 | 94.4% |
| 15-to-1 (7,3,3) | 1 | $2.91 \times 10^{-5}$ | 19.05 | 94.5% |
| 4 × 15-to-1 → 20-to-4 (15,7,9) | 2 | $1.69 \times 10^{-9}$ | 96.69 | — |
| Small-footprint 15-to-1 → 15-to-1 (21,9,11) | 2 | $1.75 \times 10^{-13}$ | 453.82 | — |

In Fig. S6, $u$ units with output face $L_{\text{out}}$, $n_{\text{out}}$ outputs each and output distance $d_{\text{out}}$ span

$$L_{\text{path}}=L_{\text{out}}(u-1)+n_{\text{out}}d_{\text{out}}. \quad \text{(S73)}$$

For odd unit counts the middle output aligns with the buffer. Averaging distances along the bus gives

$$\bar{L}_{\text{path}}=\begin{cases}\dfrac{uL_{\text{path}}}{4(u-1)}, & u\geq 2,\text{ even},\\ \dfrac{(u+1)L_{\text{path}}}{4u}, & u\geq 3,\text{ odd},\\ 0, & u=1.\end{cases} \tag{S74}$$

We approximate lattice-surgery routing by a rectilinear corridor one patch wide. A bridge connects each output to the shared bus. Ignoring detours and extra space at junctions, a state travels the mean distance along the bus plus the length of one bridge, $L_{\text{bridge}}$. In patches of code distance $d_{\text{out}}$, the mean transfer distance, including the bridge, is

$$N_{\text{path}}=\frac{\bar{L}_{\text{path}}+L_{\text{bridge}}}{d_{\text{out}}}. \tag{S75}$$

We approximate the routing error by accumulating independent logical-memory faults over this effective number of patches,

$$e_{\text{path}}=1-\{1-e_{\text{mem}}(d_{\text{out}})\}^{N_{\text{path}}}. \tag{S76}$$

Both alternatives use (S55) to combine the factory-output error, routing error and patch-enlargement error.

For the 116-to-12 protocol, Litinski[24] gives no separate Clifford rejection term, so we use the input-only acceptance probability $(1-e_{\text{in}})^{116}$ with raw input error $e_{\text{in}}=p_{\text{phys}}$. With $e_{\text{Cliff}}=e_{\text{mem}}(d_{\text{out}})$, all eight selected designs have $e_{\text{Cliff}}/e_{\text{in}}<10^{-8}$ (maximum $1.11\times10^{-9}$), suggesting little additional rejection from Clifford faults (Source Data for Fig. S8b).

For physical-qubit allocation, we count the full length of the shared bus once and reserve one bridge for each of the $un_{\text{out}}$ outputs. The bus length $L_{\text{path,area}}$ and individual bridge length $L_{\text{bridge,area}}$ use the padded patch pitch $d_{\text{out}}+1$. The routing area therefore requires

$$Q_{\text{path}}=\frac{Q_{\text{surface}}(d_{\text{out}})}{d_{\text{out}}+1}(L_{\text{path,area}}+un_{\text{out}}L_{\text{bridge,area}}). \tag{S77}$$

To compare factories, we fix the 33-step circuit, fast core layout, buffer and per-attempt error budget ($10^{-4}$), and reoptimize code distances and rotation-synthesis precision at each physical error probability to find the fewest qubits for one circuit attempt[24,32]. Note 12 gives the results.

## Supplementary Note 12 Sensitivity of the designs

We test how the designs change with the per-attempt error budget, the factory architecture, the supply policy, the qubit limit and the step count.

### Error budget and factory architecture

At the reference physical error probability of $5 \times 10^{-4}$, the minimum counts for the three factories range from 56,100 to 122,000 physical qubits (Fig. S8b). The upper end comes from the Litinski model shown by triangles. It assigns raw input T states the physical error probability and uses the same code distance at every distillation level. This design meets the $10^{-4}$ per-attempt error budget, with a modelled circuit error of $9.45 \times 10^{-5}$. At this budget, (S46) gives 815 shots for 1% relative RMS field error. Shots run one after another on the same qubits, so additional shots increase total task time while the qubit count stays fixed.

With the Silva factory and a physical error probability of $5 \times 10^{-4}$, we test four per-attempt error budgets, $10^{-2}$, $10^{-3}$, $10^{-4}$ and $10^{-6}$, for the 33-step circuit (Fig. S8a). For each budget we search all three core layouts and use (S46) to set the shot count for 1% relative RMS field error. All four fastest designs use fewer than one million physical qubits and finish in 1.33–8.66 h. As the budget tightens from $10^{-2}$ to $10^{-6}$, the minimum-qubit design grows from 38,281 to 78,172 qubits and the fastest design from 358,370 to 677,544. A 10,000-fold tighter budget roughly doubles both qubit counts and leaves the fastest design within the early fault-tolerant quantum computing (EFTQC) window of Fig. 3a.

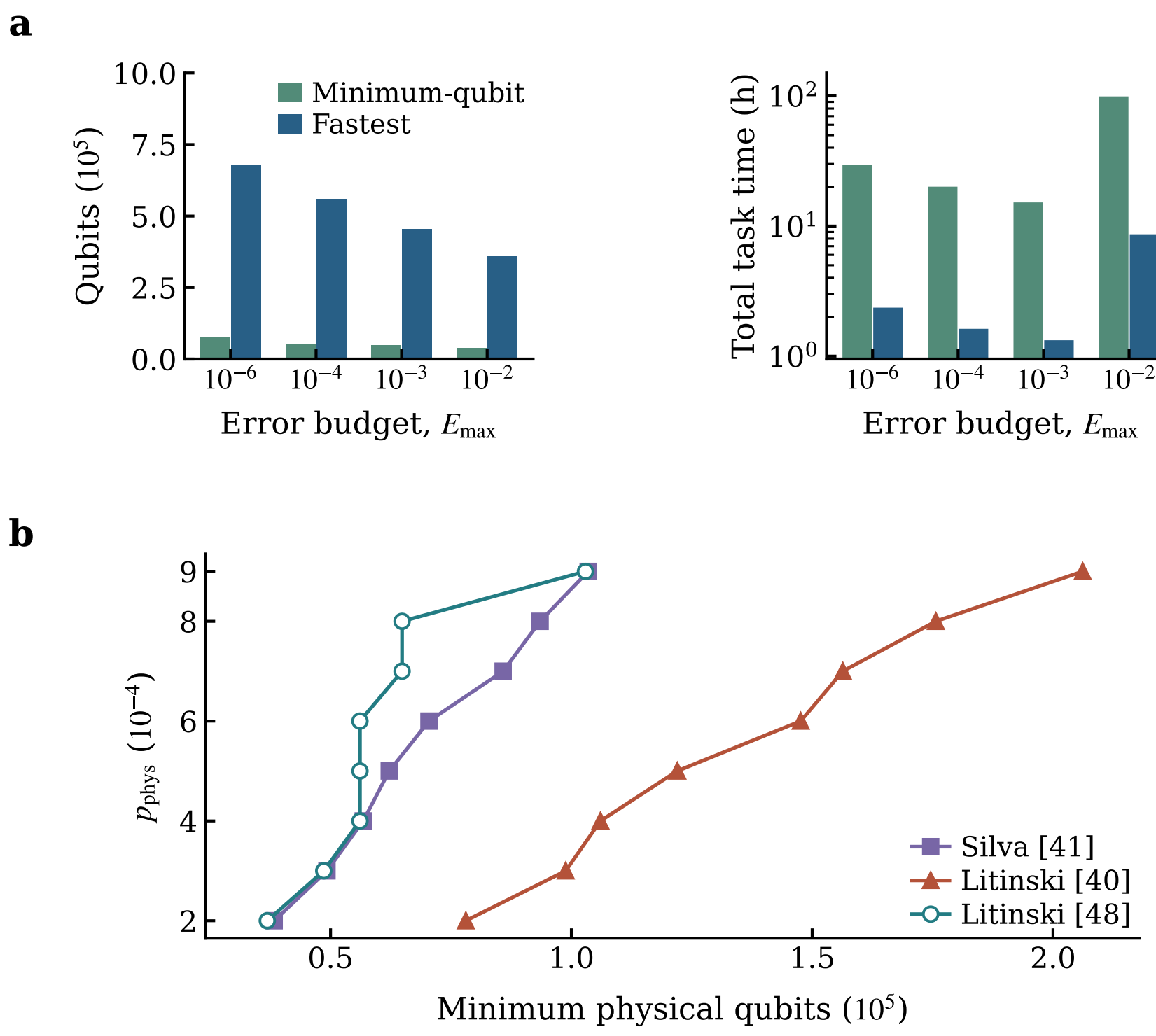

**Supplementary Figure S8 | Resource requirements for the 33-step wake under different error budgets (a) and factory architectures (b). a**, Minimum-qubit and fastest designs for the Silva factory at a physical error probability of 5 × $10^{-4}$. The fastest design is selected under a one-million-qubit cap. Raising the error budget from $10^{-3}$ to $10^{-2}$ shortens the fastest attempt (5.68 to 4.29 s) but increases the shots for 1% field error (841 to 7,266) and total task time (1.33 to 8.66 h). The benchmark uses $10^{-4}$. **b**, Minimum physical qubits at a per-attempt error budget of $10^{-4}$. The fast core is paired with the factory of Silva et al. [41] (squares) and the two Litinski layouts [40] (triangles) and [48] (open circles). Brackets cite the main bibliography.

### Qubit–time trade-offs under different factory supply policies

We compare two policies. With supply enforced, mean supply meets core demand. With waiting allowed, a single final-level unit feeds the core. We vary the native T count at fixed register width, rotation count (4,803), input, shot count and error budget, reoptimizing code distances and synthesis at each physical error probability (Fig. S9).

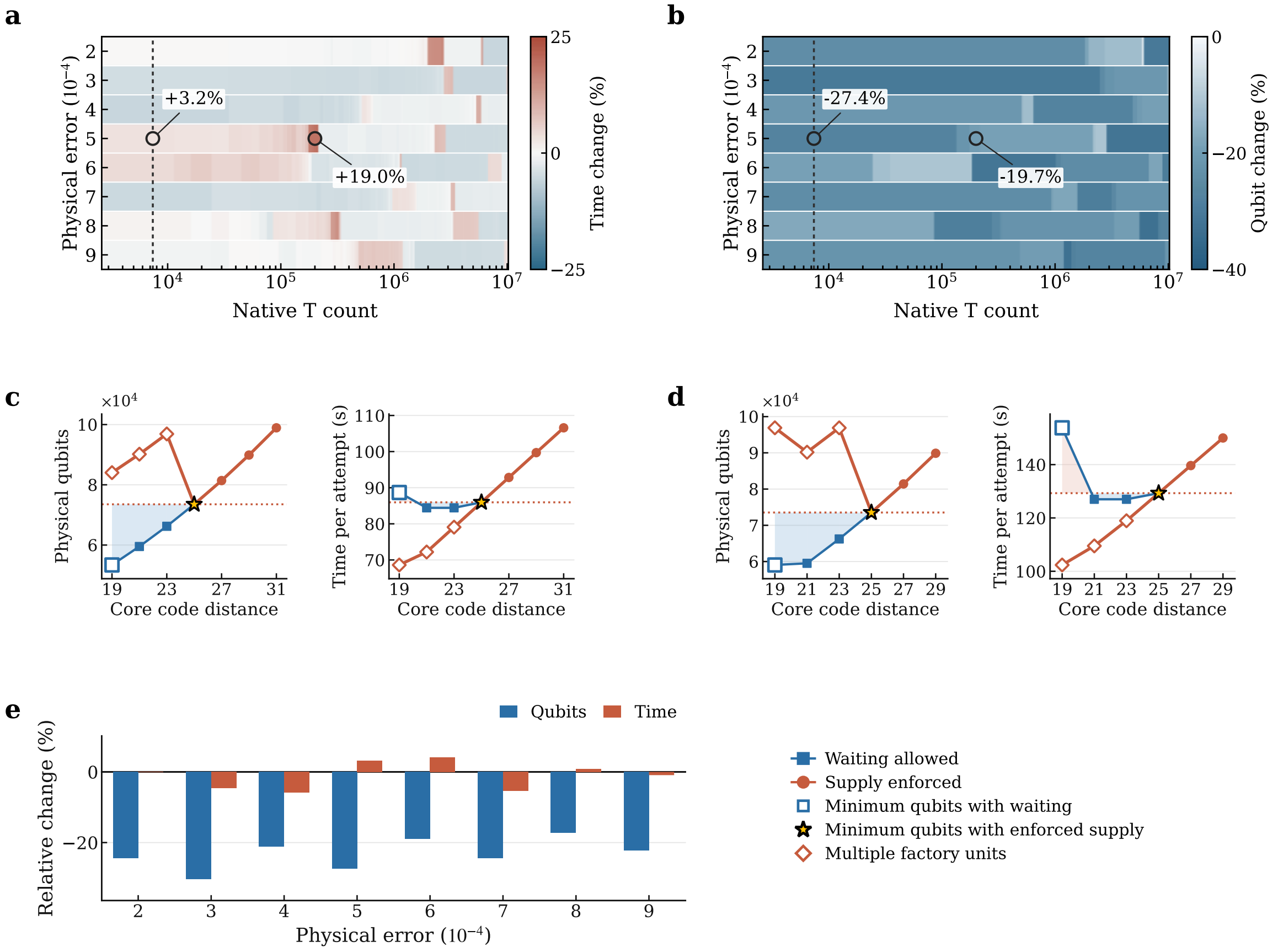

**Supplementary Figure S9 | Effect of waiting for T states on minimum-qubit designs. a**,**b**, Percentage change in time per attempt (**a**) and physical qubits (**b**) with waiting allowed, relative to designs with supply enforced. Dashed lines mark the wake's 7,392 native T gates. **c**,**d**, Qubits and time per attempt versus core code distance for the wake (**c**) and a control with the native T count raised to 200,000 (d). Circles in a and b mark these circuits at physical error probability $5 \times 10^{-4}$. **e**, Changes in wake qubits and time per attempt across physical error probabilities.

At the reference physical error of $5 \times 10^{-4}$, waiting lowers the wake's core code distance from 25 to 19 and raises the T count per rotation from 78 to 82. The wake saves 27.4% of its qubits for 3.2% more time per attempt (Fig. S9c). With supply enforced, core distances 19, 21 and 23 need two final-level units, and 25 is the smallest distance at which one unit meets core demand, so the distance-25 design (73,530 qubits) is smaller than the distance-19 design with two units (84,044 qubits). With waiting allowed, one unit feeds a distance-19 core (53,382 qubits) at nearly the original consumption interval, and the 5.0% rise in T count lengthens the attempt from 85.96 to 88.67 s. With the native T count raised to 200,000 as a control, waiting saves 19.7% of the qubits for 19.0% more time. At $9 \times 10^{-4}$ the wake saves both qubits (22.3%) and time (0.9%) (Fig. S9e and Table S8). The preferred supply policy depends on the circuit, physical error and time limit.

**Table S8 | Qubits and time per attempt with waiting allowed, relative to designs with supply enforced.**

| Physical error | Native T | Qubit change (%) | Time change (%) |
|---|---|---|---|
| $5 \times 10^{-4}$ | 7,392 | −27.4 | +3.2 |
| $9 \times 10^{-4}$ | 7,392 | −22.3 | −0.9 |
| $5 \times 10^{-4}$ | 200,000 | −19.7 | +19.0 |

## Dependence of the optimal data block on the qubit limit

We fix the per-attempt error budget at $10^{-4}$ and require 815 shots in one day. At the reference physical error probability, raising the qubit limit from $10^5$ to $10^6$ shortens the fast block's T-state consumption interval from 110.5 to 21 µs and raises its capacity from 64 to 393 consecutive steps (Fig. S10). Under the smaller limit, the intermediate block leads with 77 steps. Under the larger limit, it reaches 84, while the fast block can use 14 final-level factory units in place of two. The qubit limit therefore determines which combination of data block and factory supply delivers the most steps.

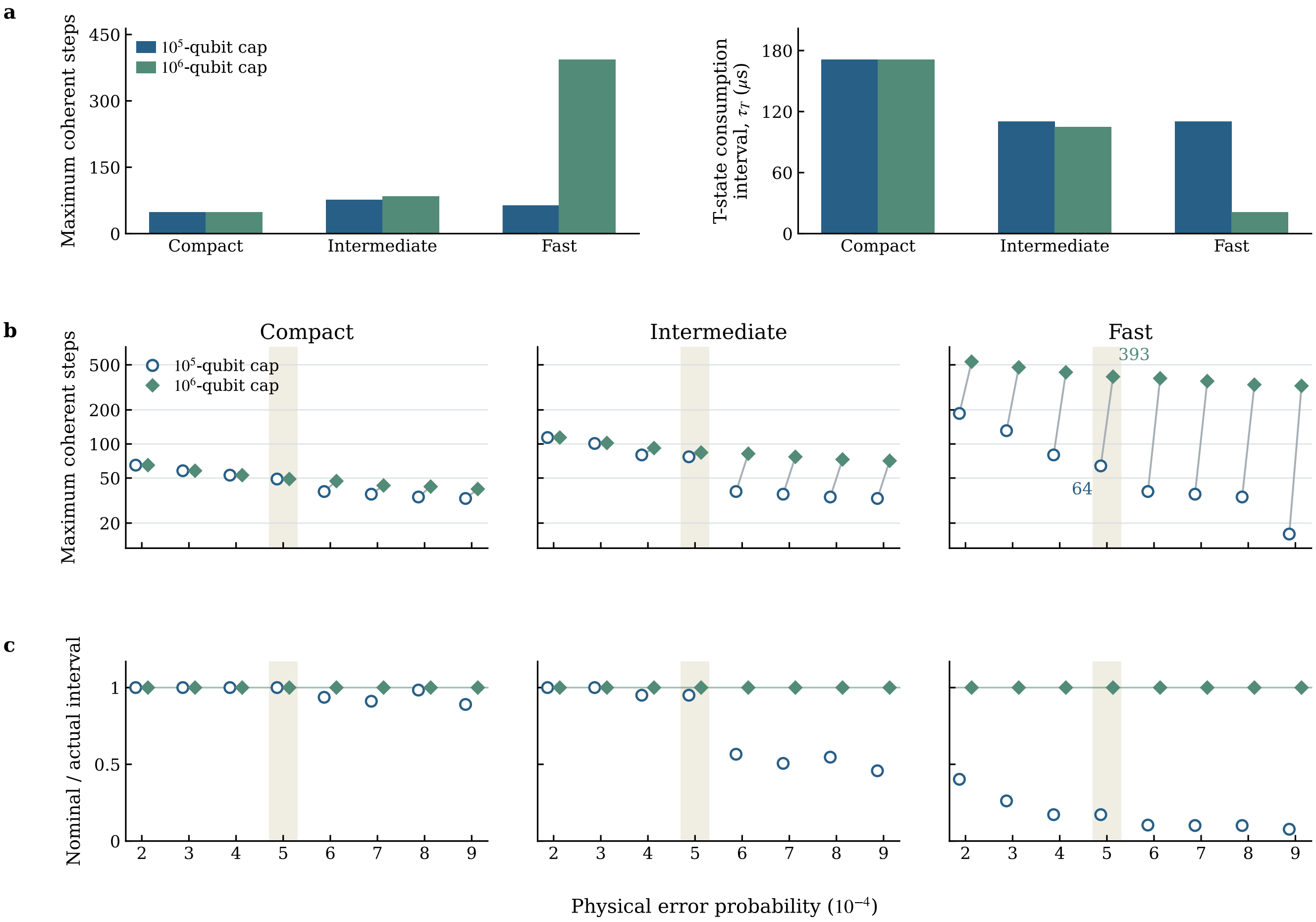

**Supplementary Figure S10 | Maximum consecutive steps for 815 shots per day. a**, Maximum steps and T-state consumption interval at physical error probability $5 \times 10^{-4}$. **b**,**c**, Maximum steps and nominal-to-actual core interval ratio across physical errors. Below one, the core waits for the factory. Shading marks the reference error. The wake accuracy test uses 33 steps.

## Dependence of total task time on step count

On the 64-cell grid, six steps are the fewest that meet the 1% target, and eight steps give the shortest total task time among the tested counts at physical error probability $5 \times 10^{-4}$ (Table S9). At fixed 64-cell grid and signed MPS input, we test every integer from 1 to 64 steps and also 66, 80, 96, 128, 132 and 256. The forecast interval, phase readout and continuum reference remain fixed. For each candidate, (S46) sets the shot count using the same per-attempt error budget of $10^{-4}$, and we reoptimize rotation synthesis and hardware. The six-step design requires 2,857 shots and 0.933 h; the eight-step design requires 1,074 shots and 0.464 h on 422,901 physical qubits. For eight steps, the noiseless field bias is 0.4893%, the Trotterization error 0.4134% and the spatial error 0.0870%. An attempt uses 1,228 arbitrary rotations and 1,792 native T gates; the selected synthesis gives 91,436 T gates in total. The effective bias and variance bounds are $b_{\mathrm{eff}} = 0.00497312$ and $v_{\mathrm{eff}} = 0.0807991$, giving a relative RMS bound of 0.999819% at 1,074 shots and 1.00017% at 1,073. The larger six-step noiseless bias (0.8383%) leaves a smaller sampling budget $\varepsilon^2 - {b_{\mathrm{eff}}}^2$ at target $\varepsilon = 0.01$, explaining its higher shot count. For the eight-step circuit, simulating the compiled gates reproduces the direct pair rotations and returns the work qubit to zero. Table S9 compares the designs for 6, 8, 16 and 33 steps (Supplementary Data 8).

**Table S9 | Radial step-count comparison on 64 cells at physical error probability 5 × 10⁻⁴.** Designs minimize total task time for the 1% final-time relative RMS field-error target.

| Steps | Shots | Time (h) | Physical qubits | Code distance* | Supply units† |
|---|---|---|---|---|---|
| 6 | 2,857 | 0.933 | 453,563 | 17 | 14 |
| 8‡ | 1,074 | 0.464 | 422,901 | 17 | 13 |
| 16 | 834 | 0.788 | 559,953 | 19 | 14 |
| 33§ | 815 | 1.62 | 559,953 | 19 | 14 |

* The core and the single distillation level use the same distance in these designs. † Number of final-level distillation units. ‡ Fastest among the tested step counts. § Reference design: its 33 steps match the late three-coordinate run over the same forecast interval. Supplementary Data 8.

## Supplementary Note 13 Resource estimates for the comparison circuits

We estimate resources for two comparison circuits. BE–KvN uses block encoding; we adapt it to the same wake input, 64-cell generator, duration and velocity readout as Walsh–KvN, which uses Walsh-synthesized rotations. Walsh–KvN requires 815 shots and BE–KvN 813 accepted shots for 1% relative RMS field error; Table S10 gives the resulting designs, and total task time includes retries (S61b). LCHS–KvN, based on a linear combination of Hamiltonian simulations, is estimated on its own benchmark, a one-dimensional sine drift without the wake phase qubits. We describe LCHS–KvN first, then the BE–KvN adaptation and its grid dependence, and finally gate-count bounds for the three circuits.

**Table S10 | Matched wake designs and hardware comparisons.**

a, Matched field estimates at physical error probability $5 \times 10^{-4}$. A denotes incrementer A (Note 3).

| Workload | Selection | Physical qubits | Time per accepted shot | Total task time |
|---|---|---|---|---|
| Walsh–KvN A field | Minimum-qubit, compact | 53,382 | 88.67 s | 20.075 h |
| Walsh–KvN A field | Fastest, fast | 559,953 | 7.17 s | 1.6226 h |
| BE–KvN field | Minimum-qubit, compact | 122,100 | 4.77 d | 10.63 yr |
| BE–KvN field | Fastest, fast | 953,563 | 9.37 h | 317.30 d |

b, Hardware markers and error measures in Fig. 3a. Reference numbers in this part follow the main bibliography.

| Marker | Main-text ref. | Quantity and comparison scope |
|---|---|---|
| Zhang | 9 | 10 active qubits; median/range two-qubit calibration error |
| Wang / Meng | 7 / 8 | 8 / 10 active qubits; reported gate-error context |
| Tiwari | 10 | 19-qubit circuit; manufacturer device-error proxy |
| Minimal QEC | 20 | Illustrative architecture examples, not a universal boundary |
| STAR / RSA | 46 / 47 | Estimated qubit counts and error rates |

### Resource estimation for LCHS–KvN

For Novikau and Joseph's[33] sine drift at $t = 1$, we reconstruct single-target multi-controlled (STMC) gates from the public QuCF code[34]. We tally gate types, angles, repetitions and controls activated by zero or one. With one clean work qubit, we use Khattar–Gidney[8] for multi-controlled $X$ gates and Zindorf–Bose[35] for general controlled one-qubit gates (Table S11).

**Table S11 | Resource estimates for the LCHS–KvN sine-drift circuit at physical error probability $5 \times 10^{-4}$ (815 accepted shots).** This shot budget matches the radial Walsh–KvN count but is applied to a different problem, the sine drift, without a matched 1% field-error target. Table S10 gives the matched 64-cell wake comparison.

| Design | Physical qubits | T per attempt | Time per accepted shot | 815-shot total |
|---|---|---|---|---|
| Min. qubits, compact | 158,001 | 44,634,906,570 | 191.60 d | 427.54 yr |
| Fastest capped, fast | 853,526 | 45,864,130,546 | 16.41 d | 36.61 yr |
| Fastest uncapped, fast | 2,983,519 | 44,634,906,570 | 15.92 d | 35.52 yr |

### BE–KvN for the same wake generator and input

To compare BE–KvN with Walsh–KvN on the same wake problem and initial distribution, we adapt the block-encoding construction of Guseynov et al.[36] in three steps. First, we encode the physical radial coordinates $r_j$ at the cell centres. The coordinate loader assumes that the squared coordinates sum to one. On our 64-cell grid they sum to 21.33. We retain the $\sqrt{2}$ block-normalization convention of the real-coordinate oracle (ref. 36, Corollary 1). In our circuit, an auxiliary Hadamard supplies the $1/\sqrt{2}$ zero-state amplitude, giving $\beta_{\mathrm{R}}\approx 6.53$ (S79). When the auxiliary qubits start and end in zero, the block acting on the radial register is $\mathbf{R}_{\mathrm{diag}}/\beta_{\mathrm{R}}$, with $\mathbf{R}_{\mathrm{diag}}=\mathrm{diag}(r_j)$ the radial-coordinate operator.

Second, we remove the periodic connection between the first and last cells because the wake drift vanishes at the physical boundaries $r=0$ and $r=1$. The centred derivative combines the grid-index increment and decrement, which translate by one cell in opposite directions. We average the two orders of drift multiplication and differentiation, so each interior face uses the mean of the two neighbouring cell-centre values. For cell width $h=1/N_x$, the corrected cubic $g_h$ recovers $\mathbf{H}_{\mathrm{r}}=\mathrm{i}\boldsymbol{\Omega}$ in (S13a)–(S13b) by giving the physical drift at each face $s$,

$$g_h(r)=(1+3h^2/4)r-r^3, \tag{S78a}$$

$$\frac{1}{2}[g_h(s-h/2)+g_h(s+h/2)]=s-s^3. \tag{S78b}$$

Finally, we incorporate the radius-dependent phase evolution by coupling the squared-radius operator to the two phase qubits. An even polynomial encodes this coupling. Combining it with radial transport gives the joint Hamiltonian

$$\mathbf{H}=\mathbf{H}_{\mathrm{r}}\otimes\mathbf{I}-\kappa\mathbf{R}_{\mathrm{diag}}^2\otimes(\hat{n}_{\mathrm{X}}+\hat{n}_{\mathrm{Y}}). \tag{S78c}$$

Here $\hat{n}_{\mathrm{X}}$ and $\hat{n}_{\mathrm{Y}}$ project onto $|1\rangle$ on their respective phase qubits. The identity $\mathbf{I}$ acts on the two phase qubits, and $\kappa=\gamma/\mu$ is the normalized phase rate. The radial Hamiltonian $\mathbf{H}_{\mathrm{r}}$ transports amplitudes between cells. The second term accumulates the residual phase $\Phi$. Reading both phase qubits and the radius in the same shot gives the radius-weighted first circular moment, $\langle re^{\mathrm{i}\Phi}\rangle=m_1+\mathrm{i}m_2$ (Note 4).

After multiplying the encoded block by its normalization, we compared all 256 columns with the target joint Hamiltonian above. The largest absolute difference in any matrix entry was $6.01\times10^{-13}$. Quantum singular value transformation (QSVT) evolves the joint Hamiltonian over the full interval with Jacobi–Anger polynomials. One round of oblivious amplitude amplification raises the success probability by applying the evolution circuit forward, in reverse and forward again, with two reflections about the success subspace[37].

**Table S12 | Gate counts and field accuracy for the adapted 64-cell BE–KvN wake circuit.**

| Logical resources | Count | Evolution and sampling | Value |
|---|---:|---|---:|
| Logical qubits | 25 | Full-interval evolutions | 1 |
| Arbitrary $R_Z$ rotations | 7,677,481 | Total block calls | 48,591 |
| Native T or T† gates | 511,970,388 | T gates per arbitrary rotation | 109 |
| Base Clifford gates | 732,189,482 | Field bias (%) | 0.08701 |
| Estimated total T gates | 1,348,815,817 | Accepted shots for 1% field error | 813 |

Normalization $\alpha\approx 17{,}695$, duration $T\approx 0.90901$ and tolerance $\varepsilon_{\mathrm{exp}}=10^{-4}$ give degree $d_{\mathrm{exp}}=16{,}197$.

The radial and phase contributions to the block normalization are

$$\alpha=\frac{\beta_{\mathrm{R}}^3-(1+3h^2/4)\beta_{\mathrm{R}}}{h}+2\kappa\beta_{\mathrm{R}}^2\simeq 17{,}417+278. \tag{S78d}$$

The first term bounds the corrected cubic (S78a) over the loader interval $|r|\le\beta_{\mathrm{R}}$ and divides by $h$. The second combines the squared-radius block normalization with $\kappa$ and the maximum total phase occupation, two.

We choose $d_{\mathrm{exp}}$ as the smallest polynomial degree $d$ satisfying

$$\varepsilon_{\text{tail}}(d) = 2\max_{\sigma\in\{0,1\}} \sum_{\substack{m>d \\ m \bmod 2=\sigma}} |J_m(\alpha T)|, \quad 6\sqrt{2}\,[\varepsilon_{\text{tail}}(d) + 10^{-12}] \le \varepsilon_{\text{exp}}. \tag{S78e}$$

Here $J_m$ is the Bessel function of the first kind of order $m$, and $\sigma$ selects the even or odd omitted orders. The factor $6\sqrt{2}$ accounts conservatively for rescaling and amplitude amplification; $10^{-12}$ is a numerical allowance. The left-hand side is $1.0750 \times 10^{-4}$ at degree 16,196 and $9.4999 \times 10^{-5}$ at degree 16,197; the latter is $\varepsilon_{\text{poly}}$ below.

Three passes each use 16,196 ordinary calls and one controlled call. Two selector qubits choose the cosine or sine polynomial and its phase-sequence branch. Together with the 22-qubit block and one shared clean work qubit, they give 25 logical qubits. Counts include preparation, readout and one arbitrary rotation for each adjustable phase outside the block. The fastest design uses $N_{\text{T}}^{\text{native}} + 109N_{R_Z}$ T gates per attempt. The task time in Table S10 uses the success probability of the ideal circuit, $P_s \approx 1 - 1.4 \times 10^{-9}$, in (S61b). The shot count uses a conservative lower bound, $(1 - \varepsilon_{\text{poly}})^2 \approx 0.99981$, in (S40), where $\varepsilon_{\text{poly}} \approx 9.5 \times 10^{-5}$ bounds the error of the polynomial approximation to the evolution.

The block-encoding normalization $\alpha$ is 671 times the spectral norm (largest singular value) of the joint generator. If an encoding attained that norm, we estimate 111 block calls and a total task time of 17.4 h at the same effective cost per call, synthesis precision, shot count, success probability and hardware architecture.

## Grid dependence of the BE–KvN call count

Grid refinement changes both cost per call and call count. In this construction, each QSVT transform uses a polynomial with only even or only odd powers. Linear and cubic terms change sign with the coordinate and share an odd transform. Squared radius keeps its sign and uses an even transform. A polynomial containing both requires two transforms. Each call[36] costs $O(N_{\text{term}}\, K\, n^2 + N_{\text{term}}\, m\, n)$ gates for $N_{\text{term}}$ terms, $n = \log_2 N_x$ bits to store a cell number, degree $K$ and derivative order $m$.

The coordinate loader assumes $\sum_j {x_j}^2 = 1$, where $x_j$ is the amplitude assigned to cell $j$. Setting $x_j$ to the physical cell centre $r_j$ cannot preserve this condition under uniform refinement of [0,1]. Enforcing it with nonuniform cells would concentrate centres near zero. Even two centres at $r \ge 0.9$ would contribute at least 1.62 to the sum, precluding refinement near the attracting radius $r = 1$. We therefore keep the uniform grid and rescale the amplitudes. Recovering the physical drift restores this scale in the polynomial and hence in the block normalization (S80a)–(S81b).

On the fixed interval, $r_j = (j + 1/2)h$ and $h = 1/N_x$ give $\sum_j {r_j}^2 = N_x/3 - 1/(12N_x)$. The oracle coordinate is $\tilde{r} = r/\beta_{\text{R}}$, where

$$\beta_{\text{R}} = \left(2\sum_{j=0}^{N_x-1} {r_j}^2\right)^{\frac{1}{2}} = \left(\frac{2N_x}{3} - \frac{1}{6N_x}\right)^{\frac{1}{2}} = \Theta\left({N_x}^{\frac{1}{2}}\right). \tag{S79}$$

For physical drift $f_K(r) = \sum_{k=0}^{K} a_k r^k$ of fixed degree $K \ge 1$ with bounded coefficients and a nonzero limiting leading coefficient, the transformed polynomial is

$$P_K(\tilde{r}) = \frac{f_K(\beta_{\text{R}}\tilde{r})}{C_K}, \tag{S80a}$$

$$C_K = \max_{|\tilde{r}|\le 1} |f_K(\beta_{\text{R}}\tilde{r})|. \tag{S80b}$$

The transform must stay within unit magnitude throughout [−1, 1], including outside the physical grid. This sets $C_K$. The first derivative contributes $1/h = N_x$, giving generator normalization $\lambda_K$.

$$C_K = \Theta(\beta_{\mathrm{R}}{}^K) = \Theta\left(N_x{}^{\frac{K}{2}}\right), \tag{S81a}$$

$$\lambda_K = \frac{C_K}{h} = \Theta\left(N_x{}^{1+\frac{K}{2}}\right). \tag{S81b}$$

For fixed term count and first derivative, each call costs $O(Kn^2)$ gates. Dividing the generator by $\lambda_K$ extends duration $T$ to $\lambda_K T$. At tolerance $\varepsilon_{\mathrm{exp}}$, degree $d_{\mathrm{exp}} = O(\lambda_K T + \log(1/\varepsilon_{\mathrm{exp}}))$ suffices, giving

$$G_{\mathrm{exp}} = O\left(Kn^2\left[TN_x{}^{1+\frac{K}{2}} + \log\left(\frac{1}{\varepsilon_{\mathrm{exp}}}\right)\right]\right). \tag{S82}$$

At fixed duration and tolerance, cubic-drift normalization contributes $\Theta(N_x{}^{5/2})$ to the call bound.

### Comparing gate counts

Counting Walsh bit parities (Note 3) for $N_{\mathrm{c}}$ coordinates, $n = \lceil \log_2 N_x \rceil$ bits each and degree $K$ gives the following upper bounds on the Walsh rotation counts per layer and per time step, respectively:

$$C_{\mathrm{layer}}^{\mathrm{rot}} = O((N_{\mathrm{c}} \log N_x)^K), \tag{S83a}$$

$$C_{\mathrm{step}}^{\mathrm{rot}} = O(N_{\mathrm{c}}(N_{\mathrm{c}} \log N_x)^K). \tag{S83b}$$

With at most $s_{\mathrm{c}} \leq N_{\mathrm{c}}$ coupled coordinates per component, each step uses $O(N_{\mathrm{c}}(s_{\mathrm{c}} \log N_x)^K)$ rotations costing $O(K+1)$ gates each. These bounds hold when the stencil and boundary treatment keep the rotation angles polynomial in the coordinates. At fixed dimension and fixed $K \geq 1$, including the decrement and increment gives $O(n^K + n)$ transport gates per step with incrementer B[9] and $O(n^K + n^2)$ with incrementer A before rotation synthesis. For $K = 1$, the bound is linear with incrementer B and quadratic with incrementer A. For $K \geq 2$, both give $O(n^K)$.

Accuracy sets the cubic wake's step count $N_t$. The following gate counts are for the full evolution before rotation synthesis, in order for Walsh–KvN, BE–KvN and optimized LCHS–KvN. Here $\mathbf{A}_{\mathrm{LCHS}}$ is the discretized evolution generator used by LCHS–KvN and $\|\mathbf{A}_{\mathrm{LCHS}}\|$ is its operator norm.

$$C_{\mathrm{Walsh}} = O(N_t n^3), \tag{S84a}$$

$$C_{\mathrm{BE}} = O\left(TN_x^{5/2} n^2\right), \tag{S84b}$$

$$C_{\mathrm{LCHS,opt}} = O(\|\mathbf{A}_{\mathrm{LCHS}}\| T \mathrm{poly}(n)). \tag{S84c}$$

For the cubic wake, the separate $r^2$ phase updates add $O(N_t n^2)$ gates over all steps. Cubic transport gives the leading bound in (S84a) with either incrementer. At dimensionless duration $T$, the optimized LCHS bound[33] assumes that each block call costs at most polynomially many elementary gates in the register width, with efficient selection and weight preparation, fixed precision and bounded retries. The LCHS representation[33] assumes that the Hermitian part of the generator is positive semidefinite. Novikau and Joseph[33] count each single-target multi-controlled gate as one operation; the bound in (S84c) counts elementary gates after decomposition.

## Supplementary Note 14 IBM hardware demonstration with MPS input

We test whether IBM hardware reproduces the outward transport predicted by $\frac{\mathrm{d}r}{\mathrm{d}t^*} = \mu(r - r^3)$ near saturation. After contrast correction, the probability in the two outermost cells rises from 0.798 at step one to 0.962 at step six; at step five it lies 3.7 bootstrap standard deviations below the ideal-circuit probability. A four-qubit MPS of bond dimension two prepares 16 cells (Note 5). Four or six equal steps span times 40–49.829. With three pair bits, the cubic drift uses all eight Walsh parities, comprising one constant, three single-bit terms, three two-bit terms and one three-bit term.

In Extended Data Fig. 1, panels a and c show the probability in the two outermost cells, which cover radii from 7/8 to 1. Panels b and d show the mean radius as the distribution moves towards the attracting state. For nonnegative cell probabilities, the largest possible mean is 31/32, the centre of the last cell.

Extended Data Fig. 1 uses 16,000 shots per circuit in the four-step run and 32,000 in the six-step run. Both jobs on ibm_cleveland (Heron r2) randomize gates and readout over 32 equivalent settings (twirling) and apply paired rotations about the positive and negative X axes during idle periods (XpXm dynamical decoupling). Of six compilation seeds, each circuit uses the one with the lowest estimated gate error.

## Noise mitigation and uncertainty

We assess how clearly the hardware reproduces the initial pattern of cell probabilities by running a reference circuit for $k$ propagation steps with all Walsh rotation angles set to zero. Its ideal output is the initial distribution $\boldsymbol{p}_0$. We fit the measured reference distribution $\boldsymbol{p}_{\text{ref,k}}$ by least squares to a mixture of $\boldsymbol{p}_0$ and the uniform distribution $\boldsymbol{u}$. The fitted weight of $\boldsymbol{p}_0$ is the reference contrast $F_k$. It equals one when the measurements reproduce the initial distribution and zero when all cells are equally likely.

$$F_k = \frac{(\boldsymbol{p}_0 - \boldsymbol{u})^{\mathrm{T}}(\boldsymbol{p}_{\text{ref,k}} - \boldsymbol{u})}{\|\boldsymbol{p}_0 - \boldsymbol{u}\|^2}. \tag{S85}$$

Assuming the reference and transport circuits have the same contrast, (S86) subtracts reference mean $O_{\text{ref,k}}$ from transport mean $O_k$, divides by contrast and restores input mean $O_0$.

$$O_{\text{corr,k}} = O_0 + \frac{O_k - O_{\text{ref,k}}}{F_k}. \tag{S86}$$

Measured points show standard errors from shot noise. For corrected points we draw 1,000 bootstrap samples, independently resampling the transport and reference counts in each, refit $F_k$ and plot the standard deviation. Step zero has only an input measurement. Bars exclude systematic bias and within-job correlations.

At step five, the corrected outer-cell probability is 0.04786 below the ideal circuit, with a bootstrap standard deviation of 0.01290. Unequal contrast between the reference and transport circuits is one possible systematic contribution to this residual. At step six the corrected mean radius is 0.998 (bootstrap standard deviation 0.009), above the largest cell centre, 31/32, because the correction permits negative weights (Extended Data Fig. 1d and its Source Data).

We checked the compiled gates by simulating them and comparing their amplitudes with direct cell-pair rotations. The amplitudes agree after removing an overall phase, which leaves measurement probabilities unchanged. The dashed lines show continuum means from classical trajectories. To initialize them, we spread each prepared cell probability uniformly within that cell. The hardware records give raw counts and circuits. The Source Data for Extended Data Fig. 1 contain the plotted 16-cell results and a separate 32-cell run over the first quarter of the interval (one step).

## Supplementary Note 15 Strang splitting error bounds

We prove that $N_t = O_\epsilon(N_x)$ steps suffice for coordinate-wise even–odd transport at fixed dimension, domain, input, duration and state-error tolerance $\epsilon$. Here $N_t$ counts steps and $N_x$ is the largest cell count along one coordinate. Error is the Euclidean norm of the difference between the circuit state and exact evolution on the same grid. The bound includes the radial transport–phase splitting; the three-coordinate circuit's Fourier-layer splitting and finite-mode truncation are outside its scope. We use the standard commutator form of the Strang error[38] and the state-vector strategy of An, Fang and Lin[39]; a related vector-norm analysis treats divergence-free transport (main-text ref. 11).

The proof has three steps. First, a double commutator applied to a grid vector with bounded two-cell difference quotients is $O(h^{-2})$, one power better than its operator norm, because neighbouring amplitudes cancel (S94). Second, this bound on the differences is preserved by exact-grid evolution and, for $\delta/h \le c_0$, by the layers inside one step (S98)–(S100). Third, inserting both into the Strang error integral gives local error $O(\delta^3 h^{-2})$ and, over $T/\delta$ steps, global error $O(T(\delta/h)^2)$ (S102)–(S104). We give the one-coordinate proof in full, then the changes for several coordinates, the radial phase and the piecewise-constant inputs of the refinement tests.

**Setting.** Divide $[0,1]$ into an even number $N_x \ge 4$ of cells of width $h = 1/N_x$, with cell $j = 0, \dots, N_x - 1$ centred at $x_j = (j + 1/2)h$. The real drift $F$ is twice continuously differentiable, with $M_0 = \| F \|_\infty$, $M_1 = \| F' \|_\infty$ and $M_2 = \| F'' \|_\infty$. Cells $j$ and $j + 1$ share the face at $(j + 1)h$ and are coupled with strength

$$a_j = \frac{F\big((j+1)h\big)}{2h}, \qquad |a_j| \le \frac{M_0}{2h}, \qquad |a_{j+m} - a_j| \le \frac{|m|\, M_1}{2}, \tag{S87}$$

where $m$ is an integer offset for which the indexed faces are defined; the last bound follows from the mean value theorem. The even layer $E_h$ collects the pairs with even $j$ and the odd layer $O_h$ those with odd $j$; each pair contributes the entries $G_{j,j+1} = -a_j$ and $G_{j+1,j} = a_j$ to its layer matrix $G$. Boundaries are either periodic, with periodic $F$ and indices modulo $N_x$, or closed, with $F(0) = F(1) = 0$. In the closed case (S87) gives $a_{-1} = a_{N_x - 1} = 0$, so no amplitude leaves the interval. Operators omit out-of-range couplings; the boundary contributions are treated below.

For duration $T$, step count $N_t$ and $\delta = T/N_t$, let

$$\Omega_h = E_h + O_h, \qquad U_h = e^{\delta \Omega_h}, \qquad S_h = e^{\delta E_h/2} e^{\delta O_h} e^{\delta E_h/2}. \tag{S88}$$

$U_h$ is one exact step on the grid and $S_h$ one circuit step, as in (S14). The generators are real and skew-symmetric, so every flow is unitary. For a grid vector $v \in \mathbb{C}^{N_x}$, with component $v_j$ the amplitude in cell $j$, the two-cell difference and the regularity norm are

$$(D_{2h} v)_j = \frac{v_{j+2} - v_j}{2h}, \qquad \| v \|_{1,h} = \| v \|_2 + \| D_{2h} v \|_2. \tag{S89}$$

The difference spans two cells because a layer rotates pairs: cell $j + 2$ sits in the same position of its pair as cell $j$. In the closed case $D_{2h} v$ has components $j = 0, \dots, N_x - 3$. Unlabelled norms are Euclidean vector norms or the induced operator norms, and $[A, B] = AB - BA$. Constants $C$ do not depend on the grid; subscripts list what they depend on.

**Operator-norm bound.** Each layer is a direct sum of two-state blocks, so $\| E_h \|, \| O_h \| \le M_0/(2h)$. With $\| [A, B] \| \le 2 \| A \| \| B \|$, each double commutator has norm at most $M_0^3/(2h^3)$. The Strang bound[38] then gives local error $O(\delta^3 h^{-3})$ and global error $O(T \delta^2 h^{-3})$, hence $N_t = O_\epsilon\left(N_x^{3/2}\right)$ for any input. The steps below remove one power of $h^{-1}$ for regular inputs.

**Step 1: cancellation in the double commutators.** Multiplying out $[E_h,[E_h,O_h]] = E_h^2 O_h - 2E_h O_h E_h + O_h E_h^2$ gives the interior rows (or all rows with periodic indices),

$$\begin{aligned}([E_h,[E_h,O_h]]v)_j &= 2a_j a_{j+1} a_{j+2}\, v_{j+3} - a_{j-1}\left(a_j^2 + a_{j-2}^2\right) v_{j-1}, \quad j \text{ even}\\ ([E_h,[E_h,O_h]]v)_j &= a_j\left(a_{j-1}^2 + a_{j+1}^2\right) v_{j+1} - 2a_{j-3} a_{j-2} a_{j-1}\, v_{j-3}, \quad j \text{ odd}\end{aligned}. \tag{S90}$$

For constant drift on a periodic grid, $a_j = a$, both coefficients in a row equal $2a^3$, and an even row is $2a^3\left(v_{j+3} - v_{j-1}\right)$: an $O(h^{-3})$ coefficient times a difference of amplitudes four cells apart, which is $O(h)$ for a regular vector. For variable drift, write the even row with $p_j = 2a_j a_{j+1} a_{j+2}$ and $q_j = a_{j-1}\left(a_j^2 + a_{j-2}^2\right)$ as

$$p_j\left(v_{j+3} - v_{j-1}\right) + \left(p_j - q_j\right) v_{j-1}. \tag{S91}$$

The first part is bounded as for constant drift, using $|p_j| \le M_0^3/(4h^3)$ and $v_{j+3} - v_{j-1} = 2h\left[(D_{2h}v)_{j+1} + (D_{2h}v)_{j-1}\right]$. In the second part the coefficients no longer cancel exactly, but their difference is one power of $h$ smaller than either. Set $a_* = M_0/(2h)$. Subtracting the products and bounding the three terms gives

$$\begin{aligned}p_j - q_j &= 2a_j a_{j+1}\left(a_{j+2} - a_{j-1}\right) + 2a_j a_{j-1}\left(a_{j+1} - a_j\right)\\ &\quad + a_{j-1}\left(a_j - a_{j-2}\right)\left(a_j + a_{j-2}\right),\\ \left|p_j - q_j\right| &\le 2a_*^2 \frac{3M_1}{2} + 2a_*^2 \frac{M_1}{2} + 2a_*^2 M_1 = \frac{3M_0^2 M_1}{2h^2}.\end{aligned} \tag{S92}$$

The three coupling differences are bounded by $3M_1/2$, $M_1/2$ and $M_1$, respectively, using (S87). Over the even rows $2 \le j \le N_x - 4$, or all even rows in the periodic case,

$$\| [E_h,[E_h,O_h]]v \|_{2,\text{even}} \le h^{-2}\left(M_0^3 \| D_{2h} v \|_2 + 3/2\, M_0^2 M_1 \| v \|_2\right). \tag{S93}$$

The odd rows have the same form, $p_{j-3}\left(v_{j+1} - v_{j-3}\right) + \left(q_{j+1} - p_{j-3}\right) v_{j+1}$, and the same bound. In the closed case at most two rows next to each end keep a single term, because $a_{-1} = a_{N_x-1} = 0$ removes its partner. Its coefficient is a product of three couplings within four faces of the endpoint, where $|a_j| \le 2M_1$ because $F$ vanishes there, so these rows contribute at most $C M_1^3 \| v \|_2$, without negative powers of $h$. Finally $[O_h,[E_h,O_h]] = -[O_h,[O_h,E_h]]$, and $[O_h,[O_h,E_h]]$ is given by (S90) with the even and odd rows exchanged. Since $h \le 1$,

$$\| [E_h,[E_h,O_h]]v \|_2 + \| [O_h,[E_h,O_h]]v \|_2 \le C_F\, h^{-2} \| v \|_{1,h}. \tag{S94}$$

**Step 2: regularity is preserved.** The bound (S94) helps only if $\|\cdot\|_{1,h}$ stays bounded along the evolution. Two kinds of flow matter: exact-grid evolution, which carries the state from step to step, and single layers, which act inside a step.

**Exact evolution.** Let $w(t) = e^{t\Omega_h} v$ and $z(t) = D_{2h} w(t)$. From $(\Omega_h w)_j = a_{j-1} w_{j-1} - a_j w_{j+1}$,

$$\begin{aligned}\dot z &= \tilde\Omega_h z + W_h w, &\quad \left(\tilde\Omega_h z\right)_j &= a_{j-1} z_{j-1} - a_{j+2} z_{j+1}\\ (W_h w)_j &= c_j w_{j+1}, &\quad c_j &= \frac{a_{j+1} + a_j - a_{j-1} - a_{j+2}}{2h}.\end{aligned} \tag{S95}$$

The differences are thus transported by $\tilde\Omega_h$, which resembles $\Omega_h$, and driven by the state through $W_h$. Neither makes $z$ grow quickly. $\tilde\Omega_h$ is nearly skew-symmetric: its symmetric part has off-diagonal entries $\left(a_j - a_{j+2}\right)/2$ of size at most $M_1/2$, with at most two per row. The maximum absolute row sum therefore bounds its spectral norm by $M_1$, giving $\mathrm{Re}\langle z, \tilde\Omega_h z\rangle \le M_1 \| z \|_2^2$. $W_h$ is bounded because $c_j$ is a second difference of $F$:

$$c_j = -\frac{1}{4h^2}\int_0^h \int_0^{2h} F''(jh + s + u)\, du\, ds, \qquad \left|c_j\right| \le \frac{M_2}{2}. \tag{S96}$$

Taking the inner product of (S95) with $z$, using $\| w(t) \|_2 = \| v \|_2$ and Grönwall's inequality,

$$\begin{aligned} &1/2\ \partial_t \| z \|_2^2 \leq M_1 \| z \|_2^2 + M_2/2 \| z \|_2 \| v \|_2 \\ \Rightarrow\quad &\| z(t) \|_2 \leq e^{M_1 t}(\| D_{2h} v \|_2 + t M_2/2 \| v \|_2). \end{aligned} \tag{S97}$$

Hence

$$\| e^{t\Omega_h} v \|_{1,h} \leq C_{F,T} \| v \|_{1,h}, \qquad 0 \leq t \leq T. \tag{S98}$$

**Single layers.** A layer $G \in \{E_h, O_h\}$ rotates each of its pairs: on the pair $(j, j+1)$, $e^{tG}$ is the real two-state rotation $R(ta_j)$. Let $u = (v_j, v_{j+1})^T$ and $u' = (v_{j+2}, v_{j+3})^T$ be neighbouring pairs. After the layer their difference is

$$R(ta_{j+2})\, u' - R(ta_j)\, u = R(ta_{j+2})\,(u' - u) + [R(ta_{j+2}) - R(ta_j)]\, u. \tag{S99}$$

The first term has the norm of $u' - u$. The second is small because neighbouring pairs rotate by nearly the same angle: $\| R(\alpha) - R(\beta) \| \leq |\alpha - \beta|$ and $|a_{j+2} - a_j| \leq M_1$. In the closed case an end cell left unpaired by $O_h$ is compared with the neighbouring pair in the same way, with rotation angle zero for the end cell; that pair's coupling is at most $M_1$ because $F$ vanishes at the endpoint. Dividing by $2h$ and summing over pairs,

$$\| D_{2h} e^{tG} v \|_2 \leq \| D_{2h} v \|_2 + C_F \frac{|t|}{h} \| v \|_2, \qquad \| e^{tG} v \|_{1,h} \leq \left(1 + C_F \frac{|t|}{h}\right) \| v \|_{1,h}. \tag{S100}$$

A layer of duration $|t| \leq \delta$ therefore preserves regularity uniformly in the grid when $\delta/h \leq c_0$, for a fixed constant $c_0 > 0$. This is where the condition on $\delta/h$ enters.

**Step 3: local and global error.** Write $A = E_h$, $B = O_h$ and $S_h(\tau) = e^{\tau A/2} e^{\tau B} e^{\tau A/2}$. Differentiating in $\tau$ and integrating twice gives the exact error of one step[38]:

$$\begin{aligned} S_h - U_h &= \int_0^\delta e^{(\delta - \tau)\Omega_h}\, K(\tau)\, S_h(\tau)\, d\tau, \\ K(\tau) &= -\frac{1}{4}\int_0^\tau \int_s^\tau e^{uA/2}\, [A, [A, B]]\, e^{-uA/2}\, du\, ds \\ &\quad -\frac{1}{2}\int_0^\tau \int_0^s e^{\tau A/2}\, e^{uB} [B, [A, B]]\, e^{-uB} e^{-\tau A/2}\, du\, ds. \end{aligned} \tag{S101}$$

Bounding the commutators in operator norm and integrating gives the weights $\delta^3/24$ and $\delta^3/12$ of the standard bound. Applied to a vector $v$, each double commutator in (S101) acts on $v$ after at most three layers of duration at most $\delta$: on $e^{(\tau-u)A/2} e^{\tau B} e^{\tau A/2} v$ in the first term and on $e^{(\tau-u)B} e^{\tau A/2} v$ in the second. By (S100) these vectors have regularity norm at most $(1 + C_F c_0)^3 \| v \|_{1,h}$, and all remaining factors are unitary. With (S94),

$$\| (S_h - U_h)\, v \|_2 \leq C_{F,c_0}\, \delta^3 h^{-2} \| v \|_{1,h}, \qquad \delta/h \leq c_0. \tag{S102}$$

Let $v_{\text{in}}$ be the initial grid state, with $\| v_{\text{in}} \|_2 = 1$ and $\| v_{\text{in}} \|_{1,h} \leq V_0$ for a constant $V_0$ independent of the grid. The telescoping identity[40]

$$S_h^{N_t} - U_h^{N_t} = \sum_{k=0}^{N_t - 1} S_h^{N_t - 1 - k}\, (S_h - U_h)\, U_h^k \tag{S103}$$

places the error of step $k+1$ on the exactly evolved state $U_h^k v_{\rm in}$, as in the proof of Theorem 4 of ref. 39. That state is regular by (S98), the step error is bounded by (S102), and the remaining circuit steps $S_h^{N_t-1-k}$ are unitary. Summing $N_t = T/\delta$ terms,

$$\epsilon_{\rm Trot} := \| \left(S_h^{N_t} - U_h^{N_t}\right) v_{\rm in} \|_2 \le N_t\, C_{F,c_0}\, \delta^3 h^{-2}\, C_{F,T}\, V_0 \\ = C_{F,T,V_0,c_0}\, T(\frac{\delta}{h})^2. \tag{S104}$$

A fixed ratio $\delta/h$ thus gives the same error bound on every grid, and $N_t = T/\delta$ grows in proportion to $N_x = 1/h$.

**Several coordinates.** In $N_{\rm c}$ coordinates on a rectangular domain with side lengths $L_\ell$, direction $\ell = 1, \dots, N_{\rm c}$ has an even number $N_\ell \ge 4$ of cells of width $h_\ell = L_\ell/N_\ell$; let $h_{\rm min} = \min_\ell h_\ell$ and $N_x = \max_\ell N_\ell$. Each drift component $F_\ell$ may depend on every coordinate and has bounded partial derivatives through order two; here $M_0 = \max_\ell \| F_\ell \|_\infty$. Each direction is periodic, or closed with zero normal drift, $F_\ell = 0$, on its two boundary faces. Cells carry a multi-index $\mathbf{j}$, with unit increment $\mathbf{e}_\ell$. Cells $\mathbf{j}$ and $\mathbf{j}+\mathbf{e}_\ell$ are coupled with strength $a_{\ell,\mathbf{j}} = F_\ell/(2h_\ell)$, evaluated at the centre of their shared face. The layer $G_{\ell,\sigma}$ collects the direction-$\ell$ pairs whose index $j_\ell$ has parity $\sigma \in \{0,1\}$, and

$$\Omega_h = \sum_{\ell=1}^{N_{\rm c}} \sum_{\sigma=0}^{1} G_{\ell,\sigma}, \qquad (D_\ell v)_{\mathbf{j}} = \frac{v_{\mathbf{j}+2\mathbf{e}_\ell} - v_{\mathbf{j}}}{2h_\ell} \\ \| v \|_{1,h} = \| v \|_2 + \sum_{\ell=1}^{N_{\rm c}} \| D_\ell v \|_2. \tag{S105}$$

Order the $2N_{\rm c}$ layers as $G_1, \dots, G_{2N_{\rm c}}$. The circuit step $S_h$ applies half steps of $G_1, \dots, G_{2N_{\rm c}-1}$, a full step of $G_{2N_{\rm c}}$ and the same half steps in reverse order; $U_h = e^{\delta\Omega_h}$.

**Reduction to two generators.** Let $\Omega_{>1} = \Omega_h - G_1$ and let $S_h{}'$ be the Strang step of the layers $G_2, \dots, G_{2N_{\rm c}}$. Then

$$S_h - U_h = e^{\delta G_1/2}\left[S_h{}' - e^{\delta\Omega_{>1}}\right]e^{\delta G_1/2} \\ + \left[e^{\delta G_1/2} e^{\delta\Omega_{>1}} e^{\delta G_1/2} - e^{\delta\Omega_h}\right]. \tag{S106}$$

The second bracket is a two-generator Strang error of the form (S101), with $A = G_1$ and $B = \Omega_{>1}$. Repeating the argument on the first bracket expresses $S_h - U_h$ through $2N_{\rm c} - 1$ such errors, the $m$th with $A = G_m$ and $B = \Omega_{>m} = \sum_{m'>m} G_{m'}$. As in one coordinate, three facts are needed: bounds on the double commutators, regularity under exact evolution, and regularity under the flows inside a step, which are now single layers and the partial sums $\Omega_{>m}$.

**Double commutators.** Expanding each $B = \Omega_{>m}$ gives $O\left(N_{\rm c}{}^3\right)$ nested commutators of three layers in total. If all three layers belong to one direction, the commutator acts along each line of cells in that direction, with the transverse coordinates as parameters, and (S94) applies line by line. Otherwise two of the layers belong to different directions. Layers of different directions pair different indices, so they would commute if their couplings were constant; their commutator contains only differences of couplings across the transverse direction. For a direction-$k$ face at position $\mathbf{x}$,

$$a_k(\mathbf{x} + h_\ell \mathbf{e}_\ell) - a_k(\mathbf{x}) = \frac{1}{2h_k} \int_0^{h_\ell} \partial_\ell F_k(\mathbf{x} + s\,\mathbf{e}_\ell)\, ds = O(h_\ell/h_k). \tag{S107}$$

Multiplying by a direction-$\ell$ coupling of size $O\left(h_\ell^{-1}\right)$ gives $O\left(h_k^{-1}\right)$, and exchanging the directions gives

$$\| \left[G_{\ell,\sigma}, G_{k,\sigma'}\right] \| \le C_F \left(h_\ell^{-1} + h_k^{-1}\right), \qquad \ell \ne k, \quad \sigma, \sigma' \in \{0,1\}. \tag{S108}$$

A further commutator with any layer multiplies by at most $M_0/h_{\min}$. If the two inner layers share a direction $k$ and the outer layer belongs to direction $\ell \neq k$, write $A_\ell = G_{\ell,\sigma}$, $E_k = G_{k,0}$ and $O_k = G_{k,1}$. Jacobi's identity moves the outer layer inside,

$$[A_\ell, [E_k, O_k]] = [[A_\ell, E_k], O_k] + [E_k, [A_\ell, O_k]], \qquad \ell \neq k, \tag{S109}$$

so that (S108) applies again. Every mixed commutator of three layers is therefore $O(h_{\min}^{-2})$ in operator norm, for any vector.

**Regularity.** For exact evolution $w(t) = e^{t\Omega_h}v$, write $\Omega_k = G_{k,0} + G_{k,1}$ and $z_\ell = D_\ell w$. The part $D_\ell \Omega_\ell w$ is given by (S95) along lines. For $k \neq \ell$, apply the discrete product rule $(D_\ell(aw))_{\mathbf{j}} = a_{\mathbf{j}+2\mathbf{e}_\ell}(D_\ell w)_{\mathbf{j}} + (D_\ell a)_{\mathbf{j}} w_{\mathbf{j}}$ for a coefficient field $a$. Set $b_+ = (D_\ell a_k)_{\mathbf{j}}$ and $b_- = (D_\ell a_k)_{\mathbf{j}-\mathbf{e}_k}$, and let $\Omega_k^{(+\ell)}$ denote direction-$k$ transport with couplings shifted by two cells in direction $\ell$. The identity $w_{\mathbf{j}+\mathbf{e}_k} - w_{\mathbf{j}-\mathbf{e}_k} = 2h_k(D_k w)_{\mathbf{j}-\mathbf{e}_k}$ then gives, at interior direction-$k$ cells,

$$\begin{aligned}(D_\ell \Omega_k w)_{\mathbf{j}} &= \left(\Omega_k^{(+\ell)} z_\ell\right)_{\mathbf{j}} + b_- w_{\mathbf{j}-\mathbf{e}_k} - b_+ w_{\mathbf{j}+\mathbf{e}_k} \\ &= \left(\Omega_k^{(+\ell)} z_\ell\right)_{\mathbf{j}} - 2h_k b_+ (D_k w)_{\mathbf{j}-\mathbf{e}_k} \\ &+(b_- - b_+) w_{\mathbf{j}-\mathbf{e}_k}.\end{aligned} \tag{S110}$$

The shifted generator is skew-symmetric, so its contribution to the derivative of the difference norm vanishes. The other two terms have bounded coefficients, $2h_k|b_+| \leq \| \partial_\ell F_k \|_\infty$ and $|b_- - b_+| \leq \| \partial_k \partial_\ell F_k \|_\infty/2$. At a closed direction-$k$ endpoint, the product-rule remainder is the single surviving term $-b_+ w_{\mathbf{j}+\mathbf{e}_k}$ or $b_- w_{\mathbf{j}-\mathbf{e}_k}$. For $\ell \neq k$, the normal drift $F_k$ vanishes along the whole closed face, so its tangential derivative $\partial_\ell F_k$ vanishes there too. At an interior face one cell width from that boundary, $|\partial_\ell F_k| \leq h_k \| \partial_k \partial_\ell F_k \|_\infty$. The surviving coefficient is therefore at most $\| \partial_k \partial_\ell F_k \|_\infty/2$, and the end rows obey the same bounds. Summing over $k$ and $\ell$ and applying Grönwall's inequality as for (S98),

$$\begin{aligned}\partial_t \sum_{\ell=1}^{N_c} \| D_\ell w \|_2 &\leq C_{F,N_c}\left(\sum_{\ell=1}^{N_c} \| D_\ell w \|_2 + \| w \|_2\right) \\ \Rightarrow \quad \| e^{t\Omega_h} v \|_{1,h} &\leq C_{F,N_c,T} \ \| v \|_{1,h}.\end{aligned} \tag{S111}$$

For the flows inside a step, let $\Lambda$ be a single layer or a partial sum $\Omega_{>m}$. Each of its layers $G$ satisfies $D_\ell G = \tilde{G}\, D_\ell + W$, with $\tilde{G}$ skew-symmetric and $\| W \| \leq C_F/h_{\min}$. Here $\tilde{G}$ uses the same pattern of disjoint pairs on the difference indices, with shifted face couplings; unpaired endpoint terms belong to $W$. Within the layer's own direction, the identity follows by differentiating the pair identity (S99) at $t = 0$. Across directions, the same discrete product rule used for (S110) gives remainder coefficients $D_\ell a_k = O(h_k^{-1})$. Summed over the layers of $\Lambda$, the shifted generator remains skew-symmetric. Denote the sums of the shifted generators and remainders by $\tilde{\Lambda}_\ell$ and $W_\ell$, respectively. Then $D_\ell \Lambda = \tilde{\Lambda}_\ell D_\ell + W_\ell$ and $\| W_\ell \| \leq C_{F,N_c}/h_{\min}$. Variation of constants gives the first line below. Unitarity bounds the integrand by the remainder norm times the conserved input norm; summing over directions and absorbing the fixed number of directions into the constant gives the last line:

$$\begin{aligned}D_\ell e^{t\Lambda} v &= e^{t\tilde{\Lambda}_\ell} D_\ell v + \int_0^t e^{(t-s)\tilde{\Lambda}_\ell} W_\ell e^{s\Lambda} v \, ds, \\ \| D_\ell e^{t\Lambda} v \|_2 &\leq \| D_\ell v \|_2 + \frac{C_{F,N_c}|t|}{h_{\min}} \| v \|_2, \\ \| e^{t\Lambda} v \|_{1,h} &\leq \left(1 + \frac{C_{F,N_c}|t|}{h_{\min}}\right) \| v \|_{1,h}.\end{aligned} \tag{S112}$$

Each two-generator error then satisfies (S102) with $h$ replaced by $h_{\min}$, for $\delta/h_{\min} \le c_0$: its double commutators are bounded by the same-direction or mixed estimates, and they act on vectors obtained from $v$ by $O(N_c)$ flows of duration at most $\delta$, which are regular by (S112). Summing the $2N_c - 1$ errors in (S106) and then the steps as in (S103)–(S104),

$$\epsilon_{\text{Trot}} \le C_{F,N_c,T,V_0,c_0}\; T(\frac{\delta}{h_{\min}})^2, \qquad \delta/h_{\min} \le c_0. \tag{S113}$$

This covers the coupled model (S19a)–(S19b).

**Radial phase.** In growth-rate time $\tau = \mu(t^* - 40)$, the radial model reads $dr/d\tau = r - r^3$ and $d\Phi/d\tau = \kappa r^2$, with $\kappa = \gamma/\mu$ from the fitted coefficients in (S8b); either sign of $\kappa$ is allowed. The step size $\delta$, duration $T$ and elapsed time $t$ below use this time scale. The drift $F(r) = r - r^3$ satisfies the closed-interval assumptions, with cell centres $r_j = x_j$. Let $\hat{n}_X$ and $\hat{n}_Y$ project onto $|1\rangle$ on the two phase qubits and let $I$ be their identity. Each phase qubit in $|1\rangle$ accumulates $\Phi$, so the phase generator and the full circuit step are

$$P_h = i\kappa\,\text{diag}(r_j^2) \otimes (\hat{n}_X + \hat{n}_Y), \qquad S_{\text{full}} = e^{\delta P_h/2}\,(S_h \otimes I)\,e^{\delta P_h/2}. \tag{S114}$$

If $\delta_*$ is the step in $t^*$, then $\delta = \mu\delta_*$ and $\kappa\delta = \gamma\delta_*$. Thus $S_{\text{full}} = U_f(\delta_*)$ in Methods equation (6b), with $D(\delta_*/2) = e^{\delta P_h/2}$. We compare $S_{\text{full}}$ with the exact step $U_{\text{full}} = e^{\delta(\Omega_h \otimes I + P_h)}$ through an intermediate step that uses exact transport:

$$\begin{aligned} S_{\text{full}} - U_{\text{full}} &= e^{\delta P_h/2}[(S_h - U_h) \otimes I]e^{\delta P_h/2} \\ &\quad + \left[e^{\delta P_h/2}(U_h \otimes I)\,e^{\delta P_h/2} - U_{\text{full}}\right]. \end{aligned} \tag{S115}$$

The first term is the even–odd transport error between two phase half steps. The second is the Strang error of splitting phase from exact transport.

For the second term, note that a diagonal matrix $D = \text{diag}(d_j)$ satisfies $[D, G]_{jk} = (d_j - d_k)\,G_{jk}$: the commutator is nonzero only because the two coupled cells accumulate phase at different rates. A layer $G \in \{E_h, O_h\}$ couples only neighbouring cells, with $|a_j| \le M_0/(2h)$, and neighbouring rates differ by $|r_{j+1}^2 - r_j^2| = h\,(r_{j+1} + r_j) \le 2h$. With $\| \hat{n}_X + \hat{n}_Y \| = 2$, the factors of $h$ cancel, giving $\| [P_h, G \otimes I] \| \le 2M_0|\kappa|$. Commuting once more with $P_h$ multiplies by another rate difference, and commuting with a transport layer multiplies by at most $M_0/h$. Hence, for $G, G' \in \{E_h, O_h\}$,

$$\| [P_h, [P_h, G \otimes I]] \| \le C_F\,\kappa^2, \qquad \| [G \otimes I, [P_h, G' \otimes I]] \| \le C_F\,|\kappa|\,h^{-1}. \tag{S116}$$

The standard Strang bound[38] then gives operator-norm local error $C_F\,\delta^3(\kappa^2 + |\kappa| h^{-1})$ and, over $T/\delta$ steps, global error $C_F\,T\delta^2(\kappa^2 + |\kappa|h^{-1}) \le C_F\,(\kappa^2 + |\kappa|)\,T(\delta/h)^2$, since $h \le 1$.

For the first term, (S102) applies once the phase half step and the exact combined evolution preserve regularity. Here $D_{2h}$ acts on the radial register. The phase factor $e^{i\kappa t r_j^2 m}$, with $m \in \{0,1,2\}$ the number of phase qubits in $|1\rangle$, differs between cells $j$ and $j+2$ by at most $8h|\kappa t|$, so $\| D_{2h} e^{tP_h} v \|_2 \le \| D_{2h} v \|_2 + 4|\kappa t|\ \| v \|_2$. For $w(t) = e^{t(\Omega_h \otimes I + P_h)} v$, (S95) gains the term $D_{2h} P_h w$: a skew-Hermitian diagonal operator acting on $z = D_{2h} w$, plus a remainder of norm at most $4|\kappa|\ \| w \|_2$. The Grönwall argument leading to (S98) therefore applies with $M_2/2$ replaced by $M_2/2 + 4|\kappa|$. Summing both terms of (S115) over the steps as in (S103)–(S104),

$$\| (S_{\text{full}}^{N_t} - U_{\text{full}}^{N_t})\,v_{\text{in}} \|_2 \le C_{F,T,V_0,c_0,\kappa}\; T(\frac{\delta}{h})^2, \qquad \delta/h \le c_0. \tag{S117}$$

**Fixed inputs.** The bounds above assume $\| v_{\mathrm{in}} \|_{1,h} \leq V_0$ on every grid. To see what this requires, represent a function $\psi$ on the grid by its cell average times $\sqrt{h}$,

$$(\Pi_h \psi)_j = h^{-1/2} \int_{jh}^{(j+1)h} \psi(x)\, dx. \tag{S118}$$

When $\psi$ equals a constant $c$ on the cell, the squared modulus of this amplitude is its cell probability, $\left|c\sqrt{h}\right|^2 = |c|^2 h$. For smooth $\psi$, let $x_j = (j + 1/2)h$ be the cell centre. A component is approximately $\psi(x_j)\sqrt{h}$, and $\psi$ changes by approximately $2h\psi'(x_j)$ across two cells. Thus $(D_{2h}\Pi_h\psi)_j \approx \psi'(x_j)\sqrt{h}$, and summing its squared magnitude gives the Riemann sum $\sum_j \left|\psi'(x_j)\right|^2 h$ for $\int |\psi'(x)|^2 dx$. Cauchy–Schwarz gives the exact inequality $\| D_{2h}\Pi_h\psi \|_2 \leq \| \psi' \|_{L^2}$ on every grid. Across a jump of height $\Delta$, however, the components differ by $\Delta\sqrt{h}$, so the difference quotient has magnitude $|\Delta|/(2\sqrt{h})$ and grows under refinement when $\Delta \neq 0$.

The refinement tests subdivide a fixed normalized piecewise-constant amplitude $\psi$ on grids aligned with its jumps. Then $v_{\mathrm{in}} = \Pi_h\psi$ is exactly the subdivided input, but its regularity norm is not bounded uniformly. We therefore compare with a smooth amplitude $\psi_\eta$, periodic in periodic directions, with $\| \psi - \psi_\eta \|_{L^2} \leq \eta$. Since $\Pi_h$ does not increase norms, $\| \Pi_h\psi \|_2 \leq \| \psi \|_{L^2}$, the grid vectors $\Pi_h\psi$ and $\Pi_h\psi_\eta$ differ by at most $\eta$. The circuit and the exact evolution are both unitary, so replacing the input changes each final state by at most $\eta$:

$$\epsilon_{\mathrm{Trot}} \leq C_\eta \left(\frac{\delta}{h_{\min}}\right)^2 + 2\eta, \qquad N_t = O_\epsilon(N_x). \tag{S119}$$

Choose $\eta \leq \epsilon/4$ first, which fixes $C_\eta$, then a grid-independent $\delta/h_{\min}$ small enough that the first term is at most $\epsilon/2$ and the local-step condition holds. Since $h_{\min}^{-1} = \max_\ell(N_\ell/L_\ell) = O(N_x)$, this gives $N_t = T/\delta = O_\epsilon(N_x)$. In several dimensions $\Pi_h$ is the tensor product of the one-dimensional projections, and a fixed normalized phase-register input is tensored with both amplitudes without changing these bounds. The constant depends on the fixed input, drift, dimension, domain, duration, phase coupling and tolerance. For the cubic radial drift, multiplying by the per-step Walsh bound gives $O_\epsilon(N_x(\log N_x)^3)$ transport rotations before synthesis.

## Supplementary references